\documentclass[aps,prb,twocolumn,floatfix,showpacs,longbibliography]{revtex4-2}

\usepackage{graphicx,amsmath,amsfonts,bm,mathtools}
\usepackage[colorlinks=true, citecolor=blue, linkcolor=blue, urlcolor=blue]{hyperref}
\usepackage[all]{hypcap}
\usepackage{bbold}
\usepackage{amssymb}
\usepackage{mathrsfs}
\usepackage{dutchcal}
\usepackage{clrscode3e,etoolbox}
\usepackage{varwidth}
\usepackage{newfloat}
\usepackage{color}
\usepackage{braket}
\usepackage[caption=false,subrefformat=parens,labelformat=empty]{subfig}
\DeclareCaptionType{Algorithm}
\usepackage[normalem]{ulem}

\usepackage[usenames,dvipsnames,svgnames,table]{xcolor}
\newcommand\sbullet[1][.5]{\mathbin{\vcenter{\hbox{\scalebox{#1}{$\bullet$}}}}}

\DeclareMathOperator {\tr}{\text{Tr}}

\begin{document}

\author{Claudia Artiaco}
\thanks{These alphabetically ordered authors contributed equally.}
\author{Christoph Fleckenstein}
\thanks{These alphabetically ordered authors contributed equally.}
\author{David Aceituno Chávez}
\author{Thomas Klein Kvorning}
\author{Jens H. Bardarson}

\affiliation{Department of Physics, KTH Royal Institute of Technology, Stockholm 106 91, Sweden}

\title{Efficient Large-Scale Many-Body Quantum Dynamics via Local-Information\\Time Evolution}

\begin{abstract}
During time evolution of many-body systems entanglement grows rapidly, limiting exact simulations to small-scale systems or small timescales.
Quantum information tends however to flow towards larger scales without returning to local scales, such that its detailed large-scale structure does not directly affect local observables.
This allows for the removal of large-scale quantum information in a way that preserves all local observables and gives access to large-scale and large-time quantum dynamics.
To this end, we use the recently introduced \textit{information lattice} to organize quantum information into different scales, allowing us to define \textit{local information} and \textit{information currents} which we employ to systematically discard long-range quantum correlations in a controlled way.
Our approach relies on decomposing the system into subsystems up to a maximum scale and time evolving the subsystem density matrices by solving the subsystem von Neumann equations in parallel.
Importantly, the information flow needs to be preserved during the discarding of large-scale information.
To achieve this without the need to make assumptions about the microscopic details of the information current, we introduce a second scale at which information is discarded while using the state at the maximum scale to accurately obtain the information flow.
The resulting algorithm, which we call local-information time evolution (LITE), is highly versatile and suitable for investigating many-body quantum dynamics in both closed and open quantum systems with diverse hydrodynamic behaviors.
We present results for energy transport in the mixed-field Ising model and magnetization transport in an open $XX$ spin chain where we accurately determine the diffusion coefficients.
The information lattice framework employed here promises to offer insightful results about the spatial and temporal behavior of entanglement in many-body systems.
\end{abstract}

\maketitle

\section{Introduction}

Simulating the time evolution of many-body quantum systems presents a significant challenge, often limiting the analysis to small-scale systems.
The obstacle lies in the rapid spreading of entanglement through the system during time evolution~\cite{calabrese2005evolution,schuch2008entropy,lauchli2008spreading}.
As entanglement can induce quantum correlations between arbitrarily distant-in-space degrees of freedom that cannot be decomposed into local parts, the exact representation of generic quantum states demands an exponentially large number of parameters.
This is related to the exponential growth of the Hilbert space with the system size.
However, for states exhibiting solely local correlations, such as local product states or Gibbs states, efficient representation becomes feasible.
This efficiency stems from the possibility of parametrizing them exclusively through local observables~\cite{cramer2010,Baumgratz2013,kim2014,klein2022time}.
Notably, the parametrization of the local observables entails a linear growth of the number of parameters with the system size, ensuring scalability.

In the paradigmatic case of thermalizing dynamics, local states are obtained at short evolution times when starting from an initially low-entangled (often product) state.
During this time, exact time evolution with matrix product states is feasible~\cite{White1992,Rommer1997,Vidal2003,Vidal2004,White2004,PAECKEL2019}.
While at long times the full pure state has volume-law entanglement, the reduced density matrices of small subsystems are thermal, consistent with the eigenstate thermalization hypothesis~\cite{DAlessio2016}, and are well described by local Gibbs states with maximum entropy subject to constraints such as the energy density in the initial state~\cite{Prosen2008a, Molnar2015}. 
Such high-temperature thermal states can be efficiently described by pure states via purification that introduces ancillary degrees of freedom~\cite{Verstraete2004, Feiguin2005,Barthel2009,Karrasch2012,Karrasch2013,Hauschild2018}.
At intermediate times, this representation is however inefficient, again because of large entanglement in the purified state.
The hope of interpolating between these two efficient descriptions is thus complicated by the presence of an \textit{entanglement barrier} at intermediate times separating the two limits, excluding efficient and exact large-scale simulations on current classical computers.

Seeking to bypass the entanglement barrier, various approximate algorithms have been proposed providing access to quantum dynamics of interacting systems beyond the system sizes accessible by exact dynamics.
The essential idea connecting different approximate time-evolution approaches is to only keep track of the most relevant features in order to represent quantum states with less than exponential (in system size) degrees of freedom. 
In the time-dependent variational principle approach, an approximate time evolution is obtained by projecting the time-evolved state into a given fixed subspace of the Hilbert space. 
A common example is that spanned by matrix product states of a fixed bond dimension~\cite{haegeman2011TDVP,haegeman2016TDVP,leviatan2017quantum,kloss2018time}, or related approaches that use a neural network ansatz for the wave function~\cite{Carleo2017,Schmitt2020,Lopez2020}.
Other approaches adopt dynamical quantum typicality~\cite{richter2019ClusterExpansion,Heitmann2020}.
Most related to our work are approaches based on density matrix product operators~\cite{white2018MPO,Pollmann2022}, approaches that trade entanglement for mixture~\cite{surace2019simulating,frias2023converting}, cluster truncated Wigner approaches~\cite{wurtz2018cluster}, and other discussion of the entanglement barrier~\cite{pastori2019disentangling,rams2020breaking}.

Here, we further develop the local-information approach for time evolution introduced in Ref.~\onlinecite{klein2022time}. 
Central to this approach is the fundamental question: how does the emergence of long-range entanglement during the entanglement-barrier regime impact local density matrices and, consequently, physical observables?
We know that at late times local density matrices closely resemble thermal density matrices.
Since there is a limited amount of information in thermal states, most of the correlations in the steady state are found on large scales of the order of the system size.
As a result, during time evolution an inherent statistical drift of quantum correlations occurs, which is consistent with the principles of the second law of thermodynamics for entanglement entropy~\cite{gogolin2016equilibration}.
This implies that information flows from smaller to larger scales, bounded solely by the system size, and generally does not return to influence the local density matrices.
Essentially, the primary role of the large-scale entanglement is to make the local density matrices mixed and thermal. 
Since there are many different long-range entanglement structures that give rise to the same local reduced density matrices, the pivotal idea of the local-information approach is to systematically discard long-range entanglement once information has reached a sufficiently large scale. 
To make this more concrete, a proper definition of information is essential, allowing us to identify its location and scale.
This crucial step has been pursued in Ref.~\onlinecite{klein2022time} by introducing the concepts of local information and the information lattice (see Secs.~\ref{sec:subsystem_lattice} and \ref{sec:subsystem_time_evolution} for a recap).

Our approach---which we refer to as local-information time evolution (LITE)---combines two essential ideas.
First, we divide the system into smaller subsystems, each characterized by a scale $\ell$ and a center $n$, which denote its extent (that is, the number of neighboring physical sites it encompasses) and position on the lattice.
Two subsystems, say subsystem $A$ and $B$, can be independently evolved in time as long as the quantum state of the combined subsystem $AB$ is not entangled. 
In this scenario, the equations of motion for the subsystems are self-contained, allowing for exact time evolution.
This self-containment is achieved by reconstructing the state of $AB$ from the individual states of $A$ and $B$ using recovery maps (see Sec.~\ref{sec:PRM}).
During time evolution, the scale of those combined subsystems $AB$ that lack entanglement continues to grow as more distant-in-space physical degrees of freedom become entangled (see Sec.~\ref{sec:exact_Petz_map_evolution}).
This growth permits the application of this reconstruction scheme only up to a time $\tau \sim \ell_\mathrm{max}/v_{\mathrm{LR}}$, where $v_{\mathrm{LR}}$ is the Lieb-Robinson speed~\cite{LiebRobinson} and $\ell_\mathrm{max}$ is the maximum manageable subsystem scale achievable through exact numerical techniques.
Second, we extend time evolution beyond times of about $\tau$ by implementing a truncation scheme that maintains entanglement spread without further increasing the subsystem scale. 
To achieve this, one needs to time evolve the subsystem density matrices on scale $\ell_\mathrm{max}$ such that the subsystems' states at lower scales and the flow of information are not altered. 
The truncation scheme proposed in Ref.~\onlinecite{klein2022time} accomplishes this by introducing a boundary condition for the information flow at scale $\ell_\mathrm{max}$ suitable for systems characterized by significant chaos and approximate translation invariance. 
To transcend these assumptions, it becomes necessary to construct boundary conditions on a case-by-case basis, thereby limiting the general applicability of that algorithm.

In this work, we devise a modified time-evolution algorithm that eliminates the necessity for assumptions about the information flow while maintaining it unaltered (see Sec.~\ref{sec:intro_LITE}). 
We introduce an additional length scale $\ell_\mathrm{min} < \ell_\mathrm{max}$, and we deliberately remove information on large scales $\ell \geq \ell_\mathrm{min}$ (see Secs.~\ref{sec:minimization-with-constraints} and \ref{sec:numerical-parameters}).
To correctly capture the information flow, the removal of information is done by minimizing information on scale $\ell_\mathrm{min}$ under certain constraints determined by the state of the subsystems at all scales $\ell \leq \ell_\mathrm{max}$. 
The resulting LITE algorithm preserves all local constants of motion with a range of $\ell \leq \ell_\mathrm{min}$ physical lattice sites.
By circumventing the need for assumptions on the flow of information, the algorithm can be applied across a wide variety of models, potentially including those with unknown hydrodynamic behavior.

As a benchmark example, we apply the algorithm to translation-invariant spin chains---specifically, the mixed-field Ising model.
By injecting a finite amount of energy in a small spatial region of the system, we investigate the energy transport in an infinitely extended system up to timescales much longer than previously obtained (see Sec.~\ref{sec:numerical_simulations}).
Our results are compatible with those obtained in other works for the same model~\cite{Pollmann2022,thomas2023comparing,wang2023diffusion}.
We carefully examine the convergence of our algorithm for relevant numerical parameters such as $\ell_\mathrm{min}$ and $\ell_\mathrm{max}$.
Importantly, LITE is also well suited to simulate the dynamics of dissipative systems governed by the Lindblad master equation in the presence of local dissipators (see Sec.~\ref{sec:dissipative_dynamics}).
The reason is that local dissipators selectively remove information enhancing the convergence properties of the algorithm.
We demonstrate this by simulating the magnetization transport in the $XX$ spin chain subject to local dephasing. 
Perfect convergence of the diffusion constant to the analytic result in Refs.~\cite{znidarivc2010exact, turkeshi2021diffusion,jin2022exact} is achieved already with truncation scales $(\ell_\mathrm{min}, \ell_\mathrm{max})$ for which the algorithm computational complexity is considerably smaller than the capacities of modern computers.

\section{Exact time evolution on the subsystem lattice}
\label{sec:exact_time_evolution}

\subsection{Subsystems, the subsystem lattice, and the information lattice}
\label{sec:subsystem_lattice}

To time evolve density matrices for large systems and long timescales, we decompose the entire system into smaller subsystems and solve the corresponding time evolution on each subsystem in parallel.
To achieve this task, we introduce the \textit{subsystem lattice}.

\begin{figure}[t]
\centering
\includegraphics[scale=0.5]{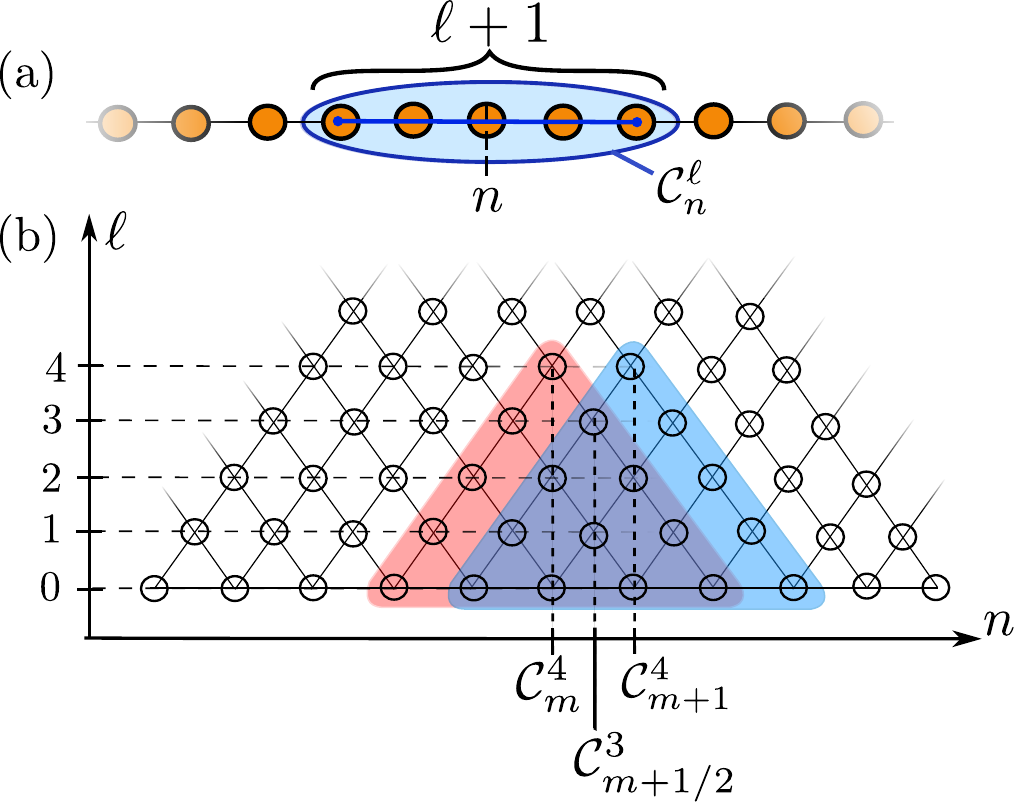}
\caption{(a) Schematic of the subsystem $\mathcal{C}_n^\ell$ extending over $\ell +1$ physical sites and centered around $n$. (b) Schematic of the subsystem lattice. 
The vertical axis labels different levels $\ell$, while the horizontal axis labels the subsystem centers $n$: $n$ takes integer values on even levels $\ell$ and half-integer values on odd levels. 
The red and blue colored regions depict the triangles associated with the subsystems $\mathcal{C}_m^{4}$ and $\mathcal{C}_{m+1}^4$, respectively, where $m \in \mathbb{Z}$ labels the physical site index.
Such triangles remark that a subsystem $\mathcal{C}^\ell_n$ contains the lower-level subsystems defined on a subset of the contiguous physical sites $\mathcal{C}^\ell_n$ includes.
The two neighboring subsystems $\mathcal{C}_m^{4}$ and $\mathcal{C}_{m+1}^4$ share the subsystem $\mathcal{C}_{m+1/2}^3$ (in purple).
}
\label{fig:subsys_lattice}
\end{figure}

Let us consider a system composed of a chain of sites, each representing some physical degree of freedom.
We define the subsystem $\mathcal{C}_n^{\ell}$ as the set of $\ell + 1$ contiguous physical sites centered around $n$. 
In this convention, subsystems with $\ell=0$ describe single physical sites.
From the pictorial representation in Fig.~\ref{fig:subsys_lattice}(a), we see that if $\ell$ is odd $n$ is a half-integer; e.g., a subsystem constituting two sites $m$ and $m+1$ (where $m \in \mathbb{Z}$ is the physical site index) is denoted as $\mathcal{C}_{m+1/2}^{1}$. If $\ell$ is even, $n$ is an integer; e.g., $\mathcal{C}_{m}^{2}$ indicates the subsystem composed of the three sites $m-1$, $m$, and $m+1$.

Each subsystem $\mathcal{C}_n^{\ell}$ is uniquely determined by the labels $(n,\ell)$. 
We order $(n,\ell)$ in a two-dimensional triangular structure, shown in Fig.~\ref{fig:subsys_lattice}(b), that we call the \textit{subsystem lattice}.
Black circles represent the subsystem-lattice points $(n,\ell)$. 
The horizontal axis of the subsystem lattice corresponds to the subsystem center $n$, while the vertical axis increases with the number of physical sites within subsystems. 
In the following, we refer to $\ell$ as ``level'' or ``scale''.
For a finite system of size $L$, the subsystem lattice is a triangle of base and height of length $L$, as there are fewer and fewer possible subsystems for increasing values of $\ell$. 
By increasing $\ell \rightarrow \ell + \ell'$ there are $\ell'$ fewer subsystems on level $\ell + \ell'$ compared with level $\ell$. 
The base of the subsystem lattice corresponds to $\ell = 0$ and the topmost label to $\ell = L-1$.

The subsystem lattice is a hierarchical structure: higher-level subsystems contain a triangle of lower-level subsystems that extends all the way down to level zero. 
Fig.~\ref{fig:subsys_lattice}(b) illustrates two neighboring subsystems and the corresponding hierarchy by means of the red and blue triangles.
The subsystem with label $(n,\ell)$ contains at one lower lever the two subsystems with labels $(n-1/2,\ell-1)$ and $(n+1/2,\ell-1)$. 
The subsystem $(n-1/2,\ell-1)$, for example, in turn contains at the next lower level the two subsystems $(n-1,\ell-2)$ and $(n,\ell-2)$, which are of course also subsystems of the top level $(n,\ell)$.
This hierarchy continues all the way down to level zero. 
Moreover, two neighboring subsystems at level $\ell$, say $(n,\ell)$ and $(n+1,\ell)$, share some lower-level subsystems starting with $(n+1/2,\ell-1)$ [see the red, blue, and purple triangles in Fig.~\ref{fig:subsys_lattice}(b)].

So far, the subsystem lattice is just a collection of labels of subsystems.
We wish to endow this lattice with a further structure by associating these labels with quantum states and quantum information.
To that end, we define $\bar{\mathcal{C}}_n^{\ell}$ as the complement of $\mathcal{C}_n^{\ell}$, i.e., the set of all the physical sites that do not belong to $\mathcal{C}_n^{\ell}$.
We then define the subsystem density matrix as
\begin{equation}
    \rho_n^{\ell} \coloneqq \tr_{\bar{\mathcal{C}}_n^{\ell}}(\rho),
\end{equation}
and the subsystem Hamiltonian as~\footnote{
The definition of the subsystem Hamiltonian $H_n^\ell$ in Eq.~\eqref{eq:subsystem-hamiltonian} assumes that the full-system Hamiltonian is traceless. In the case of a finite trace Hamiltonian, one has to shift the local Hamiltonians by a constant to ensure that the sum of the local energies of non-overlapping subsystems adds up to the total energy---expressed as $\tr(H\rho)=\sum_{n\in S}\tr(H_n^\ell \rho_n^\ell)$ where $\cup_{n\in S}\mathcal C_n^\ell$ represents the entire system and $\mathcal C_n^\ell\cap \mathcal C_{n^\prime}^\ell=0$ for $n, n^\prime\in S$ and $n \neq n^\prime$. Importantly, these constant shifts would not affect the time evolution.
}
\begin{equation}
    \label{eq:subsystem-hamiltonian}
	H_n^\ell \coloneqq \frac{\tr_{\bar{\mathcal{C}}_n^{\ell}}(H)}{D(\bar{\mathcal{C}}_n^{\ell})},
\end{equation}
where $\tr_{\bar{\mathcal{C}}_n^{\ell}}$ is the trace operator over the complement $\bar{\mathcal{C}}_n^{\ell}$, $D(\bar{\mathcal{C}}_n^{\ell})$ is  the Hilbert space dimension of $\bar{\mathcal{C}}_n^{\ell}$, and $\coloneqq$ denotes ``defined to be equal to''.
To ease the notation, let us assume that the Hilbert space dimension is the same for all the physical sites: $D(\mathcal{C}^0_n) = d$ for any $n$.
Furthermore, we define the von Neumann information of the subsystem density matrix $\rho_n^{\ell}$ as
\begin{equation}
\label{eq:vonNeumann_information}
I_n^{\ell} \coloneqq I(\rho^\ell_n) = \ln(d^{\ell+1})+ \mathrm{Tr}\left(\rho_n^{\ell} \ln(\rho_n^{\ell})\right).
\end{equation} 
$I(\rho_n^\ell)$ quantifies the amount of information in the state $\rho_n^\ell$ of the subsystem $\mathcal{C}_n^\ell$~\cite{von2013mathematische}. 
Thus, if $(\rho^\ell_n)^2 = \rho^\ell_n$ then in principle we have access to $\ln (d^{\ell+1})$ bits of information. 
Instead, if $\rho^\ell_n \propto \mathbb{1}_{d^{\ell+1}}$ (where $\mathbb{1}_{d^{\ell+1}}$ is the identity matrix of dimension $d^{\ell+1} \times d^{\ell+1}$) we have access to 0 bits of information.

We can now associate $\rho_n^\ell$, $H_n^\ell$, and $I_n^\ell$ with the subsystem-lattice label $(n,\ell)$.
These quantities are all \textit{global} for the subsystem $\mathcal{C}_n^\ell$, as they comprise (via partial trace) the same quantities on the lower-level subsystems contained in the triangle having as topmost site $(n,\ell)$ and as base the $\ell+1$ contiguous physical sites constituting the subsystem [see Fig.~\ref{fig:subsys_lattice}(b)].

One can also have \textit{local} quantities that are instead exclusively assigned to a single label $(n, \ell)$. 
By knowing the value of a local quantity on $(n, \ell)$, one cannot infer any information about its value on any other point in the lattice. 
For our purposes, a central example of a local quantity is mutual information.
This is defined as the information in the state of the union system $AB \coloneqq A\cup B$ that is neither in the state of $A$ nor of $B$:
\begin{equation}
\label{eq:mutual_information}
i(A; B) \coloneqq I(\rho_{AB})-I(\rho_A)-I(\rho_B)+I(\rho_{A\cap B}),
\end{equation}
with $A\cap B$ is the overlap region of $A$ and $B$. As in Ref.~\onlinecite{klein2022time}, we are interested in the \textit{local information} of the state of the subsystem $\mathcal{C}_n^{\ell}$ that is not present in any of the lower-level states of $\mathcal{C}_{n-1/2}^{\ell-1}$ or $\mathcal{C}_{n+1/2}^{\ell-1}$:
\begin{equation}
\label{eq:local_information}
i^\ell_n \coloneqq
    i(\mathcal{C}_{n-1/2}^{\ell-1}; \mathcal{C}_{n+1/2}^{\ell-1}).
\end{equation}
Here, it is implicit that the von Neumann information of empty subsystems is zero.
When we endow the subsystem lattice with the local information, we refer to it as the \textit{information lattice}.

\begin{figure}
    \centering
    \includegraphics[scale=0.29]{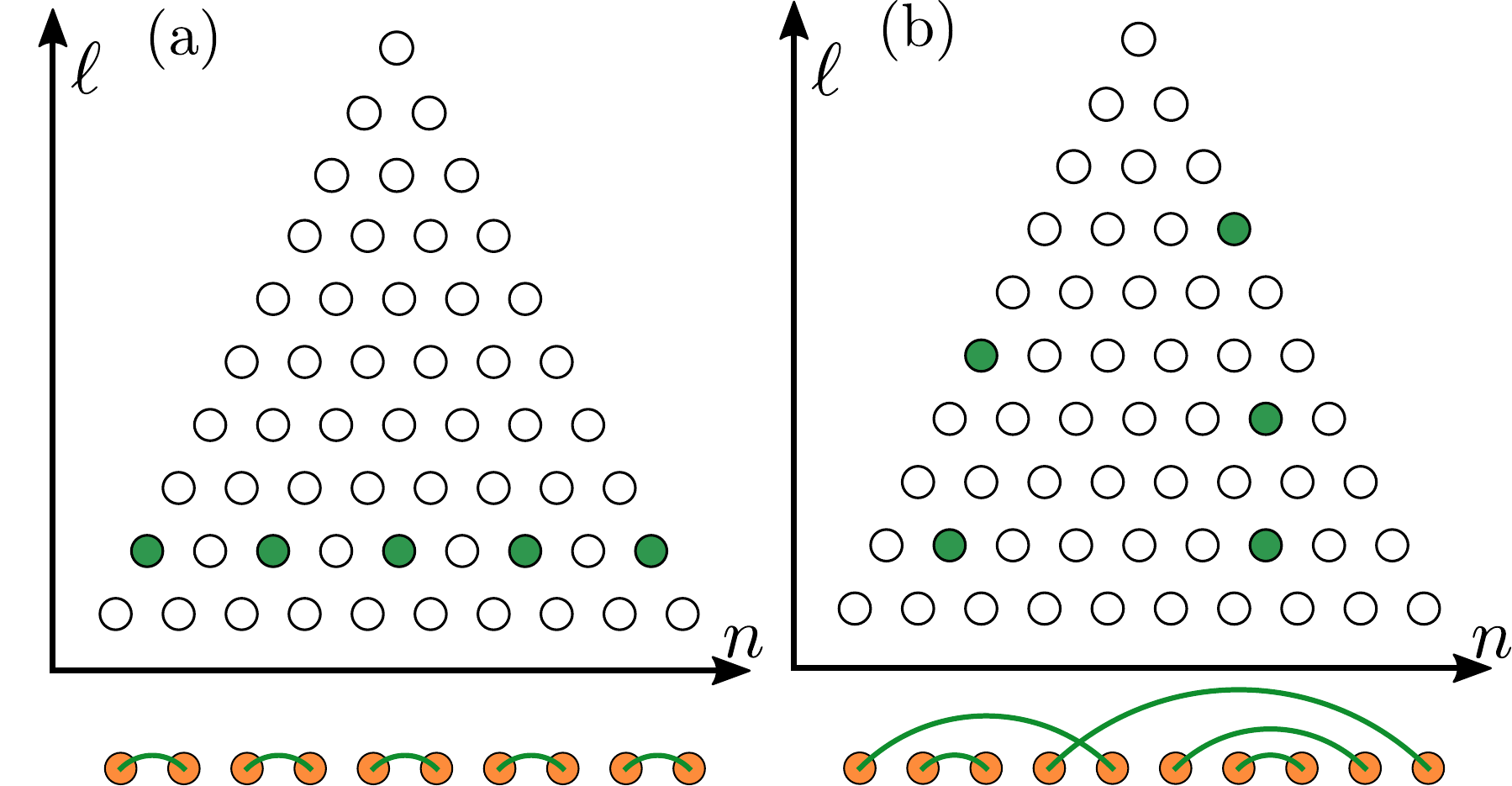}
    \caption{Schematic of local information [see Eq.~\eqref{eq:local_information}] on the information lattice. 
    The intensity of the colors quantifies the amount of local information: intense green corresponds to $i_n^\ell = \ln(4)$, white dots to $i_n^\ell=0$.
    (a) Local information for a singlet product state on nearest-neighbor sites (bottom orange dots with green connection lines) for $L=10$. Local information is located only on $\ell=1$ and every second lattice point $n$.     
    (b) Local information for the product state of random singlets depicted at the bottom. 
    }
    \label{fig:singlet}
\end{figure}

In Fig.~\ref{fig:singlet}, we plot the information lattice for two simple cases. 
Fig.~\ref{fig:singlet}(a) depicts the local information in the product state of singlets on consecutive pairs of sites (more intense colors correspond to larger amounts of local information). 
All the information is located on level $\ell=1$ where only every other lattice site $n$ carries nonzero local information, reflecting the singlet coupling.
In the case of random singlets where pairs are not necessarily between consecutive sites, as in Fig.~\ref{fig:singlet}(b), the configuration of green points containing information is rearranged. 
In both cases, $i_n^\ell = \ln(4)$ for $n = (m_0 + m_1)/2$ and $\ell = m_1 - m_0$ where $m_0$ ($m_1$) is the physical site index of the first (second) spin of the singlet; otherwise $i_n^\ell = 0$.

Importantly, local information is additive, and one can show~\cite{klein2022time} that the von Neumann information in the state of $\mathcal{C}^\ell_n$ is given by the sum of the local information $i^\ell_n$ on all the labels in the triangle with topmost site $(n,\ell)$; in equation form
\begin{equation}
    I^\ell_n = \sum_{(n,\ell) \in S^\ell_n} i^\ell_n
\end{equation}
with $S^\ell_n = \{ (n',\ell')\, | \,\mathcal{C}^{\ell'}_{n'} \subseteq \mathcal{C}^{\ell}_{n} \}$.

\subsection{Time evolution on the subsystem lattice}
\label{sec:subsystem_time_evolution}

To study time evolution on the subsystem lattice we need the equation of motion for the reduced density matrix $\rho^\ell_n$ defined on the subsystem $\mathcal{C}^\ell_n$. 
This is contained in the unitary time evolution of the full system density matrix $\rho$ given by the von Neumann equation ($\hbar = 1$)
\begin{equation}
\label{eq:full_vonNeumann}
    \partial_t \rho = -i [H, \rho],
\end{equation}
where $[\sbullet,\sbullet]$ denotes the commutator.
We consider a generic short-range, one-dimensional Hamiltonian $H$ with a maximum range of $r$ (if $r=0$, $H$ contains only onsite terms; for $r=1$ there are at most nearest-neighbor couplings; and so on). 
By tracing out the complement subsystem $\bar{\mathcal{C}}^\ell_n$ from Eq.~\eqref{eq:full_vonNeumann}, we obtain the equation of motion for the density matrix of the subsystem $\mathcal{C}^\ell_n$:
\begin{multline}
\label{eq:subsys_vonNeumann}
\partial_t \rho^{\ell}_n = -i \left[H^{\ell}_n, \rho_n^{\ell}\right] \\
-i \mathrm{Tr}_{L}^r\left(  \left[H_{n-r/2}^{\ell+r}- H_n^{\ell},\rho_{n-r/2}^{\ell+r} \right] \right) \\
-i \mathrm{Tr}_{R}^r\left(  \left[H_{n+r/2}^{\ell+r}- H_n^{\ell},\rho_{n+r/2}^{\ell+r} \right] \right), 
\end{multline}
with $\mathrm{Tr}_{L}^r$ ($\mathrm{Tr}_{R}^r$) is the partial trace operator tracing over the $r$ leftmost (rightmost) sites of the subsystem $\mathcal{C}^\ell_n$.
Here, it is implicit that, to subtract from $H_{n-r/2}^{\ell+r}$ ($H_{n + r/2}^{\ell+r}$) the lower-level density matrix $H_n^{\ell}$, one has to take the tensor product of the latter with a proper identity matrix; in this case, it reads $\mathbb{1}_{d^r} \otimes H_n^{\ell}$ ($H_n^{\ell} \otimes \mathbb{1}_{d^r}$).
Proper identity matrices to be used are evident from the specific equations. 
Thus, we use this convention throughout this work, unless identity matrices are explicitly stated.
The first term on the right-hand side of Eq.~\eqref{eq:subsys_vonNeumann} originates from all the terms of the Hamiltonian that have no overlap with $\bar{\mathcal{C}}_n^{\ell}$; in this case, the partial trace over $\bar{\mathcal{C}}_n^{\ell}$ is easily performed, leaving us only with the commutator of the subsystem Hamiltonian $H_n^\ell$ and subsystem density matrix $\rho_n^{\ell}$.
The second and third terms are given by the terms in the full Hamiltonian that couple $\mathcal{C}_n^\ell$ to $\bar{\mathcal{C}}_n^{\ell}$: if the full Hamiltonian $H$ has coupling terms with maximum range $r$, then $\mathcal{C}^\ell_n$ and $\bar{\mathcal{C}}^\ell_n$ are coupled via $H_{n-r/2}^{\ell+r}$ and $H_{n+r/2}^{\ell+r}$. 
Note that, to avoid double counting, one has to subtract $H_n^{\ell}$ in both the second and third terms.

The time evolution in Eq.~\eqref{eq:subsys_vonNeumann} can be visually understood with the help of Fig.~\ref{fig:subsys_lattice}(b) for the case $r=1$.
The time evolution of the state of subsystem $\mathcal{C}^3_{m+1/2}$ (purple triangle) is determined by the subsystem Hamiltonian $H^3_{m+1/2}$ (containing all the terms within the physical sites at the bottom of the purple triangle), and by the Hamiltonian terms of range $r=1$ that couple those physical sites with the two (left and right) nearest neighbors.
Such coupling terms, in the framework of the subsystem lattice, are included in the subsystem Hamiltonians of $\mathcal{C}^{4}_m$ (topmost site of the red triangle) and $\mathcal{C}^4_{m+1}$ (topmost site of the blue triangle).

Eq.~\eqref{eq:subsys_vonNeumann} implies that, for solving the time evolution of the subsystem density matrix $\rho^\ell_n$, one also needs to solve the time evolution for the higher-level subsystem density matrices $\rho^{\ell + r}_{n-r/2}$ and $\rho^{\ell + r}_{n+r/2}$. 
This gives rise to a hierarchy of equations of motion that, in principle, only closes when one solves the time evolution of all the subsystems or, equivalently, the full-system equation~\eqref{eq:full_vonNeumann}. 
To make this point clearer, let us consider the situation in which we know the subsystem density matrices for any $\ell$ and $n$ at the initial time $t = 0$, and we want to solve the time evolution of the subsystems at level $\ell^*$, denoted as $\mathcal{C}^{\ell^*}$. 
Then, for the infinitesimal time increment $\delta t$, we can use Eq.~\eqref{eq:subsys_vonNeumann} to compute the exact, time-evolved density matrices $\rho^{\ell^*}_n(\delta t)$. 
However, to further increase time by $\delta t$ and obtain the exact $\rho^{\ell^*}_n(2\delta t)$, we need to know the higher-level density matrices   $\rho^{{\ell^*} + r}_{n-r/2}(\delta t)$ and $\rho^{{\ell^*} + r}_{n+r/2}(\delta t)$ at time $\delta t$. 
In summary, in general, the exact time evolution of the subsystem density matrices $\rho^{\ell^*}_n$ can only be obtained by knowing all the density matrices at each level at any time.

\subsection{Recovery of higher-level subsystem density matrices from lower-level ones}
\label{sec:PRM}

\begin{figure}[]
\centering
\includegraphics[scale=0.48]{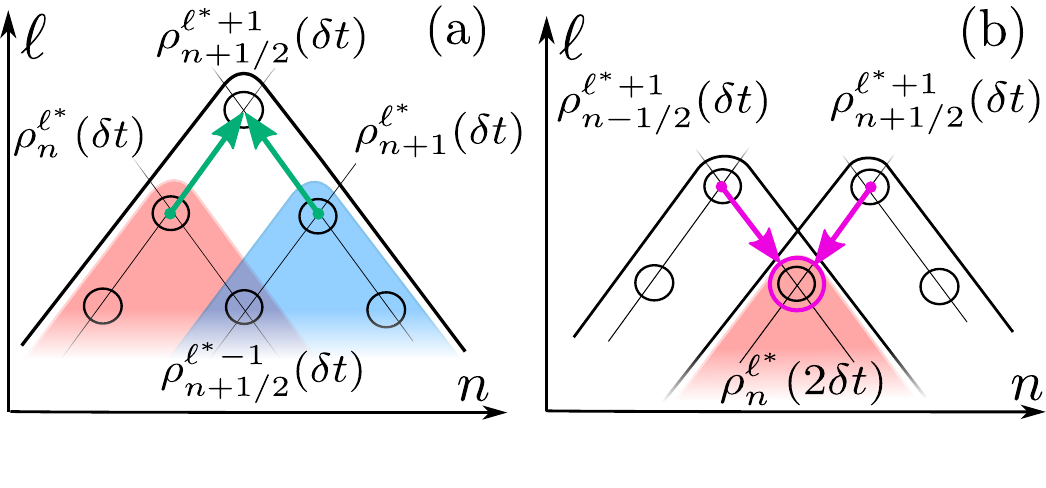}
\caption{(a) Schematic of the twisted Petz recovery map [see Eq.~\eqref{eq:Petz_map_subsys_lattice}]. The density matrices of two neighboring subsystems, $\rho^{\ell^*}_n$ and $\rho^{\ell^*}_{n+1}$, can be used to compute the density matrix of the higher-level subsystem, $\rho^{\ell^*+1}_{n+1/2}$, at the same time, as pictorially illustrated by the two green arrows. The recovery map is exact only if $i^{\ell^*+1}_{n+1/2}=0$.
(b) Schematic of the time evolution of $\rho^{\ell^*}_n$ by means of the Petz recovery map. This requires knowledge of the density matrices at level $\ell^* + r$, as represented by the magenta arrows and circle. Here, we illustrate the simplest non-trivial case with $r=1$.
}
\label{fig:Petz_map}
\end{figure}

Knowledge of the density matrices for all subsystems $\mathcal{C}^{\ell^*}$ on level $\ell^*$ allows us to construct the subsystem density matrices at all lower levels $\ell < \ell^*$ by suitable partial trace operations. 
The inverse operation is generally not possible, that is, recovering the density matrices on higher levels $\ell > \ell^*$ only from density matrices of level $\ell^*$.
As an illustrating example, consider a two-spin density matrix $\rho_{A B} = \vert \psi \rangle \langle \psi \vert$ with $\vert\psi\rangle$ the Bell state $\vert\psi\rangle = 1/\sqrt{2}\left( \vert \uparrow \rangle_A \vert \uparrow\rangle_B +\vert \downarrow \rangle_A \vert \downarrow\rangle_B \right)$.
In this case, the subsystem density matrices $\rho_{A}$ and $\rho_{B}$ are maximally mixed and take the form $\rho_{A}= \frac{\mathbb{1}_{A}}{D(A)}$, where $\mathbb{1}_{A}$ denotes the identity matrix on the Hilbert space of spin $A$ and $D(A)= 2$ denotes its dimension.
Clearly, $\rho_{AB} = \vert \psi \rangle \langle \psi \vert $ is not the only two-spin density matrix for which the subsystem density matrices $\rho_A$ and $\rho_B$ are maximally mixed.
The same result is, for instance, obtained for the maximally mixed two-spin density matrix, $\rho_{AB}=\frac{\mathbb{1}_{AB}}{D(AB)} $.
Therefore, in general, given only the lower-level density matrices $\rho_A$ and $\rho_B$, the correct density matrix of the enlarged system $\rho_{A B}$ cannot be determined.

An important exception is the case when there is no mutual information between $A$ and $B$, $i(A;B)=0$. 
In this case, it is possible to reconstruct the state $\rho_{AB}$ from the reduced states $\rho_A$ and $\rho_B$ via the so-called Petz recovery maps (see App.~\ref{app:sec:PRM}). 
An example is the twisted Petz recovery map~\cite{petz1986sufficient}, 
\begin{equation}
\label{eq:Petz_map}
\rho^\mathrm{TPRM}_{AB} \coloneqq \exp \left[ \, \ln (\rho_A) + \ln (\rho_B) - \ln (\rho_{A\cap B}) \, \right]. 
\end{equation}
For nonzero $i(A;B)$, the twisted Petz recovery map has the error bound~\cite{zhang2014lower}
\begin{equation}
\label{eq:Petz_error}
\mathrm{Tr}\sqrt{\left(\rho_{A B} - \rho^\mathrm{TPRM}_{AB} \right)^2} \leq 2 \sqrt{i(A;B)}.
\end{equation}

Recovery maps can be used to compute the state of higher-level subsystems on the subsystem lattice, as sketched in Fig.~\ref{fig:Petz_map}(a). 
Let us assume that we know the subsystem density matrices $\rho_n^{\ell^*}$ at level $\ell^*$ for any $n$, and that there is no local information on level $\ell^*+1$: $i^{\ell^*+1}= \sum_n i^{\ell^* + 1}_n = 0$. 
Then, all the density matrices at level ${\ell^*+1}$ can be computed by using the twisted Petz recovery map as 
\begin{equation}
\label{eq:Petz_map_subsys_lattice}
\rho_{n+1/2}^{\ell^*+1} = \exp \left[ \,  \ln \left(\rho_n^{\ell^*}\right)+ \ln \left(\rho_{n+1}^{\ell^*}\right) - \ln \left(\rho_{n+1/2}^{\ell^*-1}\right) \, \right]. 
\end{equation} 
Eq.~\eqref{eq:Petz_map_subsys_lattice} involves three subsystem density matrices: $\rho_n^{\ell^*}$ and $\rho_{n+1}^{\ell^*}$ that are known by assumption, and $\rho_{n+1/2}^{\ell^*-1}$ that is easily obtained by tracing out either the leftmost physical site from $\rho_n^{\ell^*}$ (i.e., $\rho_{n+1/2}^{\ell^*-1} = \tr^1_L(\rho_n^{\ell^*})$) or the rightmost physical site from $\rho_{n+1}^{\ell^*}$ (i.e., $\rho_{n+1/2}^{\ell^*-1} = \tr^1_R(\rho_{n+1}^{\ell^*})$).

Provided that there is no local information on higher levels as well, that is $i^{\ell^*+1}=i^{\ell^*+2}=\dotsb=i^{\ell^*+r}=0$, one can iterate recovery maps reconstructing all the density matrices at level $\ell^* + r$.

\subsection{Exact time evolution on the subsystem lattice via Petz recovery maps}
\label{sec:exact_Petz_map_evolution}

\begin{figure}[]
\centering
\includegraphics[width=0.49\columnwidth]{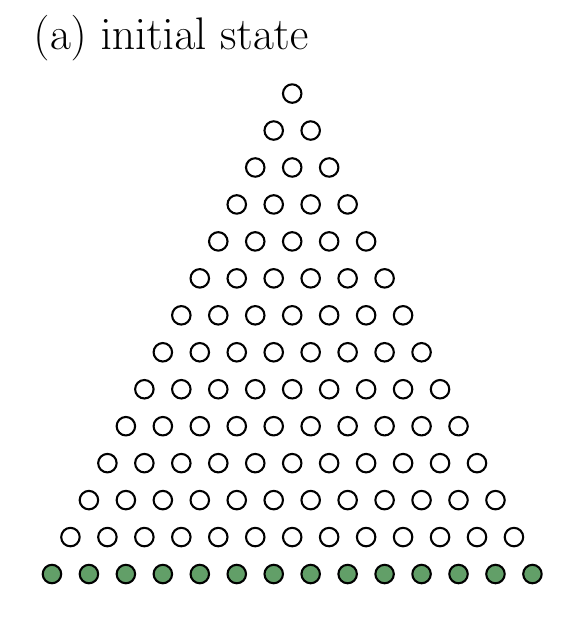}
\includegraphics[width=0.49\columnwidth]{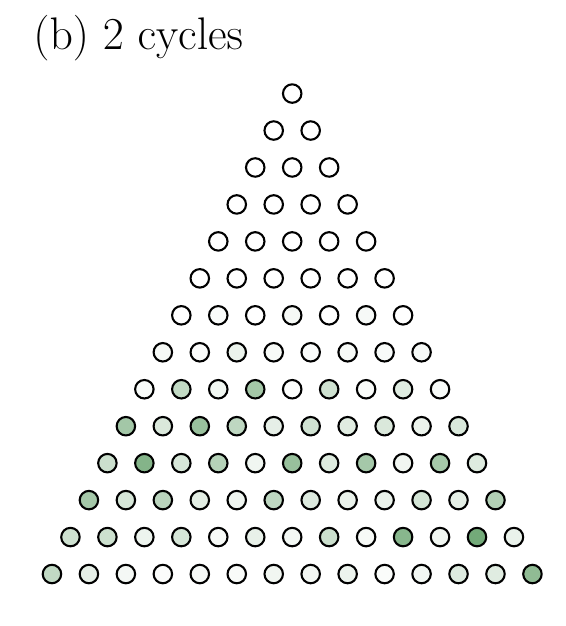}
\includegraphics[width=0.49\columnwidth]{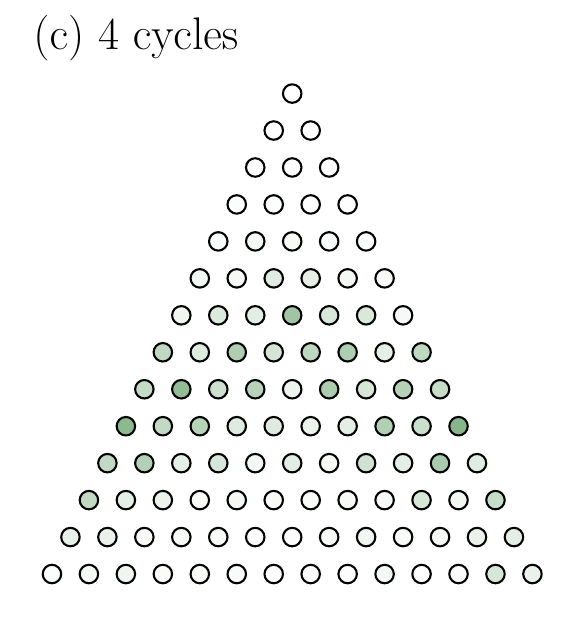}
\includegraphics[width=0.49\columnwidth]{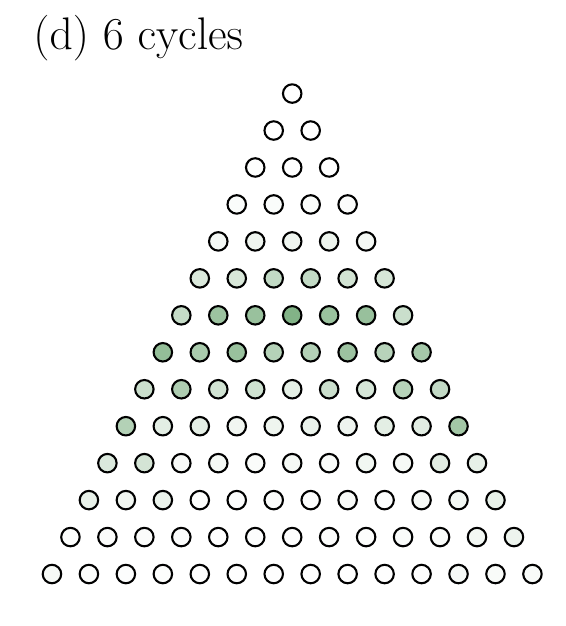}
    \caption{Four different snapshots in time of the information lattice for a system of 14 spins evolved with random unitary matrices $U=U_\mathrm{odd}U_\mathrm{even}$, where $U_\mathrm{odd}$ ($U_\mathrm{even}$) acts on all pairs of spins with indexes $m$ and $m+1$ where $m$ is odd (even). A single application of $U$ defines a cycle. The amount of local information in a site is indicated by the color intensity. The system is initialized in a product state $\bigotimes_{\mathrm{all~sites}~n} \vert  \uparrow \rangle \langle \uparrow  \vert$ where local information is located only the physical sites at $\ell=0$ [see panel (a)]. As $U$ is applied, local information builds up on increasing scales [see panels (b),(c),(d)]. Notice the presence of finite-size effects at the boundaries which only experience $U_\mathrm{odd}$.}
        \label{fig:random_circuit}
\end{figure}

\begin{figure}
\centering
\includegraphics[scale=0.5]{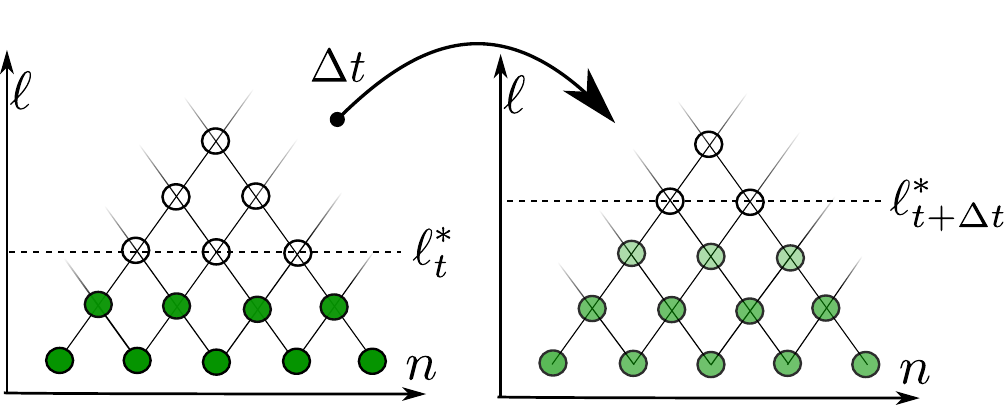}
\caption{Schematic of the exact time evolution on the subsystem lattice via Petz recovery maps. The color intensity of the information-lattice points quantifies the local information located at a given site. $\ell^*$ represents the smallest level with vanishing local information. We carry out the time evolution for each subsystem on $\ell^*$ individually without errors. After a typical timescale $\Delta t$, it is necessary to reassign $\ell^*$ to the updated minimum level where local information vanishes. For simplicity, here we consider $\ell^*_{t+\Delta t}= \ell^*_t + 1$.
} 
\label{fig:petz_map_algorithm}
\end{figure}

Recovering higher-level density matrices from lower-level ones allows us to close the subsystem equation of motion~\eqref{eq:subsys_vonNeumann} at level $\ell^* < L-1$. 
Through the Petz recovery maps, in fact, we can perform the exact time evolution of the subsystem density matrices.

Let us consider again the situation described at the end of Sec.~\ref{sec:subsystem_time_evolution} in which, by knowing the state of the full system at the initial time $t=0$, we can compute $\rho^{\ell^*}_n(\delta t)$ exactly. 
As we have discussed, to time evolve further, we need knowledge of $\rho^{\ell^*+r}_{n-r/2}(\delta t)$ and $\rho^{\ell^*+r}_{n+r/2}(\delta t)$. 
Now, provided that at time $\delta t$ there is no information on levels $\ell^*+1, \ell^*+2, \dots \ell^*+r$, that is, $i^{\ell^*+1}=i^{\ell^*+2}=\dotsb=i^{\ell^*+r}=0$, thanks to the recovery maps we can obtain the higher-level density matrices at time $\delta t$ [see Fig.~\ref{fig:Petz_map}(a)], and use them to compute $\rho^{\ell^*}_n(2\delta t)$, as schematically shown in Fig.~\ref{fig:Petz_map}(b).

After a few time steps, the assumption of zero information on levels $\ell^*+1,...,\ell^*+r$ will generically no longer hold. 
Typically, in ergodic quantum dynamics, correlations build up and spread throughout the system, adhering to the principles outlined by the Lieb-Robinson bounds~\cite{LiebRobinson}. 
On the information lattice, this is visualized by the fact that increasing levels acquire nonzero local information as time progresses. 
In Fig.~\ref{fig:random_circuit}, we illustrate a prototypical example of this generically expected behavior. 
A system of 14 spins is initialized in a product state $\bigotimes_{\mathrm{all~sites}~n} \vert  \uparrow \rangle \langle \uparrow  \vert$. 
Consequently, all local information at $t=0$ is located on the physical sites, i.e., at the information lattice sites with $\ell=0$ [see Fig.~\ref{fig:random_circuit}(a)]. 
Subsequently, we evolve this system in time by applying random unitary matrices from the circular unitary ensemble that pairwise couple neighboring sites $U=U_\mathrm{odd} U_\mathrm{even}$, where $U_\mathrm{odd}$ ($U_\mathrm{even}$) acts on all pairs of spins with index $m$ and $m+1$ where $m$ is odd (even)~\cite{nahum2017quantum,chan2018solution}. 
A single application of $U$ on the state of the system defines one random unitary evolution cycle (for each cycle we apply a different random unitary $U$). 
In each cycle, the largest level with nonzero local information can increase by a maximum of 4. 
This induces a quick buildup of local information at increasing scales [compare Fig.~\ref{fig:random_circuit}(b)-(c)]. 
Note that here, the boundaries only experience $U_\mathrm{odd}$, which is why the spread of local information at the boundary happens slower as compared to the bulk of the system.

In this example the system is relatively small and the dynamics can therefore be solved exactly.
Nonetheless, the same exact dynamics could be described by using the subsystem time evolution based on recovery maps.
At each time $t$, we fix a level $\ell^*$ such that $i^{\ell}=0$ for any $ \ell \geq \ell^*$. 
Then, the scheme starts by performing the exact time evolution of the states of the subsystems $\mathcal{C}^{\ell^*}$ having zero local information [see Fig.~\ref{fig:petz_map_algorithm}(a)].
Their equations of motion are closed via the recovery maps, as described at the beginning of this section.
Whenever an infinitesimal amount of local information reaches level $\ell^*$ due to spreading correlations, we update $\ell^*$ to the next smallest level with zero local information [see Fig.~\ref{fig:petz_map_algorithm}(b)].
As an example, in the random-unitary evolution described above, $\ell^*$ can increase by a maximum of 4 in each cycle.
Therefore, the exact subsystem time-evolution scheme requires a substantial updating of $\ell^*$ towards increasing levels and, eventually, $\ell^*=L-1$ that corresponds to the topmost site of the subsystem lattice, that is, the entire system. 
As the level $\ell^*$ increases, the computational resources needed to perform the time evolution grow exponentially. 
This is how the challenge of capturing entanglement growth appears on the information lattice.

Obviously, given a limited amount of computational time and resources, employing the exact subsystem time-evolution algorithm for large systems at long timescales becomes infeasible.
To circumvent this problem, once $\ell^*$ has reached a maximum value $\ell^*=\ell_\mathrm{max}$ set by the resources at our disposal, we require a suitable \textit{truncation method} to continue the time evolution at $\ell_\mathrm{max}$ without further increasing $\ell^*$.

\section{Approximate time evolution on the subsystem lattice: The LITE algorithm}
\label{sec:approximate_time_evolution}

\subsection{Introduction to LITE}
\label{sec:intro_LITE}

Given a maximum level $\ell_\mathrm{max}$ on which the numerical resources allow us to perform the time evolution, one might be tempted to adopt the naive approach to closing the equation of motion~\eqref{eq:subsys_vonNeumann} by simply ignoring the error term in the Petz recovery map~\eqref{eq:Petz_error}.
One would then continue to apply Petz recovery maps to compute the density matrices up to level $\ell_\mathrm{max}+r$ while keeping $\ell^* = \ell_\mathrm{max}$ even though there is nonvanishing local information on $\ell_\mathrm{max}$. 
This approach, however, is problematic.
The first problem is that the density matrices at level $\ell_\mathrm{max}+r$ obtained by recovery maps, which we denote as $\tilde{\rho}^{\ell_\mathrm{max}+r}_n$, may not preserve the lower-level density matrices (for instance, $\mathrm{Tr}_L^r (\tilde{\rho}^{\ell_\mathrm{max}+r}_n) \neq \rho^{\ell_\mathrm{max}}_{n+r/2}$).
As a result, errors are introduced on all length scales. 
In consequence, the algorithm fails to preserve the local constants of the motion, which leads to unphysical results.

To remedy this problem, we can use the \textit{projected} Petz recovery map (see Ref.~\onlinecite{klein2022time} and App.~\ref{app:sec:PPRM}), which projects the recovered density matrix to have the correct reduced density matrices, but this gives rise to two equally severe issues. 
First, the projection step may violate the positivity of the density matrices, leading to an immediate breakdown of the time evolution. 
Second, the density matrices produced by recovery maps generically have almost minimal local information.
This leads to an underestimation of local information at levels larger than $\ell_\mathrm{max}$ and, in turn, to an underestimation of the local information current from level $\ell_\mathrm{max}$ to higher levels~\cite{klein2022time}.
Then, the spuriously small information current leads to a significant buildup of local information at scale $\ell_\mathrm{max}$, which eventually distorts the dynamics.

In Ref.~\onlinecite{klein2022time}, the issue of the buildup of local information at level $\ell_\mathrm{max}$ was addressed by fixing the local information current from level $\ell_\mathrm{max}$ to higher levels through an information-current boundary condition that was empirically motivated by translational invariance and ergodic diffusive dynamics. 
Here, we aim to develop a time-evolution algorithm that does not depend on any specific assumptions about the information current and is suitable for diverse hydrodynamic behaviors.
Therefore, instead of attempting to generalize the boundary condition ansatz of Ref.~\onlinecite{klein2022time}, we adopt a different strategy that involves \textit{removing local information} directly.
The removal of information is justified by statistical reasons: we expect that during the out-of-equilibrium dynamics information flows from smaller to larger scales and does not return to influence the lower-level density matrices.
This guiding principle accounts for the second law of thermodynamics for entanglement entropy~\cite{gogolin2016equilibration}.
The removal of information must be approached with caution, as it has the potential to alter the true dynamics of the system. 
Any modification to the local information distribution will necessarily change the local information currents and thus impact the dynamics. 
Our goal is to remove local information in a way that artificial effects on the overall dynamics are minimal.

\begin{figure}[htb]
\centering
\includegraphics[scale=0.44]{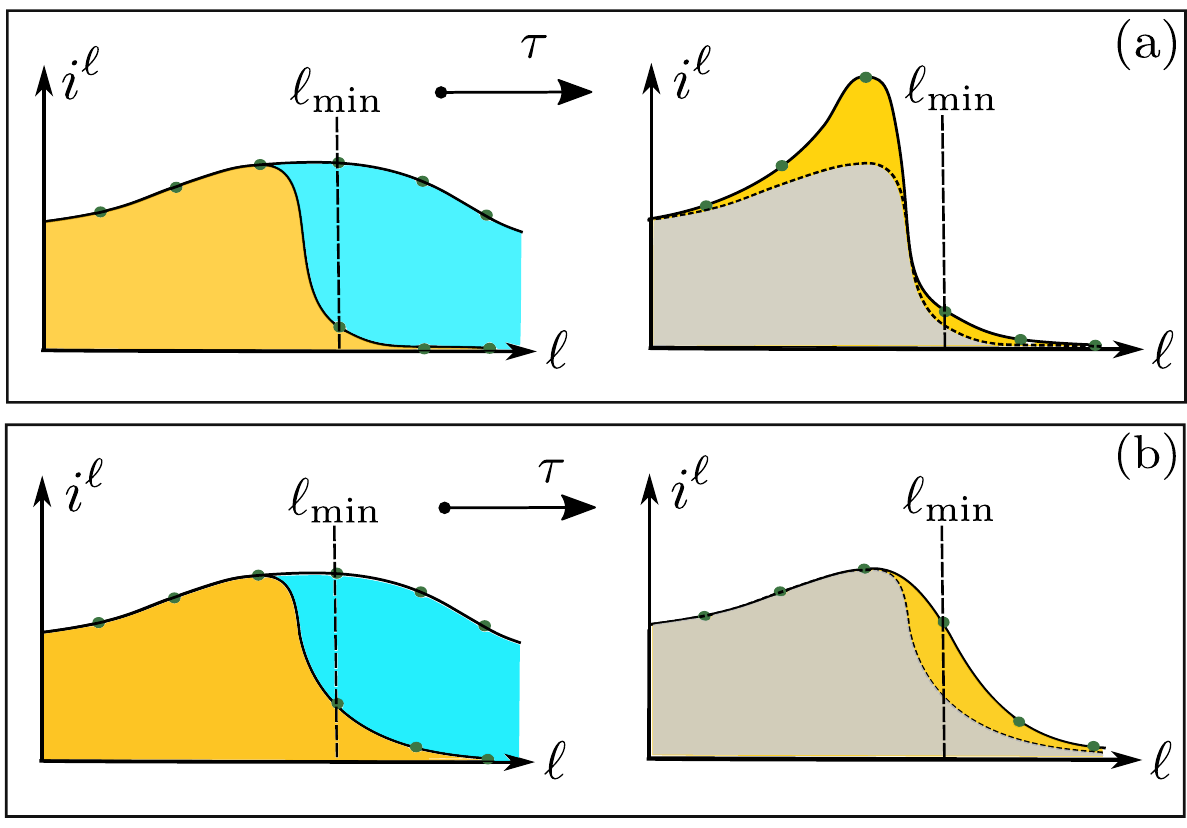}
\caption{Schematic of the local information distribution when the minimization on $\ell_\mathrm{min}$ is performed. 
(i) Minimization under the only constraint that lower-level density matrices are unaltered [see Eq.~\eqref{eq:lower-density-constraint-minimization}.] 
When the minimization is performed, the azure distribution is modified into the yellow one. 
After a typical time $\tau$, local information accumulates on levels $\ell < \ell_\mathrm{min}$. 
(ii) Minimization under the additional constraint of leaving unchanged local-information currents from $\ell_\mathrm{min}-1$ to $\ell_\mathrm{min}$. 
When the minimization is performed, one obtains the yellow distribution. 
Notice that in this case local information on $\ell_\mathrm{min}$ is, in general, larger than in panel (i). 
After time $\tau$, local information tends to accumulate on $\ell_\mathrm{min}$.}
\label{fig:information-minimization}
\end{figure}

\begin{figure}[htb]
\centering
\includegraphics[scale=0.44]{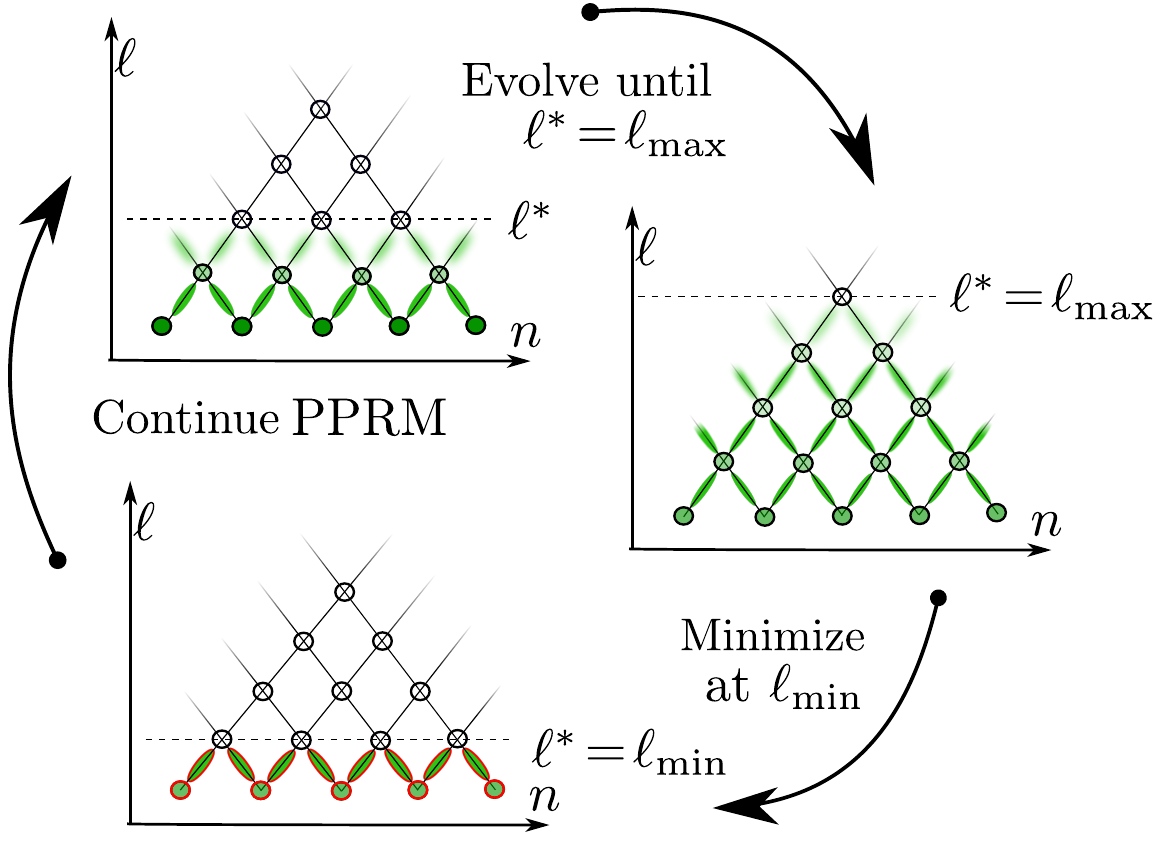}
\caption{Schematic of the LITE algorithm. 
The color intensity encodes the local information on the information-lattice sites. 
The green connections between the dots illustrate the information currents. 
We use projected Petz recovery maps (PPRM) (see App.~\ref{app:sec:PPRM}) to evolve subsystems until $\ell^*$ reaches $\ell_{\mathrm{max}}$. 
Once a small amount of information has accumulated at level $\ell_{\mathrm{max}}$, we jump back to level $\ell_{\mathrm{min}}$ where we minimize the information by preserving all density matrices on $\ell < \ell_{\mathrm{min}}$, as well as the currents flowing in between level $\ell_{\mathrm{min}}-1$ and $\ell_{\mathrm{min}}$. 
The minimization constraints are highlighted in the picture by the red contours.}
\label{fig:algorithm}
\end{figure}

The key to overcoming this issue is the introduction of a new length scale $\ell_\mathrm{min} < \ell_\mathrm{max}$ on which we remove local information. 
In order to keep the distribution of local information as stable as possible while removing parts of it, we \textit{minimize} local information on level $\ell_\mathrm{min}$ with the constraint of keeping the density matrices at lower levels $\ell<\ell_\mathrm{min}$ and the information currents on $\ell \leq \ell_\mathrm{min}$ \textit{fixed}. 
In practice, this is achieved by defining the new subsystem density matrix~\footnote{Notice that the projected Petz recovery maps discussed above can also be interpreted as a redefinition of the density matrices similar to Eq.~\eqref{eq:fixed_info}.}:
\begin{equation}
\label{eq:fixed_info}
\overline{\rho}_n^{\ell_\mathrm{min}} \coloneqq \rho_n^{\ell_\mathrm{min}} +\chi_n^{\ell_\mathrm{min}} ,
\end{equation}
where $\chi_n^{\ell_\mathrm{min}}$ is a Hermitian matrix acting on the Hilbert space associated with $\mathcal{C}_n^{\ell_\mathrm{min}}$. 
First, we impose that~\footnote{Notice that the condition $\tr_R^1\bigl(\chi_n^{\ell_\mathrm{min}}\bigr) {=} 0$ (or, equivalently, $\tr_L^1\bigl(\chi_n^{\ell_\mathrm{min}}\bigr) {=} 0$) implies that $\tr\bigl(\chi_n^{\ell_\mathrm{min}}\bigr) {=} 0$. Moreover, the conditions~\eqref{eq:lower-density-constraint-minimization} leave unchanged the information currents on small scales, up to the one between $\ell_\mathrm{min}-2$ and $\ell_\mathrm{min}-1$.}
\begin{equation}
    \label{eq:lower-density-constraint-minimization}
    \tr_R^1\left(\chi_n^{\ell_\mathrm{min}}\right)=\tr_L^1\left(\chi_n^{\ell_\mathrm{min}}\right)=0.
\end{equation}
These two conditions ensure that shifting the density matrix by $\chi_n^{\ell_\mathrm{min}}$ does not change the density matrices on lower levels. 
One can in principle try to minimize information solely under the conditions~\eqref{eq:lower-density-constraint-minimization}.
The minimization would consist in finding the optimal $\chi_n^{\ell_\mathrm{min}}$ satisfying conditions~\eqref{eq:lower-density-constraint-minimization} such that the corresponding local information $i_n^{\ell_\mathrm{min}}$ is minimized.
However, while this naive minimization scheme does not distort the distribution of local information on levels $\ell < \ell_\mathrm{min}$ at the time it is applied, this is not guaranteed at later times since such minimization modifies the local information currents.
Specifically, just after the minimization (up to the characteristic timescale $\tau$ of the system), the local information currents flowing into level $\ell_\mathrm{min}$ are suppressed, which causes an erroneous buildup of information on lower levels.
This is similar to the erroneous buildup of information caused by the Petz recovery maps discussed above.
Fig.~\ref{fig:information-minimization}(a) schematically depicts the distribution of local information before (azure) and after (yellow) the minimization under only the constraints~\eqref{eq:lower-density-constraint-minimization}, as well as after evolving for an additional time $\tau$ (yellow).

This problem is circumvented by imposing a second constraint on $\chi_n^{\ell_\mathrm{min}}$: we enforce that the local information currents between $\ell_\mathrm{min}-1$ and $\ell_\mathrm{min}$ remain unchanged under the minimization (see Sec.~\ref{sec:minimization-with-constraints} below for details). 
In this way, after the current-constraint minimization, there will be no buildup of information at levels $\ell < \ell_\mathrm{min}$ [see Fig.~\ref{fig:information-minimization}(b)].
The reason is that local information tends to accumulate at level $\ell_\mathrm{min}$. However, since the local information has just been minimized at level $\ell_\mathrm{min}$, the accumulation does not induce strong artificial local information gradients between different levels; therefore, the dynamics on lower levels is much less affected.

The current-constraint minimization drastically suppresses local information on levels $\ell >\ell_\mathrm{min}$.
In turn, projected Petz recovery maps performed to compute the density matrices on levels $\ell > \ell_\mathrm{min}$ become more accurate.
This allows us to continue the subsystem time evolution on level $\ell^* = \ell_\mathrm{min}$ by applying recovery maps up to $\ell_\mathrm{min}+r$. 
As the dynamics proceeds and entanglement spreads, local information reaches higher and higher levels. 
As described before, we increase $\ell^*$ as soon as a non-negligible amount of local information has reached it (this amount should be set as small as possible; see Sec.~\ref{sec:numerical-parameters} below for more details) up to $\ell^* = \ell_\mathrm{max}$.
When local information has again substantially spread up to level $\ell_\mathrm{max}$, we need to repeat the minimization of local information at level $\ell_\mathrm{min}$. 
The use of this two-level scheme endures throughout the entire time evolution.
Importantly, it can be shown that this algorithm conserves local constants of motion at $\ell < \ell_\mathrm{min}$ by construction (as discussed in App.~\ref{app:sec:local-conserved-quantities}).
A schematic of the algorithm is depicted in Fig.~\ref{fig:algorithm}.

Given the pivotal role of the \textit{local-information time evolution} in the design of the algorithm, we denote it LITE.
The remaining part of this section is devoted to providing mathematical details of the LITE algorithm.
Readers not interested in these details can proceed directly to Sec.~\ref{sec:energy_dist}.

\subsection{Removal of local information at large scales}
\label{sec:minimization-with-constraints}

\subsubsection{Time evolution of information and information currents}
\label{sec:currents}

\begin{figure}
\centering
\includegraphics[scale=0.35]{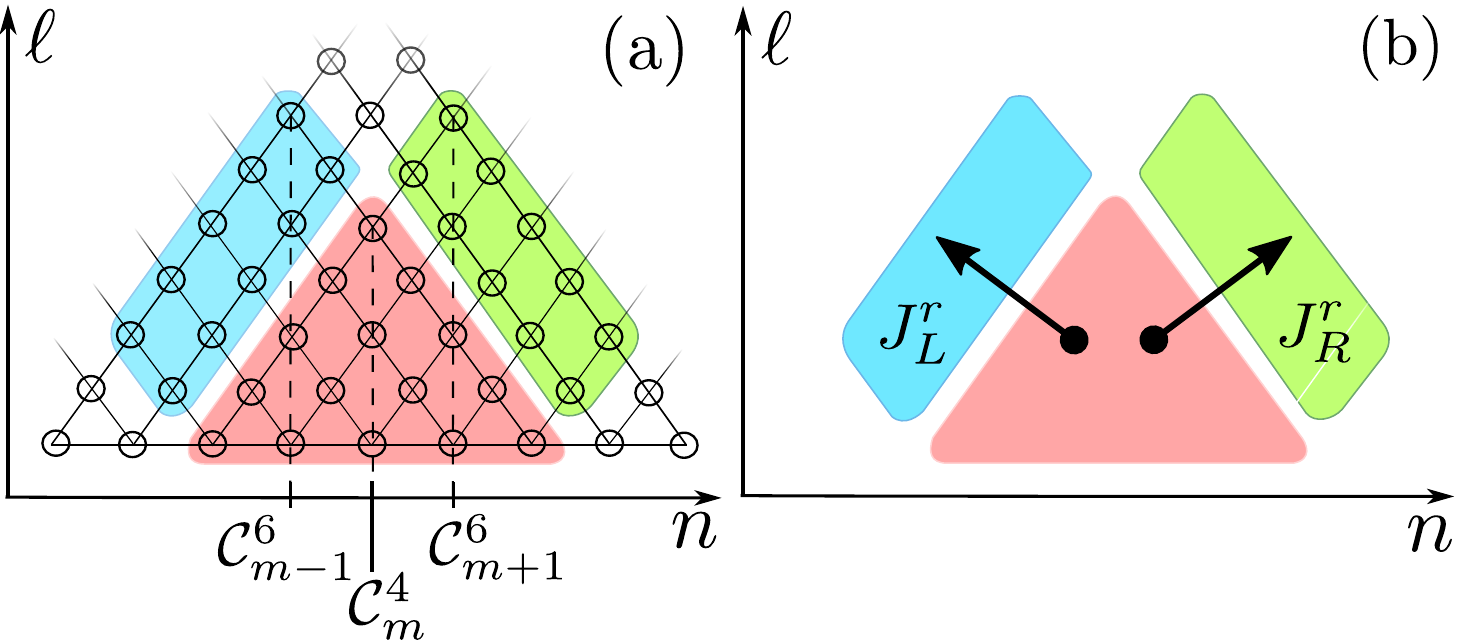}
\caption{Schematic of the von Neumann information currents of the subsystem $\mathcal{C}_m^{4}$ for $r=2$. (a) Schematic of the subsystems involved in the von Neumann information currents $J^r_L(\rho^6_{m-1})$ and $J^r_R(\rho^6_{m+1})$. The regions of interest are marked by colors. (b) The currents out of the subsystem $\mathcal{C}_m^{4}$ flow into the blue and green regions that belong to the subsystems $\mathcal{C}_{m-1}^{6}$ and $\mathcal{C}_{m+1}^{6}$.
}
\label{fig:current}
\end{figure}

To formalize the above discussion, we require a notion of information currents. 
The dynamics generated by time-evolving subsystem density matrices according to Eq.~\eqref{eq:subsys_vonNeumann} leads to a corresponding dynamics of the information~\eqref{eq:local_information}. 
The time derivative of the von Neumann information of the state of the subsystem $\mathcal{C}_n^\ell$ [see Eq.~\eqref{eq:vonNeumann_information}] reads
\begin{equation}
\partial_t I_n^\ell = \mathrm{Tr}\left( \nabla_\rho  S_n^{\ell}~\partial_t\rho_n^{\ell} \right),
\end{equation}
where $\nabla_\rho  S_n^{\ell}= -\ln\left(\rho_n^\ell\right)-\mathbb 1_{d^{\ell+1}}$.
By expanding the right-hand side, we decompose this current into two parts, one flowing left and one flowing right.
From Eq.~\eqref{eq:subsys_vonNeumann} and the cyclic property of the trace, we obtain
\begin{multline}
\label{eq:current_02}
\partial_t I_n^\ell = -i\, \mathrm{Tr}\left(\left[ \nabla_\rho S_n^{\ell},H^{\ell}_n \right] \rho_n^{\ell} \right) \\
-i\, \mathrm{Tr}\left( \left[ \nabla_\rho S_n^{\ell},  H_{n-r/2}^{\ell+r} - H_n^{\ell}\right] \rho_{n-r/2}^{\ell+r} \right)  \\
-i\, \mathrm{Tr}\left( \left[ \nabla_\rho S_n^{\ell}, H_{n+r/2}^{\ell+r}- H_n^{\ell} \right]\rho_{n+r/2}^{\ell+r} \right).
\end{multline}
The first term on the right-hand side originates from terms in the Hamiltonian within $\mathcal{C}_n^\ell$ only; therefore it cannot contribute to the information flow into or out of $(n,\ell)$. 
On a technical level, it vanishes because $[\nabla_\rho S_n^l,\rho^l_n]=0$.
The second term originates only from terms in the Hamiltonian within the subsystem $\mathcal{C}_{n-r/2}^{\ell+r}$; thus it cannot contribute to the change of the von Neumann information $I_{n-r/2}^{\ell+r}$ of the state of the subsystem $\mathcal{C}_{n-r/2}^{\ell+r}$.
Pictorially, it does not change the von Neumann information in the union of the red and blue regions of Fig.~\ref{fig:current}(a).
We conclude that it must be the left current from the red to the blue region.
By a similar argument for the third term, we get
\begin{equation}
\label{eq:current_04}
J^r_L(\rho_{n-r/2}^{\ell +r}) \coloneqq 
-i\,\mathrm{Tr}\left( \left[ \nabla_\rho S_n^{\ell},  H_{n-r/2}^{\ell+r} - H_n^{\ell}\right] \rho_{n-r/2}^{\ell+r} \right)
\end{equation}
and
\begin{equation}
\label{eq:current_05}
J_R^r(\rho_{n+r/2}^{\ell+r}) \coloneqq 
-i\,\mathrm{Tr}\left( \left[ \nabla_\rho S_n^{\ell}, H_{n+r/2}^{\ell+r}- H_n^{\ell} \right]\rho_{n+r/2}^{\ell+r} \right),
\end{equation}
with
\begin{equation}
\label{eq:current_03}
\partial_t I_n^\ell = J_L^r(\rho_{n-r/2}^{\ell +r}) + J_R^r(\rho_{n+r/2}^{\ell+r}).
\end{equation}
We schematically show $J_L^r$ and $J_R^r$ in Fig.~\ref{fig:current}(b).
The currents are linear functions of the higher-level subsystem density matrices $\rho_{n-r/2}^{\ell +r}$ and $\rho_{n+r/2}^{\ell +r}$, respectively (which motivates the notation). 
Similar to the subsystem von Neumann information $I^\ell_n$, the von Neumann currents $J^r_L(\rho_{n-r/2}^{\ell +r})$ and $J_R^r(\rho_{n+r/2}^{\ell+r})$ take values on the subsystem lattice. 
They are global properties of the subsystem, which implies that they are not assigned to individual lines that connect lattice points but to the full subsystem triangle on the subsystem lattice, and flow between subsystems.

\subsubsection{Projection onto a constrained subspace}
\label{sec:minimization_constraints}

\begin{figure}
    \centering
    \includegraphics[scale=0.33]{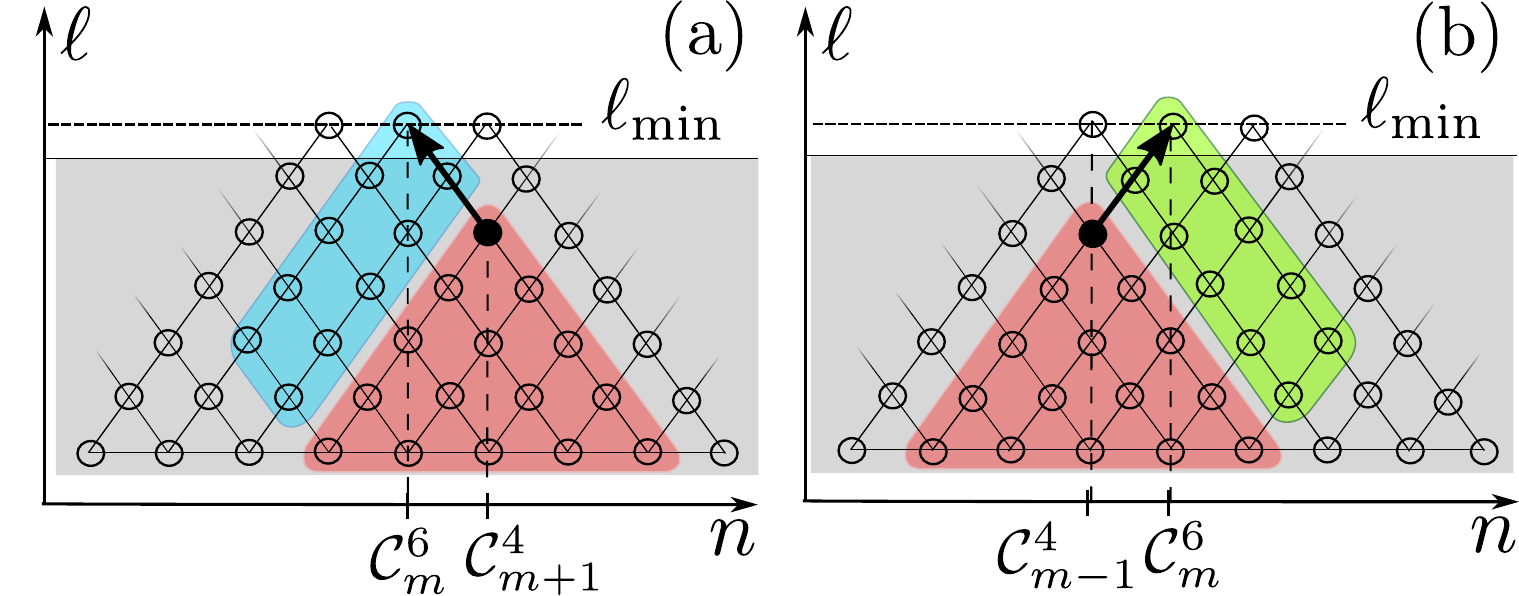}
    \caption{Schematic of information minimization at level $\ell_\mathrm{min}$ with $r=2$. 
    All subsystem density matrices on levels $\ell < \ell_\mathrm{min}$ are kept fixed (indicated by gray background). 
    Additionally, we demand the currents (black arrows) out of the red region to the left [see (a)] and right [see (b)] remain unchanged. 
    In the region of fixed density matrices (gray), currents are also automatically preserved.
    }
    \label{fig:minimization}
\end{figure}

The constraints on $\chi_n^\ell$ discussed in Sec.~\ref{sec:intro_LITE} can now be condensed into four equations:
\begin{eqnarray}
\label{eq:condition1}
\mathrm{Tr}_L^1(\chi_n^{\ell})&=&0,~~~~ \mathrm{Tr}_R^1(\chi_n^{\ell})=0,\\
\label{eq:condition2}
J_L^r(\chi_n^{\ell})&=&0,~~~~ J_R^r(\chi_n^{\ell})=0.
\end{eqnarray}
If Eqs.~\eqref{eq:condition1} and \eqref{eq:condition2} are fulfilled, the density matrices $\overline{\rho}_n^\ell \coloneqq \rho_n^\ell + \chi_n^\ell$ and $\rho_n^\ell$ have the same lower-level subsystem density matrices and identical currents on all levels below $\ell$. 
Conditions~\eqref{eq:condition1} and \eqref{eq:condition2} are schematically depicted in Fig.~\ref{fig:minimization} when applied on level $\ell_\mathrm{min}$ and $r=2$.
If constraints~\eqref{eq:condition1} are satisfied, all density matrices for $\ell < \ell_\mathrm{min}$ are kept fixed (indicated by the gray background). 
Furthermore, constraints~\eqref{eq:condition2} guarantee that the currents that flow out of the red region (black arrows) into the blue [see (a)] and green [see (b)] regions remain unchanged.
Contributions to the currents at $\ell < \ell_\mathrm{min}$ are already preserved due to the imposed trace condition (specifically, the currents within the gray background).
Thus, effectively, the current constraints~\eqref{eq:condition2} add only an additional condition on the currents that flow directly into level $\ell_\mathrm{min}$ (ensuring that no extra currents are generated by the shift matrix $\chi_n^{\ell}$).

Importantly, Eqs.~\eqref{eq:condition1} and \eqref{eq:condition2} represent linear operations on $\chi_n^\ell$. 
Thus, we can impose Eqs.~\eqref{eq:condition1} and \eqref{eq:condition2} by \textit{projecting} $\chi_n^\ell$ onto the respective kernel of the linear operators $\mathrm{Tr}_L^1$, $\mathrm{Tr}_R^1$, $J_L^r$ and $J_R^r$. 
Given an arbitrary matrix $\chi^\ell_n$, the projectors onto the (partial) trace-free spaces are defined as
\begin{eqnarray}
\label{eq:project_L}
\mathrm{P}^{\mathrm{Tr}}_L \chi^\ell_n &\coloneqq& \chi^\ell_n - \frac{\mathbb{1}_d}{d} \otimes \mathrm{Tr}_L^1(\chi^\ell_n),\\
\label{eq:project_R}
\mathrm{P}^{\mathrm{Tr}}_R \chi^\ell_n &\coloneqq& \chi^\ell_n - \mathrm{Tr}_R^1(\chi^\ell_n) \otimes \frac{\mathbb{1}_d}{d}.
\end{eqnarray}
To merge the projections of Eqs.~\eqref{eq:project_R} and \eqref{eq:project_L} with the desired conditions on current~\eqref{eq:condition2} in a combined projector, we use the definitions~\eqref{eq:current_04} and~\eqref{eq:current_05}. 
The current $J_L^r$ of an arbitrary, (partial) trace-free matrix $\mathrm{P}^{\mathrm{Tr}}_R\mathrm{P}^{\mathrm{Tr}}_L \chi_n^\ell$ with subsystem coordinates $(n,\ell)$ is given by
\begin{multline}
J_L^r(\mathrm{P}^{\mathrm{Tr}}_R\mathrm{P}^{\mathrm{Tr}}_L\chi_n^\ell) = \\
-i\,\mathrm{Tr}\left( \left[ \nabla_\rho S_{n + r/2}^{\ell-r},  H_{n}^{\ell} - H_{n + r/2}^{\ell-r}\right] \mathrm{P}^{\mathrm{Tr}}_R\mathrm{P}^{\mathrm{Tr}}_L\chi_n^\ell \right).
\end{multline}
The Hermiticity of the operators (including $\mathrm{P}^{\mathrm{Tr}}_R\mathrm{P}^{\mathrm{Tr}}_L$) and the cyclic property of the trace allows us to rewrite the right-hand side as
\begin{equation}
J_L^r(\mathrm{P}^{\mathrm{Tr}}_R\mathrm{P}^{\mathrm{Tr}}_L\chi_n^\ell) =
-i\,\mathrm{Tr}\left( f_n^\ell \chi_n^\ell \right),
\end{equation}
where
\begin{equation}
f_n^\ell\coloneqq\mathrm{P}^{\mathrm{Tr}}_R\mathrm{P}^{\mathrm{Tr}}_L \left[ \nabla_\rho S_{n + r/2}^{\ell-r},  H_{n}^{\ell} - H_{n + r/2}^{\ell-r}\right].
\end{equation}
Combining the projectors to the kernels of $\mathrm{Tr}_L^1$, $\mathrm{Tr}_R^1$ and $J_L^r$, denoted $\mathrm{P}^J_L$, we get
\begin{equation}
\mathrm{P}^J_L \mathrm{P}^{\mathrm{Tr}}_R\mathrm{P}^{\mathrm{Tr}}_L \chi_n^\ell = \chi_n^\ell - \frac{\mathrm{Tr}\left(f^\ell_n \chi_n^\ell\right)}{\mathrm{Tr}\left((f_n^\ell)^2\right)}f^\ell_n.
\end{equation}
Finally, the current $J^r_R$ of $\mathrm{P}^J_L P^{\mathrm{Tr}}_R\mathrm{P}^{\mathrm{Tr}}_L \chi_n^\ell$ is given by
\begin{equation}
J^r_R\left(\mathrm{P}^J_R P^{ \mathrm{Tr}}_R\mathrm{P}^{\mathrm{Tr}}_L \chi_n^\ell\right) = -i\, \mathrm{Tr}\left( g^\ell_n \chi_n^\ell \right),
\end{equation}
where
\begin{equation}
g_n^\ell \coloneqq \mathrm{P}^J_L\mathrm{P}^{\mathrm{Tr}}_R\mathrm{P}^{\mathrm{Tr}}_L \left[  \nabla_\rho S_{n - r/2}^{\ell-r},  H_{n}^{\ell} - H_{n - r/2}^{\ell-r}\right].
\end{equation}
Thus, the concatenated total projector $\mathrm{\textbf{P}}$ to the kernel of the system of equations defined by Eqs.~\eqref{eq:condition1} and \eqref{eq:condition2} reads
\begin{equation}
\label{eq:total-projector}
    \mathrm{\textbf{P}} \chi_n^\ell = \chi_n^\ell - \frac{\mathrm{Tr}\left(g^\ell_n \chi_n^\ell\right)}{\mathrm{Tr}\left((g_n^\ell)^2\right)}g^\ell_n.
\end{equation}
Given an arbitrary $\chi_n^\ell$, $\mathrm{\textbf{P}}$ projects it to the subspace of matrices that preserve all subsystem density matrices on $\ell<\ell_\mathrm{min}$ and information currents on $\ell\leq\ell_\mathrm{min}$.

\subsubsection{Minimization of local information under constraints}
\label{sec:minimization}

Given $\mathrm{\textbf{P}}$ that projects onto the subspace of interest, the next step is to find the optimal $\chi^\ell_n$ that minimizes the information, or, equivalently maximizes the von Neumann entropy, at level $\ell$. 
We want to find $\xi^\ell_n \coloneqq \mathrm{\textbf{P}}\chi^\ell_n$ such that $S(\rho^\ell_n+\xi^\ell_n)=-\mathrm{Tr}\left[(\rho^\ell_n+\xi^\ell_n) \ln(\rho^\ell_n+\xi^\ell_n)\right]$ is maximized. 
For ease of notation, we drop the indices $n$ and $\ell$ in this section.
Expanding the von Neumann entropy up to second order in $\xi$ yields (see App.~\ref{app:sec:math-details-minimization} for a detailed derivation)
\begin{multline}
\label{eq:entropy_expansion_main_text}
S(\rho + \xi) =  S(\rho + \mathrm{\textbf{P}}\chi) \\
= S(\rho) + \mathrm{Tr}\left( \mathrm{\textbf{P}}\nabla_\rho S \, \chi \right)  + 
\frac{1}{2}\mathrm{Tr}\left(\chi \, \mathrm{\textbf{P}}\mathscr{H}_\rho  \mathrm{\textbf{P}} \, \chi \right)  + O(\xi^3)
\end{multline}
where $\nabla_\rho S = - \ln(\rho) - \mathbb{1}$ is the entropy gradient, and 
\begin{equation}
\label{eq:hessian_main_text}
\mathscr{H}_\rho \, \sbullet \coloneqq U_\rho \bigl( \mathrm{h} \star \left( U_\rho^{\dagger} \sbullet U_\rho \right) \bigr) U_\rho^{\dagger}
\end{equation} 
is the entropy Hessian at $\rho$, $U_\rho $ diagonalizes $\rho$, that is, its columns are given by the eigenvectors of $\rho$, and $\mathrm{h}$ is a matrix with elements 
\begin{equation}
\label{eq:hessian_elements_main_text}
\mathrm{h}_{ij} =-\frac{\mathrm{arctanh}\left(\frac{\kappa_i-\kappa_j}{\kappa_i+\kappa_j}\right)}{\kappa_i-\kappa_j},
\end{equation}
where $\kappa_i$ are the eigenvalues of $\rho$. 
Finally, $A \star B$ denotes the elementwise multiplication (or Hadamard product) of two matrices $A$ and $B$. 
Importantly, since the eigenvalues of the density matrix $\rho$ are always positive, $\kappa_i\geq 0 $, from Eq.~\eqref{eq:hessian_elements_main_text} we see that all the elements of $\mathrm{h}$ are strictly negative.
Thus, we can apply Newton's method for optimization to find $\chi^*$ such that the von Neumann entropy $S(\rho +  \mathrm{\textbf{P}} \chi^*)$ is maximal (or, equivalently, such that the von Neumann information is minimal) given the imposed constraints. 
The optimal shift $\chi^*$ that maximizes the entropy is 
\begin{equation}
\label{eq:chi_update}
  \chi^* = -\left( \mathrm{\textbf{P}} \mathscr{H}_\rho  \mathrm{\textbf{P}} \right)^+ \mathrm{\textbf{P}} \nabla_\rho S,
\end{equation}
where $(\sbullet)^+$ denotes the pseudoinverse~\cite{ben2003generalized}.
To avoid the nontrivial computation of the pseudoinverse, instead of using Eq.~\eqref{eq:chi_update} we numerically solve the related linear equation
\begin{equation}
\label{eq:preconditioned}
 \mathrm{\textbf{P}} \mathscr{H}_\rho  \mathrm{\textbf{P}}\, \chi^* = -\mathrm{\textbf{P}} \nabla_\rho S
\end{equation}
via the \textit{preconditioned conjugate gradient method}~\cite{barrett1994templates} (see App.~\ref{app:sec:newton-method} for details).
Therefore, we iteratively update $\chi$ until convergence is reached and Eq.~\eqref{eq:preconditioned} is satisfied. 
To minimize the information at level $\ell_{\mathrm{min}}$ on the subsystem lattice, we perform the above scheme for each lattice coordinate $(n,\ell_{\mathrm{min}})$ individually.
Since the lower-level density matrices on $\ell<\ell_\mathrm{min}$ and information currents on $\ell \leq \ell_\mathrm{min}$ are kept fixed, the order in which the subsystem density matrices $\rho^{\ell_\mathrm{min}}_n$ are minimized does not change the result.
Finally, to lighten the notation, we replace $\overline{\rho}^{\ell_\mathrm{min}}_n$ with $\rho^{\ell_\mathrm{min}}_n$.
This ensures that the symbol $\rho$ consistently represents the density matrices used in the approximate time-evolution scheme.

\section{Numerical simulations}
\label{sec:numerical_simulations}

\subsection{Local observables: energy distribution and diffusion coefficient}
\label{sec:energy_dist}

Time evolving with a time-invariant Hamiltonian ensures energy conservation. 
However, if the initial state's energy distribution lacks translational invariance, energy tends to be redistributed as time progresses.
This energy redistribution trajectory hinges on the model's hydrodynamics, primarily characterized by the Hamiltonian $H$.
Energy transport can range from localized to subdiffusive, diffusive, ballistic, or even superdiffusive.
By initializing the dynamics from a state featuring localized energy accumulation, one can distinguish the diverse transport regimes from the energy distribution's variance over time.
For instance, in a diffusive system, the energy distribution's variance grows as $\propto t$, an outcome deduced directly from the diffusion equation.
On the other hand, ballistic transport is expected to yield a variance proportional to $t^2$.

Assuming that the system Hamiltonian $H$ is local with a maximum interaction range $r$, we can write it as the sum of local terms:
\begin{equation}
\label{eq:local-hamiltonian-0}
    H = \sum_m h_m^r
\end{equation}
with $m$ a physical site index.
Note that this partition is not unique. 
Note furthermore that $H \neq \sum_n H_n^{r}$, with $ H_n^{r}$ defined in Eq.~\eqref{eq:subsystem-hamiltonian}.
Indeed, the $H^r_n$ are global quantities on the subsystem lattice covering a range of $r + 1$ physical sites; hence, different subsystem Hamiltonians overlap. 
A local partition of the Hamiltonian has the advantage that the expectation value $\langle H\rangle $ becomes the sum of local expectation values, and $E_m^{r} \coloneqq \langle h_m^r\rangle $ quantifies the local energy.

Given the distribution of the local energy, the corresponding variance is given by
\begin{equation}
\label{eq:energy_variance}
    \sigma_E^2 \coloneqq  \sum_m (m-\overline{m})^2 \frac{E_m^{r}}{\langle H \rangle} - \left(\sum_m (m-\overline{m})\frac{E_m^{r}}{\langle H \rangle}\right)^2,
\end{equation}
where $\overline{m}= \sum_m m E_m^{r} /\langle H \rangle$ is the first moment of the distribution.
Furthermore, the diffusion coefficient is
\begin{equation}
\label{eq:diff_const}
    D \coloneqq \frac{1}{2}\partial_t \sigma_E^2.
\end{equation}
Even though for time-invariant Hamiltonians $\partial_t\langle H \rangle = 0$, the local energies are in general not constant.
We obtain
\begin{equation}
    \partial_t E_m^{r} = \mathrm{Tr}\left( h_m^r \partial_t \rho \right) = i \langle \left[H ,h_m^r\right] \rangle  ;
\end{equation}
thus,
\begin{multline}
\label{eq:diff_const_explicit}
    D = \frac{i}{2}\sum_m (m-\overline{m})^2\frac{\langle \left[H, h_m^r \right] \rangle}{\langle H\rangle } \\
    - i \sum_m (m-\overline{m}) \frac{E_m^{r}}{\langle H \rangle} \sum_k (k-\overline{m}) \frac{\langle \left[H, h_k^{r} \right] \rangle}{\langle H \rangle} .
\end{multline}
In diffusive systems, $D$ is expected to be constant as $\sigma^2_E \propto D t$ and is typically referred to as the diffusion constant.
Generically, the scaling of $\sigma_E^2$ is not linear in time and the diffusion coefficient $D$ depends on time.

Note that similar quantities as those defined in this section can be introduced in the presence of other conserved charges, for instance magnetization.

\subsection{Initial states}
\label{sec:init_state}

For a thorough investigation of the transport properties through the temporal scaling of the diffusion coefficient, the energy diffusion should occur unimpeded throughout the system, in particular, without boundary reflections.
Consequently, the analysis necessitates working with either very large systems or, ideally, systems that are \textit{infinitely} extended. 
In the general case where the Hamiltonian is not translation invariant, managing infinitely extended systems is only feasible when the initial state approaches thermal equilibrium asymptotically in space, thus remaining static~\footnote{For translation-invariant Hamiltonians, one can easily perform time evolution starting from translation-invariant initial states, as done in Ref.~\onlinecite{klein2022time}.}.
A simple state in this family of states is given by
\begin{equation}
\label{eq:init_state}
    \rho_\mathrm{init} = \left (\bigotimes_{m<n-\ell/2} \rho_{m,\infty} \right)~ \otimes \mathcal{\rho }_{n,\mathrm{init}}^\ell \otimes ~ \left( \bigotimes_{m>n+\ell/2} \rho_{m,\infty}\right), 
\end{equation}
where $\rho_{m,\infty}= \mathbb{1}_{d}/d$ is the infinite-temperature single-site density matrix located at site $m$, and $d$ is the Hilbert space dimension of the physical sites.
If the Hamiltonian is traceless, the product state formed solely from $\rho_{m,\infty}$ possesses zero total energy; if it is not traceless one can always redefine the energy by subtracting the trace such that this condition holds.
The initial state~\eqref{eq:init_state} thus contains a finite amount of energy located within a range of $\ell +1$ physical sites centered at $n$ (determined only by $\rho_{n,\mathrm{init}}^{\ell}$).

In practice, Eq.~\eqref{eq:init_state} allows us to effectively simulate a finite system at all times while performing time evolution of the infinitely extended one.
Indeed, the initial state~\eqref{eq:init_state} is asymptotically time-invariant, as infinite-temperature density matrices do not evolve in time.
We stress that this important property holds true for any Hamiltonian, including those that are non-translation-invariant or disordered.
Consequently, the algorithm exclusively performs time evolution within the central region of the system, encompassing a finite number of sites around $n$.
This effective region includes both the physical sites whose state deviates from the infinite-temperature background at time $t$, and a finite number of physical sites at the boundaries in the infinite-temperature state.
After each time-evolution step, we check the state of the boundary sites. 
If the difference (in norm) between their single-site density matrix and the infinite-temperature single-site density matrix exceeds a chosen tolerance, we enlarge the effective system by adding further physical sites (at infinite temperature) at each end of it (see App.~\ref{app:sec:asymptotic-initial-states} for more details).
By increasing the size of the effective region, we ensure that energy, while spreading over time, can \textit{freely} propagate throughout the system.
Given $\ell_\mathrm{max}$, utilizing Eq.~\eqref{eq:init_state} makes the required computational resources scale \textit{linearly} with the effective system size.

One convenient choice for $\rho_{n,\mathrm{init}}^\ell$ (used in Sec.~\ref{sec:results_mixed_field_Ising}) is a thermal density matrix with respect to the subsystem Hamiltonian $H_n^\ell$
\begin{equation}
\label{eq:init_state_thermal_part}
\begin{split}
    \rho_{n,\mathrm{init}}^\ell= \frac{1}{Z}\exp \left( - \beta H_n^\ell\right), \\
    Z = \mathrm{Tr}\left(\exp\left(-\beta H_n^\ell\right)\right).
\end{split}
\end{equation}

\subsection{Numerical parameters of LITE}
\label{sec:numerical-parameters}

To numerically implement the two-level scheme of LITE depicted in Fig.~\ref{fig:algorithm}, one needs to introduce some threshold parameters.
Let us imagine we initialize the system in a state with local information only on small levels $\ell$ where $\ell < \ell^* < \ell_\mathrm{min} < \ell_\mathrm{max}$.   
We then start the time evolution from density matrices on level $\ell^*$ by closing the equation of motion~\eqref{eq:subsys_vonNeumann} via the projected Petz recovery map (see App.~\ref{app:sec:PPRM}). 
As soon as the local information on $\ell^*$ reaches a small critical threshold value, $q^*$, we update $\ell^*$ to $\ell^* + r$ by computing the higher-level density matrices via the recovery map. 
When $\ell^*$ reaches $\ell_{\mathrm{max}}$ and a critical amount of information has accumulated at $\ell_{\mathrm{max}}$, $q_\mathrm{max}$, we perform the current-constraint minimization at level $\ell_{\mathrm{min}}$ while keeping all lower-level density matrices and information currents unaltered. 
Subsequently, we continue the time evolution at level $\ell^* = \ell_{\mathrm{min}}$ as described above.
We refer to the application of one minimization as one evolution cycle.

While $\ell_\mathrm{min}$ and $\ell_\mathrm{max}$ should be chosen as large as possible (and such that $\ell_\mathrm{min}<\ell_\mathrm{max}$), and $q^*$ should be set as small as possible, a similar extremal condition should not be applied to $q_\mathrm{max}$. 
Since the minimization removes information over time, the total information in the system (specifically, the sum of local information over all the information-lattice sites) shrinks with the number of evolution cycles.
Thus, we define $q_\mathrm{max}$ as the percentage of local information that we allow on level $\ell_\mathrm{max}$ measured with respect to the total information currently in the system. 
This implies the need to recompute the total information in the system at the beginning of each cycle. 
Choosing too large values of $q_\mathrm{max}$ is problematic as local information starts to accumulate at $\ell_\mathrm{max}$, making Petz recovery maps less accurate and possibly distorting the dynamics on the lower levels. 
On the other hand, infinitely small values for $q_\mathrm{max}$ trigger overly frequent minimizations of local information at $\ell_\mathrm{min}$. 
In this case, local information has no time to travel beyond $\ell_\mathrm{min}$, and the intrinsic dynamics of the system is altered.
Thus, there is an optimal range for $q_\mathrm{max}$ that has to be determined empirically for the system under consideration.

\subsection{Diffusive dynamics in the mixed-field Ising model}
\label{sec:results_mixed_field_Ising}

\begin{figure}
\centering
\includegraphics[width=\columnwidth]{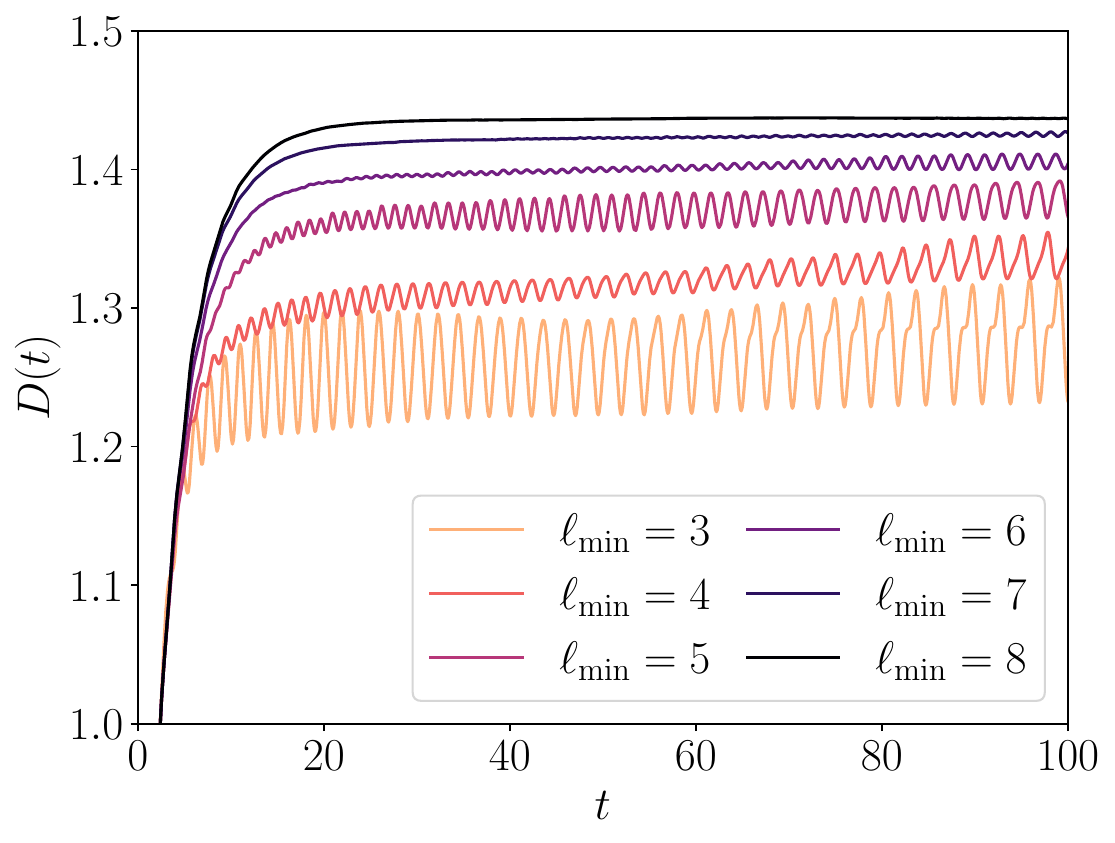}
\caption{Diffusion coefficient as a function of time $D(t)$ for different $\ell_\mathrm{min}$, at fixed $\ell_{\mathrm{max}}=9$ and $q_{\mathrm{max}}=0.5\%$. At short times, the system has a ballistic behavior, and $D(t)$ increases with time. $D(t)$ then saturates to a plateau for $t \gtrsim 20$. The oscillations of $D(t)$ along the plateau are associated with algorithmic artifacts and shrink to zero as $\ell_\mathrm{min}$ increases. The (average) plateau value of $D(t)$ increases with $\ell_\mathrm{min}$, and tends to an asymptotic value for $\ell_\mathrm{min}\to\infty$ (see Fig.~\ref{fig:benchmark_diffusion_constant_ave}).}
\label{fig:benchmark_diffusion_constant}
\end{figure}

\begin{figure}
\centering
\includegraphics[width=\columnwidth]{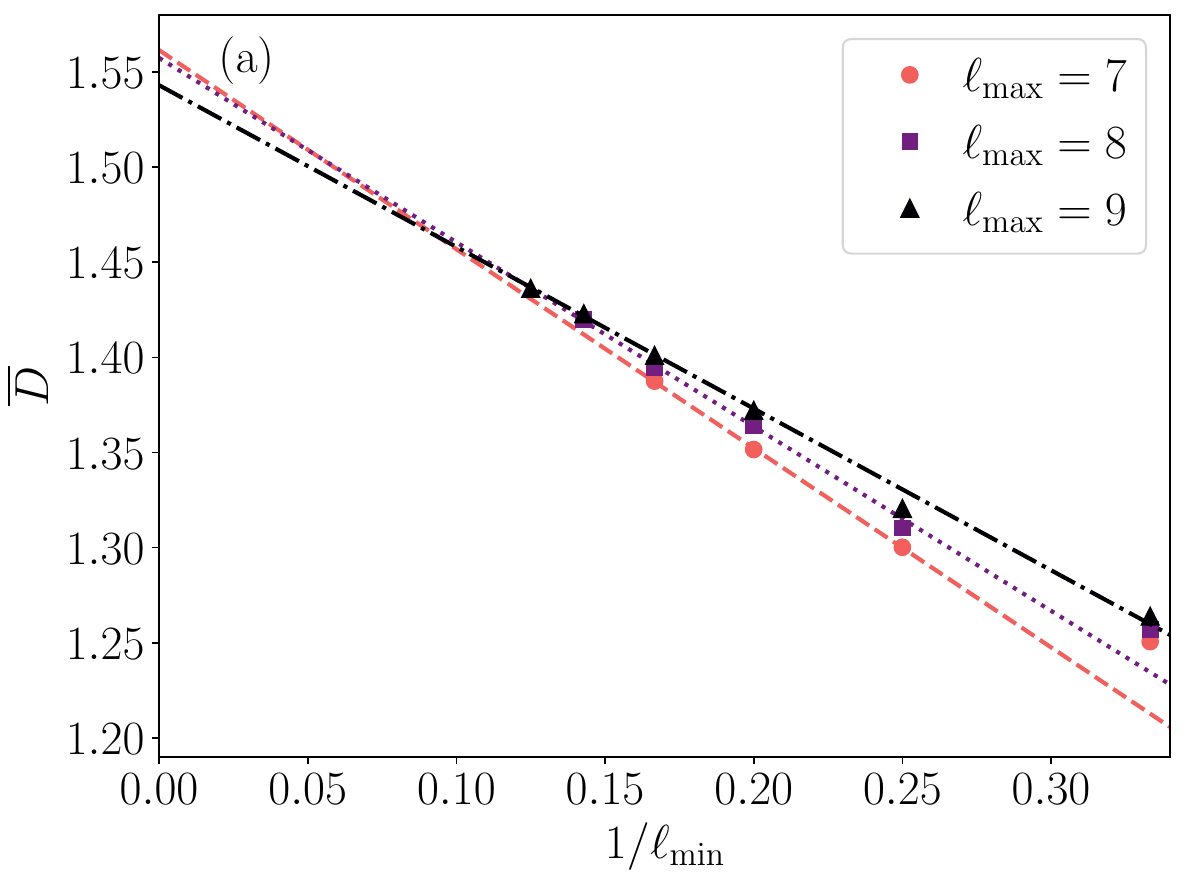}
\includegraphics[width=\columnwidth]{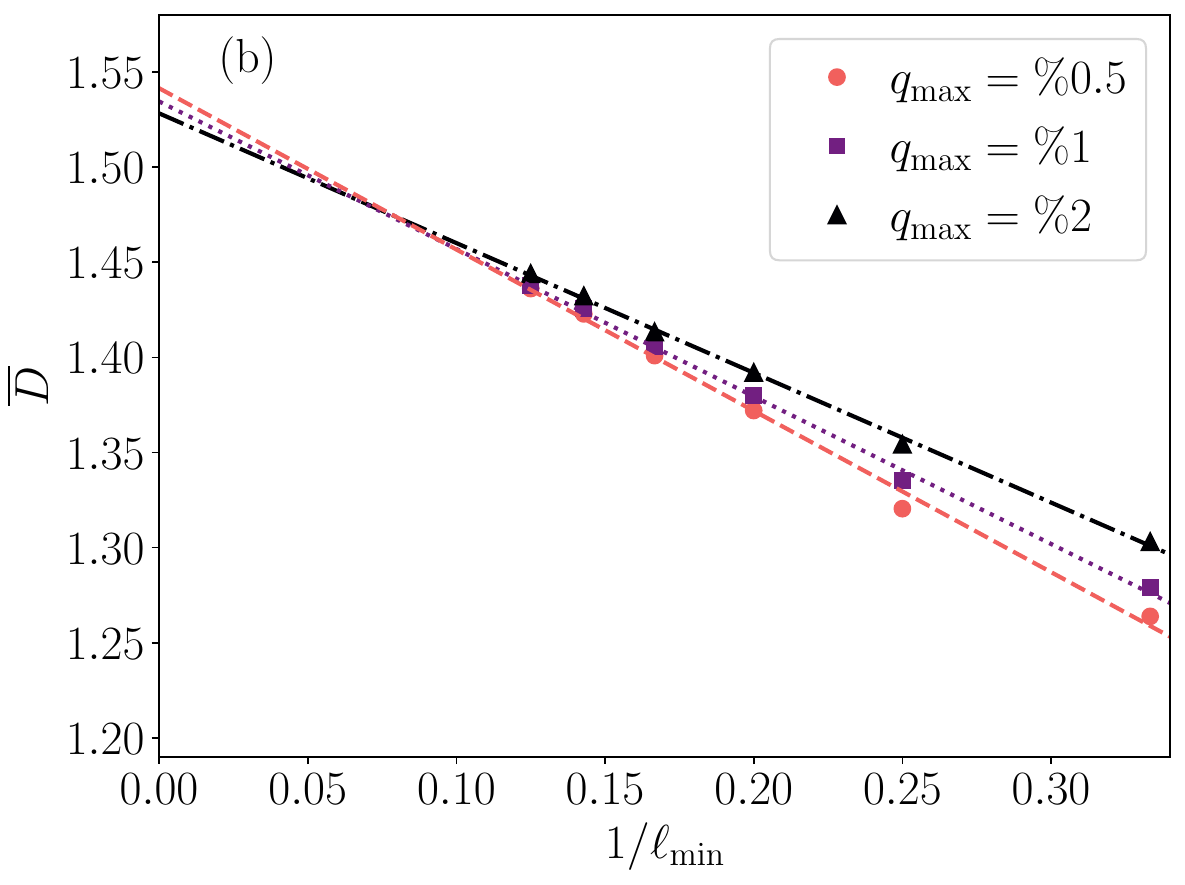}
\caption{Diffusion coefficient $\overline{D}$ averaged over the time window $t \in [20,100]$ as a function of $1/\ell_\mathrm{min}$, for different $\ell_\mathrm{max}$ [panel (a)], and $q_\mathrm{max}$ [panel (b)].
The dashed lines show a linear extrapolation of the asymptotic value of $\overline{D}$ for $\ell_\mathrm{min} \to \infty$, obtained considering the three data points at largest $\ell_\mathrm{min}$ at fixed $\ell_\mathrm{max}$ [panel (a)], and all the data points at fixed $q_\mathrm{max}$ [panel (b)]}.
This gives the rough estimate $\lim_{\ell_\mathrm{min} \to \infty} \overline{D} \approx 1.55$.
\label{fig:benchmark_diffusion_constant_ave}
\end{figure}

\begin{figure}
\centering
\includegraphics[width=\columnwidth]{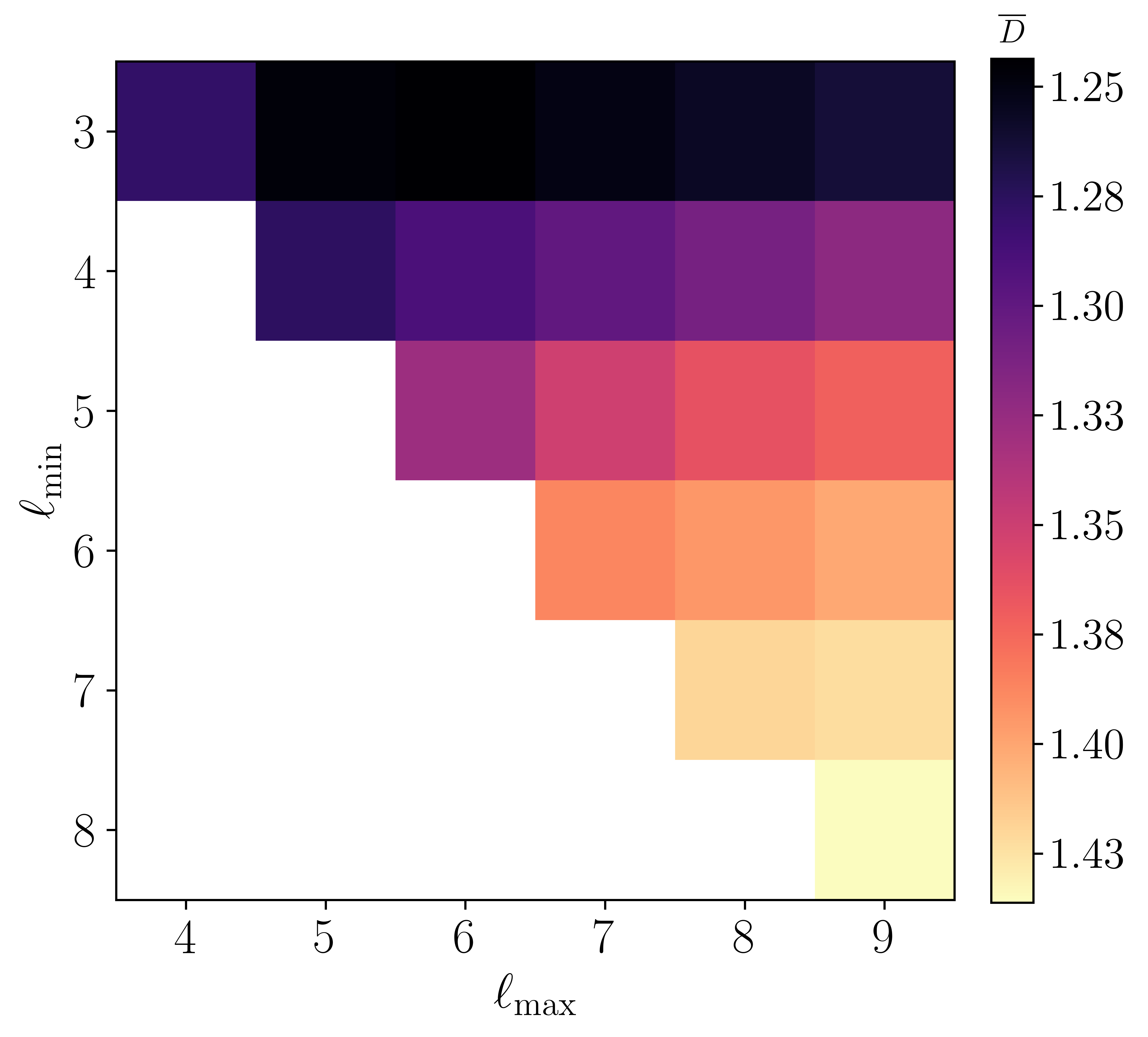}
\caption{Colormap of the diffusion coefficient $\overline{D}$ averaged over the time window $t \in [ 20,100 ]$ for different $\ell_\mathrm{min}$ and $\ell_\mathrm{max}$, at $q_{\mathrm{max}}=0.5\%$. Colors remain almost constant by moving on the horizontal axis, while they differ by moving on the vertical axis. This implies that $\overline{D}$ does not depend largely on $\ell_\mathrm{max}$, while it strongly depends on $\ell_\mathrm{min}$. Moving both from left to right and from up to down, $\overline{D}$ increases and approaches the asymptotic value.
}
\label{fig:benchmark_diffusion_constant_colormap}
\end{figure}

\begin{figure}
\centering
\includegraphics[width=\columnwidth]{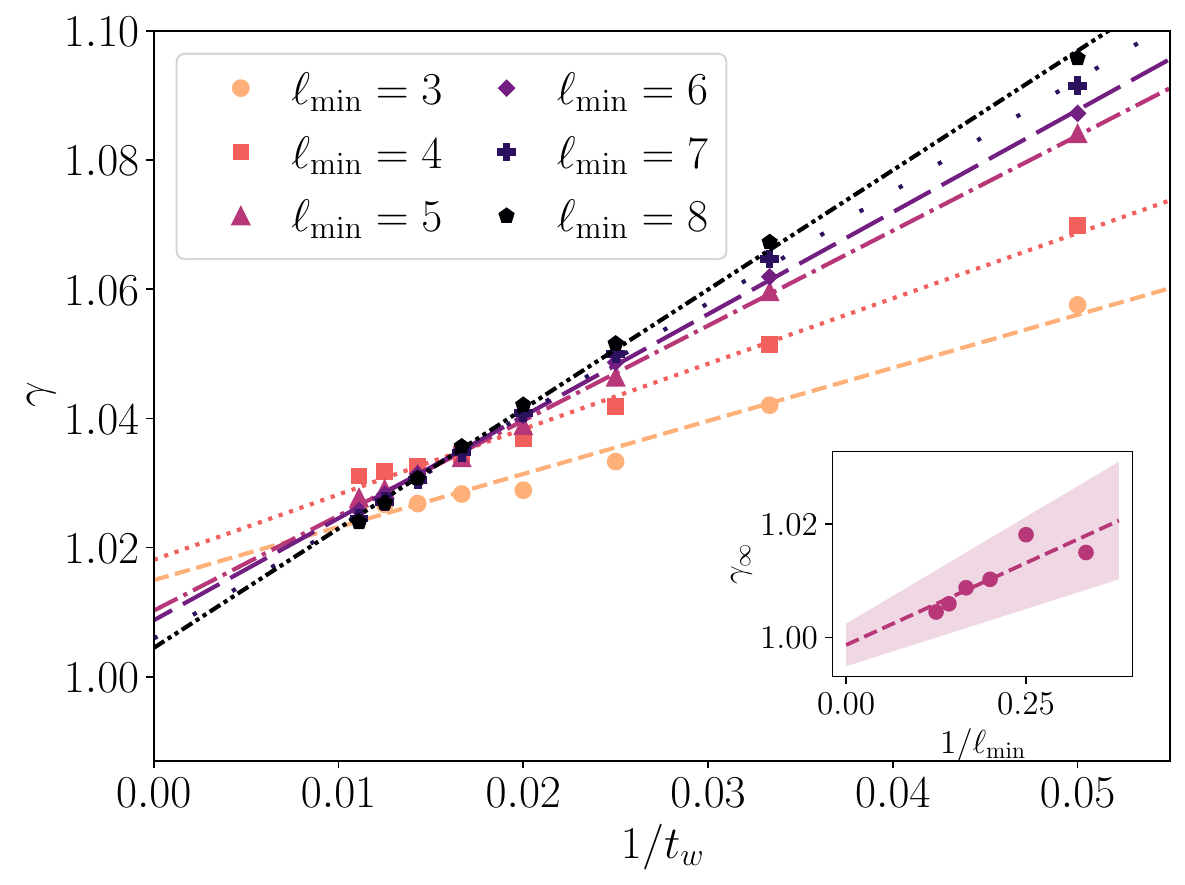}
\caption{Power-law exponent $\gamma$ of $\sigma^2_E(t)$ as a function of the window time $1/t_w$ for several $\ell_\mathrm{min}$, at $\ell_\mathrm{max}=9$ and $q_\mathrm{max}=0.5\%$. By linear fits, we extrapolate $\gamma_\infty = \lim_{t_{w} \to \infty} \gamma$. (Inset) $\gamma_\infty$ as a function of $1/\ell_\mathrm{min}$. By a linear fit, we extrapolate $\gamma_\infty$ in the limit of $\ell_\mathrm{min} \to \infty$, obtaining $\gamma^*_\infty = 0.999 \pm 0.004$, compatible with $1$ which is the expected power-law exponent for a diffusive system.}
\label{fig:benchmark_slope_drift}
\end{figure}

To demonstrate the efficiency of the LITE algorithm, we apply it to an established model for which the expected dynamics is known.
We show that LITE is able to access previously unreachable long times with excellent convergence properties.
Specifically, we consider the one-dimensional, mixed-field Ising spin chain with Hamiltonian
\begin{equation}
\label{eq:Ising_Hamiltonian}
    H = \sum_{m}J \sigma_m^z\sigma_{m+1}^z + h_T\sigma_m^x+h_L\sigma_m^z,
\end{equation}
where $\sigma_m^\eta$ (with $\eta \in \{x,y,z\}$) are Pauli matrices acting on the physical site $m$.
Hamiltonian~\eqref{eq:Ising_Hamiltonian} represents a nearest-neighbor, translation-invariant model. 
A number of recent works have discussed the same Hamiltonian in the context of diffusive dynamics~\cite{leviatan2017quantum,klein2022time,Pollmann2022}.
For Hamiltonian~\eqref{eq:Ising_Hamiltonian}, we define the local energy as
\begin{equation}
    h^r_m \coloneqq J \sigma_m^z\sigma_{m+1}^z + \frac{1}{2} \biggl( h_T \bigl(\sigma_m^x + \sigma_{m+1}^x)+h_L(\sigma_m^z + \sigma_{m+1}^z) \biggr).
\end{equation}
Given that energy is a globally conserved quantity, and considering an initial state of the form~\eqref{eq:init_state}, at late times one expects the variance of the energy distribution to scale as $\sigma^2_E = D t$, with $D$ a constant. 
The exact value of $D$ can only be determined in large-scale systems and at very long times.

We set $J=1$, $h_T=1.4$, and $h_L = 0.9045$, ensuring fast entanglement growth and chaotic dynamics~\cite{Huse2013}.
We initialize the system in the state~\eqref{eq:init_state} where the part that deviates from the infinite-temperature density matrices $\rho_{m,\infty}$ is spanned by three sites and given by Eq.~\eqref{eq:init_state_thermal_part} with $\beta=0.05$.
Using the LITE approach, we time evolve this infinitely extended initial state for different configurations of the algorithm parameters. 
After each time-evolution step, we measure the distribution of energy $E^r_m$, the corresponding variance according to Eq.~\eqref{eq:energy_variance}, and the diffusion coefficient according to Eq.~\eqref{eq:diff_const_explicit}.

In Fig.~\ref{fig:benchmark_diffusion_constant}, we depict the diffusion coefficient as a function of time for different values of $\ell_\mathrm{min}$ with fixed $\ell_\mathrm{max}=9$ and $q_\mathrm{max}=0.5\%$.
At short times, up to the first occurrence of minimization, all the curves are identical.
In this first regime of the dynamics, the system has a ballistic behavior and $D(t)$ increases with time approaching its final plateau value.
The saturation process is however longer than the timescales at which minimization appears; thus, we find a dependence of the saturated value of $D(t)$ on $\ell_\mathrm{min}$: increasing values of $\ell_\mathrm{min}$ are associated with an increasing (average) plateau value of $D$ reached at times $Jt \gtrsim 20$.
Another noticeable feature of Fig.~\ref{fig:benchmark_diffusion_constant} is the oscillations of $D$ forming at increasing times.
We attribute these oscillations to algorithmic artifacts emerging from the minimization and the removal of information at $\ell_\mathrm{min}$ that can affect information flow.
As $\ell_\mathrm{min}$ is increased, the magnitude of oscillations shrinks consistently.
Therefore, they can be interpreted as finite-scale effects associated with $\ell_\mathrm{min}$.
Intuitively, this depends on the fact that expectation values of operators with support on a few sites should not be influenced by the dynamics of local information happening at much larger scales.

Importantly, we find a clear convergence of the average plateau values of $D$ as a function of $1/\ell_\mathrm{min}$.
This convergence is better analyzed in Fig.~\ref{fig:benchmark_diffusion_constant_ave} depicting the diffusion coefficient $\overline{D}$ averaged in the interval $t \in [20,100]$ as a function of $1/\ell_\mathrm{min}$ for different $\ell_\mathrm{max}$, and $q_\mathrm{max}$.
Using a linear extrapolation of the asymptotic value of $\overline{D}$ for $\ell_\mathrm{min} \to \infty$, we find that $\lim_{\ell_\mathrm{min} \to \infty} \overline{D} \approx 1.55$.
This value is within a $5{-}10 \%$ margin from the values $1.4{-}1.46$ found for the same model in recent works~\cite{Pollmann2022,thomas2023comparing,wang2023diffusion}.
These discrepancies can be attributed to different reasons.
For instance, we consider longer timescales than Ref.~\onlinecite{Pollmann2022} that predicts the value $1.4$; while we time evolve up to $Jt \sim 100$ and average over $t \in [20,100]$, in Ref.~\onlinecite{Pollmann2022} the maximum evolution time is $Jt \sim 20 $ at which the diffusion coefficient has not yet converged to its long-time plateau. 
In addition, other works~\cite{thomas2023comparing} do not provide an asymptotic extrapolation as that in Fig.~\ref{fig:benchmark_diffusion_constant_ave}.
As the exact value of the diffusion constant for the mixed-field Ising model is not known, in Sec.~\ref{sec:numerical_open} below we further benchmark our approach by applying it to a model for which the exact value of the diffusion constant has been derived.

Clearly, $\ell_\mathrm{min}$ is the most decisive parameter of our algorithm as it determines the scale above which the conservation of local constants of motion is not guaranteed (see App.~\ref{app:sec:local-conserved-quantities}).
Given a fixed value of $\ell_\mathrm{min}$, the remaining parameters $\ell_\mathrm{max}$ and $q_\mathrm{max}$ only have a (similar) weak influence on the resulting value of $D$, as exemplified in Figs.~\ref{fig:benchmark_diffusion_constant_ave}-\ref{fig:benchmark_diffusion_constant_colormap}.
There is, however, a caveat: while $\ell_\mathrm{max}$ sets the maximum value of $\ell_\mathrm{min}$ and should be chosen as large as possible, no such line of reasoning exists for $q_\mathrm{max}$.
In fact, lowering $q_\mathrm{max}$ too much makes the removal of information via the minimization scheme ineffective, while too large values of $q_\mathrm{max}$ lead to an accumulation of information at level $\ell_\mathrm{max}$ before the minimization at level $\ell_\mathrm{min}$ is triggered, which might distort the information flow and the system dynamics.
Thus, we expect the presence of an optimal range of values for $q_\mathrm{max}$.
Empirically, we do not find significant differences in the range $q_\mathrm{max}\sim 0.5-2\%$ (see Fig.~\ref{fig:benchmark_diffusion_constant_ave}).

Fig.~\ref{fig:benchmark_diffusion_constant} contains a barely visible yet important subtlety: even in the saturated regime ($Jt \gtrsim 20$), the diffusion coefficient $D$ is not entirely constant over the time of evolution.
Instead, we observe a slight time-dependent increase in $D$.
To quantify the drift, we assume the functional form $\sigma_E^2(t) \propto D t^\gamma$ and apply a power-law fit to different time windows in the saturated regime of the time evolution.
We fix the window length to $\Delta t = 10$, and let $t_w$ denote the initial time of the window.
The power-law exponent $\gamma$ as a function of the time window is shown in Fig.~\ref{fig:benchmark_slope_drift} for different values of $\ell_\mathrm{min}$.
We find that the flow of the exponent $\gamma$ supports diffusive dynamics at late times and large $\ell_\mathrm{min}$: $\lim_{\ell_\mathrm{min}\rightarrow \infty} \lim_{t_w \rightarrow \infty} \gamma = 0.999 \pm 0.004$.
Importantly, this shows that the algorithm is able to capture the correct long-time behavior of local observables. 
Indeed, for the current model, any deviation from purely diffusive behavior would hint at a systematic error in the algorithm that distorts the dynamics.

\section{Lindblad dissipative dynamics via local-information time evolution}
\label{sec:dissipative_dynamics}

\subsection{The Lindblad master equation within LITE}

The von Neumann equation~\eqref{eq:full_vonNeumann} and the subsystem equation of motion~\eqref{eq:subsys_vonNeumann} describe the dynamics of \textit{closed} quantum systems.
However, in many practical cases, achieving a reliable description of a quantum-system dynamics requires considering its interaction with the external environment~\cite{breuer2002theory}.
This interaction introduces quantum dissipation.
Unlike closed systems, the dynamics of \textit{open} quantum systems cannot be represented by unitary time evolution. 
Nevertheless, one can often formulate it in terms of a quantum master equation.
Among those, the Lindblad master equation~\cite{benatti2005open,rivas2012open} (or Markovian master equation) holds a significant role.
It characterizes the evolution of a system that is coupled to a thermal bath which is memoryless, implying that its timescale is considerably shorter than any other timescale in the problem.
The Lindblad master equation is a first-order linear differential equation for the system's density matrix; it reads
\begin{equation}
    \label{eq:lindblad}
    \partial_t \rho = -i \left[H,\rho\right] + \sum_j \gamma_j \left(L_j \rho L_j^{\dagger} - \frac{1}{2}\lbrace L_j^{\dagger}L_j, \rho \rbrace \right).
\end{equation}
Here, $\lbrace \sbullet , \sbullet \rbrace$ is the anticommutator, and $L_j, L_j^{\dagger}$ are the Lindblad jump operators modeling the dissipative dynamics.
The jump operators describe how the environment influences the system and must in principle be derived from the full microscopic Hamiltonian that accounts for the system and the environment. 
The coupling constants $\gamma_j\geq0$ quantify the strength of dissipation.

The LITE approach can be readily extended to open quantum systems governed by the Lindblad master equation.
This extension is particularly straightforward when dealing with Lindblad jump operators acting on individual physical sites.
In this scenario, the subsystem equation of motion takes the form:
\begin{align}
\label{eq:subsystem_lindblad}
    \partial_t \rho^{\ell}_n = &-i \left[H^{\ell}_n, \rho_n^{\ell}\right] \notag\\
    &+ \sum_{m \in \mathcal{C}_n^\ell} \gamma_m\left(L_m \rho_n^\ell L_m^{\dagger} -\lbrace L_m^{\dagger}L_m, \rho_n^\ell \rbrace \right) \notag \\
    &-i \mathrm{Tr}_{L}^r\left(  \left[H_{n-r/2}^{\ell+r}- H_n^{\ell},\rho_{n-r/2}^{\ell+r} \right] \right) \notag \\
    &-i \mathrm{Tr}_{R}^r\left(  \left[H_{n+r/2}^{\ell+r}- H_n^{\ell},\rho_{n+r/2}^{\ell+r} \right] \right).
\end{align}
Notice that $m$ denotes the physical sites within the subsystem $\mathcal{C}^\ell_n$.
Remarkably, the second term on the right-hand side exclusively involves $\rho^{\ell}_n$, allowing for the seamless inclusion of onsite dissipators.
In fact, this addition does not necessitate any modifications to the LITE approximate time-evolution scheme.

As previously discussed, the fundamental idea of the LITE approach is to selectively remove local information while preserving the local dynamics.
This strategy enhances the accuracy of recovery maps, facilitating the closure of the subsystem equation of motion~\eqref{eq:subsys_vonNeumann}.
From this perspective, the introduction of Lindblad dissipators is expected to improve the convergence properties of the algorithm.
Consequently, LITE emerges as a particularly well-suited approach for addressing systems with dissipative dynamics.
In a concurrent (mainly experimental) work~\cite{harkins2023nanoscale}, some of the present authors applied this approach to reproduce experimental results for driven nitrogen-vacancy centers in diamonds.
Specifically, in Ref.~\onlinecite{harkins2023nanoscale} thermalization is used to engineer and stabilize mesoscopic shell-like spin textures characterized by spins exhibiting opposite polarization on either side of a critical radius.
These textures encompass about $O(100)$ spins and are stable for several minutes.
In the following section, we discuss in detail another application of LITE to open quantum systems.

\subsection{Numerical results for diffusive dynamics in the presence of onsite dephasing}
\label{sec:numerical_open}

\begin{figure}
\centering
\includegraphics[width=\columnwidth]{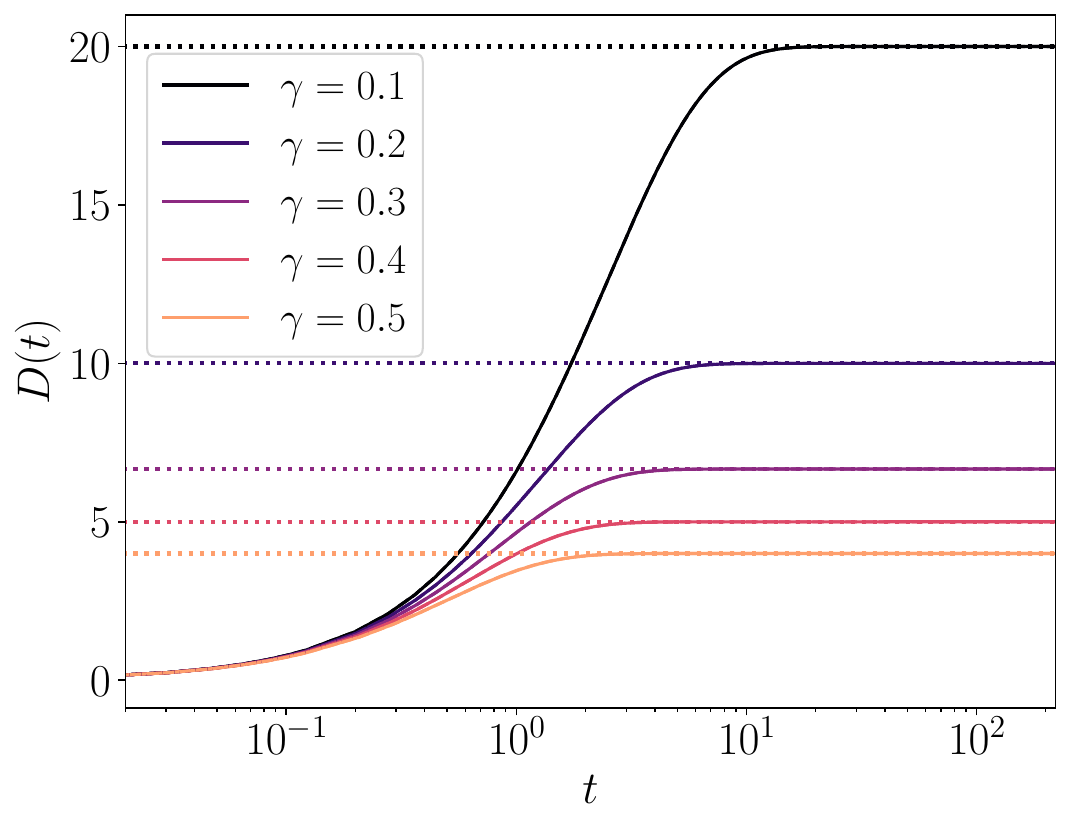}
\caption{
For each $\gamma$, we plot the diffusion coefficient of magnetization transport in Eq.~\eqref{eq:diff_const_explicit_open} at $(\ell_{\mathrm{min}},\ell_{\mathrm{max}})= (3, 5)$ (solid line) and $(4, 6)$ (dashed line), for fixed $J=1$.
The curves at different $(\ell_\mathrm{min}, \ell_\mathrm{max})$ differ at most by $\sim O(10^{-8})$, indicating the fast convergence of LITE in the presence of dissipation.
After a short-time ballistic behavior, $D(t)$ saturates to a plateau dependent on $\gamma$.
The long-time diffusion coefficient perfectly agrees with the exact value in Eq.~\eqref{eq:diffusion-coefficient-open} (dotted lines) (see also Fig.~\ref{fig:diffusion_constant_scaling_open}).}
\label{fig:diffusion_constant_open}
\end{figure}

\begin{figure}
\centering
\includegraphics[width=\columnwidth]{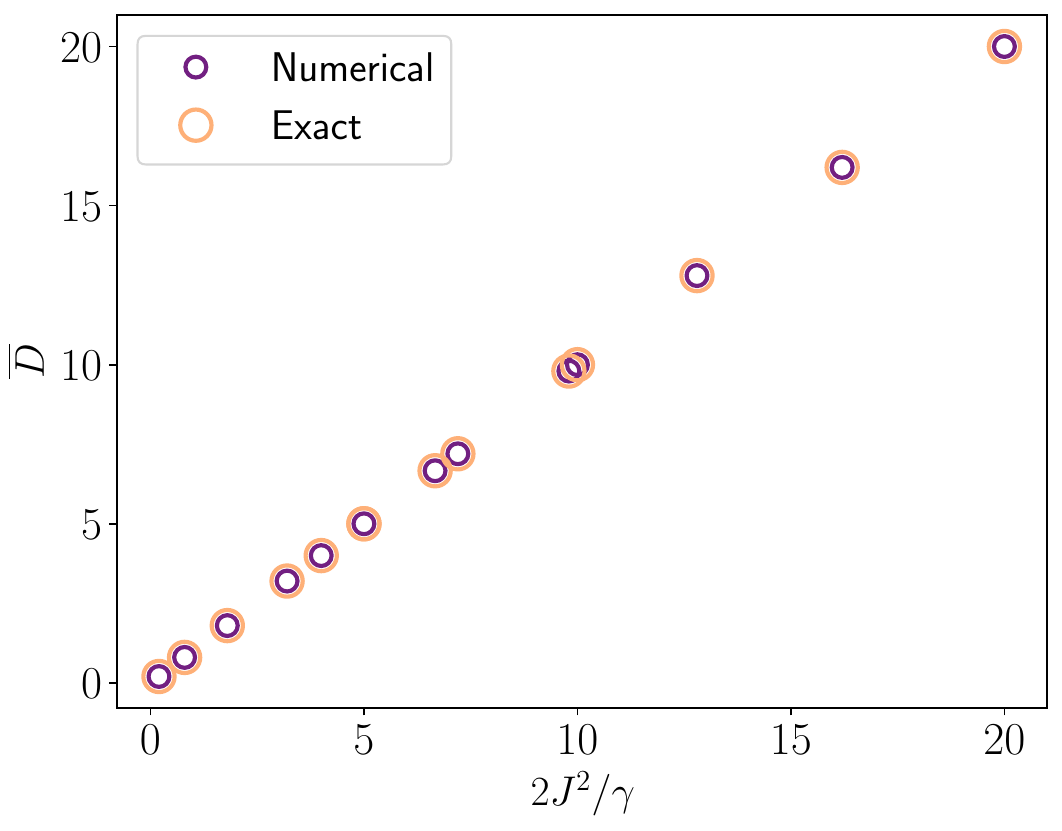}
\caption{
Diffusion constant of magnetization transport in Eq.~\eqref{eq:diff_const_explicit_open} averaged over the time window $t \in [30,220]$ for $(\ell_\mathrm{min}, \ell_\mathrm{max}) = (4,6)$ and $q_\mathrm{max} = 1 \%$ as a function of $2 J^2/\gamma$ (purple circles).
Data points correspond to $J = 0.1, 0.2, 0.3, 0.4, 0.6, 0.7, 0.8, 0.9$ for $\gamma = 0.1$, and $\gamma = 0.1, 0.2, 0.3, 0.4, 0.5$ for $J = 1$.
The long-time diffusion constant, $\overline{D}$, agrees with the exact result in Eq.~\eqref{eq:diffusion-coefficient-open} (orange circles); they differ at most by $\sim O (10^{-6})$.
}
\label{fig:diffusion_constant_scaling_open}
\end{figure}

To showcase the efficiency of LITE for open systems within the Lindblad framework, we employ it for a model for which the exact diffusion coefficient has been analytically derived~\cite{znidarivc2010exact,turkeshi2021diffusion,jin2022exact}.
Our analysis shows that in this case LITE is able to obtain the diffusion coefficient with high precision, even at small scales $\ell_\mathrm{min}$ and $\ell_\mathrm{max}$.
Specifically, we simulate an $XX$ spin chain subject to onsite dephasing Lindblad operators.
The dynamics of the system is determined by the Lindblad master equation~\eqref{eq:subsystem_lindblad}, with Hamiltonian
\begin{equation}
\label{eq:Ising_Hamiltonian_open}
    H = \sum_{m}J (\sigma_m^x\sigma_{m+1}^x + \sigma_m^y \sigma_{m+1}^y),
\end{equation}
and dephasing Lindblad operators acting on each physical site with coupling strength $\gamma$:
\begin{equation}
    \label{eq:Lindblad-open}
    L_{m} = \sigma^z_i; \quad \gamma_m = \gamma \; \forall m. 
\end{equation}
The presence of local dephasing effectively introduces interactions in the system that change the transport characteristics from ballistic to diffusive, as previously observed via matrix-product operator techniques~\cite{znidarivc2010dephasing,znidarivc2010exact,znidarivc2013transport}.
The exact value of the diffusion coefficient for the open chain in Eqs.~\eqref{eq:Ising_Hamiltonian_open}-\eqref{eq:Lindblad-open} was derived to be~\cite{znidarivc2010exact,turkeshi2021diffusion,jin2022exact}
\begin{equation}
    D = \frac{2 J^2}{\gamma}.
    \label{eq:diffusion-coefficient-open}
\end{equation}

For the system described by Eqs.~\eqref{eq:Ising_Hamiltonian_open}-\eqref{eq:Lindblad-open} we conduct a series of simulations using LITE to study magnetization transport.
We initialize the system in the state~\eqref{eq:init_state} where $\rho_{n,\mathrm{init}}^\ell$ is a product state spanned by 11 sites with a finite negative magnetization, $\rho_{n,\mathrm{init}}^\ell = \bigotimes_m \rho^0_m$ with $ p_m := \langle \sigma^z_m \rangle = -0.2$.
As in Sec.~\ref{sec:results_mixed_field_Ising}, we time evolve this effectively infinitely extended system for different parameters of the model.
At each time-evolution step, we compute the diffusion coefficient according to
\begin{multline}
\label{eq:diff_const_explicit_open}
    D = \frac{i}{2}\sum_m (m-\overline{m})^2\frac{\langle \left[H, \sigma^z_m \right] \rangle}{P} \\
    - i \sum_m (m-\overline{m}) \frac{p_m}{P} \sum_k (k-\overline{m}) \frac{\langle \left[H, \sigma^z_k \right] \rangle}{P},
\end{multline}
where $P = \sum_m p_m$.

Results are reported in Figs.~\ref{fig:diffusion_constant_open} and \ref{fig:diffusion_constant_scaling_open}.
In Fig.~\ref{fig:diffusion_constant_open}, we show the diffusion coefficient as a function of time, $D(t)$, for different coupling strengths $\gamma$, $(\ell_\mathrm{min}, \ell_\mathrm{max}) = (3,5)$ (solid lines) and $(\ell_\mathrm{min}, \ell_\mathrm{max}) = (4,6)$ (dashed line).
The diffusion coefficient $D(t)$ shows remarkable features.
First, the results are almost independent of the scales  $(\ell_\mathrm{min}, \ell_\mathrm{max})$, even at those modest values.
Moreover, the artificial oscillations present in the closed case (see Fig.~\ref{fig:benchmark_diffusion_constant}) are now completely absent, and the diffusion coefficient is highly stable and constant in the long-time limit.
Finally, we recover the analytical result in Eq.~\eqref{eq:diffusion-coefficient-open}, shown as dotted lines.
In Fig.~\ref{fig:diffusion_constant_scaling_open}, we further demonstrate the high agreement of the LITE results (purple circles) with the analytical ones (orange circles) by computing $\overline{D}$, that is, the diffusion constant averaged over the time window $t \in [30,220]$, as a function of $2J^2/\gamma$.
The points are aligned on a straight line with slope 1, as implied by Eq.~\eqref{eq:diffusion-coefficient-open}.
The difference between the LITE results and the exact ones is at most about $O(10^{-6})$.

Our findings demonstrate that onsite Lindblad operators enhance the convergence properties of LITE.
We attribute this enhancement to the removal of information operated by dephasing, which in turn improves the accuracy of the subsystem time evolution via recovery maps.
In the simulations, we have indeed observed that the minimization protocol is activated only at short timescales, due to the presence of information at level $\ell_\mathrm{max}$.
At intermediate and long timescales, the total amount of information in the system is very small.
Consequently, efficient time evolution at a reduced scale $\ell_\mathrm{max}=3$ is feasible without the need for additional minimization steps or increasing time-evolution scales $\ell_\mathrm{max}$.
Thanks to this, Figs.~\ref{fig:diffusion_constant_open} and \ref{fig:diffusion_constant_scaling_open} have been obtained by using very modest computational resources.

\section{Conclusion and outlook}
\label{sec:conclusion}

We have proposed a novel algorithm (LITE) for the approximate time evolution of generic local many-body quantum Hamiltonians.
Our approach is based on statistical arguments concerning the unidirectional flow of quantum information, which primarily progresses to larger scales without returning to smaller scales to influence local observables.
By leveraging the concepts of local information and information currents, we systematically discard long-range correlations in a controlled manner.
This allows us to obtain an accurate description of local states at any given time.

LITE operates by decomposing the system into subsystems and solving their von Neumann equations in parallel.
For closing the (in principle infinite) hierarchy of subsystem equations of motion, we have introduced two scales, $\ell_\mathrm{min}$ and $\ell_\mathrm{max}$.
Parameter $\ell_\mathrm{min}$ defines the scale at which we systematically remove local information while preserving lower-level density matrices (for $\ell<\ell_\mathrm{min}$) and information currents (for $\ell\leq\ell_\mathrm{min}$).
Parameter $\ell_\mathrm{max}$ is the maximum scale on which time evolution is performed, and is constrained by available computational resources.
The knowledge of the subsystem states at the maximum scale $\ell_\mathrm{max}$ is used to accurately determine the information flow on smaller scales.
By construction, the LITE approach conserves all local constants of motion at scales $\ell<\ell_\mathrm{min}$. 
Scale $\ell_\mathrm{min}$ also controls the accuracy of the results.
While the computational complexity of LITE scales exponentially with the subsystem level $\ell_\mathrm{max}$, it increases only \textit{linearly} with the total system size, which allows the investigation of large-scale systems and long timescales.
Crucially, LITE does not require any external assumptions on the information currents nor does it require the presence of symmetries, such as translation invariance.
Therefore, this approach is highly versatile and suitable for exploring the thermalization dynamics (or its absence) across various physical contexts, encompassing systems with diverse hydrodynamic behaviors.

Within the LITE framework, we can initialize time evolution from diverse initial states, including domain walls in finite-size systems or infinitely extended translation-invariant states~\cite{klein2022time}. 
Here, we have demonstrated its excellent convergence properties when starting from asymptotically time-invariant states (that is, states in which the state of the asymptotic region commutes with the Hamiltonian), particularly those near infinite temperature.
This enables effective simulations of infinitely extended systems.
Starting with such states, we have investigated the dynamics of the mixed-field Ising model for a set of parameters in which the system is highly chaotic; thus far, other time-evolution methods, such as matrix product states with finite bond dimensions, have not obtained concluding results.
Remarkably, we have been able to perform time evolution up to very long times and get an accurate estimate of the power-law exponent for energy diffusion and of the energy diffusion constant.

The LITE approach is especially well suited for investigating Lindblad dissipative dynamics with local dissipators.
In that case, the algorithm converges even faster than for closed systems due to the additional removal of information operated by the dissipators.
Harkins \textit{et al.}~\cite{harkins2023nanoscale} demonstrated this by reproducing experimental results for the magnetization transport driven by nitrogen-vacancy centers in diamonds. 
In addition, here we have shown results for an open $XX$ spin chain subject to onsite dephasing that presents diffusive transport in the long-time limit and for which the exact value of the diffusion constant has been analytically derived~\cite{znidarivc2010exact,turkeshi2021diffusion,jin2022exact}.
The diffusion constant calculated with LITE perfectly agrees with the exact value, even when computed for small scales $\ell_\mathrm{min}$ and $\ell_\mathrm{max}$.
This shows that, in the case of Lindblad dissipative dynamics, LITE provides accurate results at extremely modest computational costs. 
This open doors for unprecedented accurate investigations of long-time and large-scale open many-body quantum systems.

Since LITE relies solely on quantum thermalization and the consequent entanglement growth, we expect it to also be particularly appropriate for simulating subdiffusive and superdiffusive transport in quantum systems. 
The investigation of these transport regimes is a growingly interesting research direction at the forefront of theoretical and experimental physics~\cite{znidaric2016diffusive,wei2022quantum}.
Moreover, while here our focus has primarily been on one-dimensional nearest-neighbor Hamiltonians, the LITE approach can be applied to generic finite-range Hamiltonians and potentially extended to higher-dimensional systems.
In the future, we anticipate exploring connections between the LITE approach and tensor network algorithms for time evolution, with the potential for mutual insights and efficiency improvements.
Additionally, combining a tensor-network ansatz with the LITE approach for compressing high-level density matrices could yield further algorithmic enhancements.
Furthermore, we anticipate the possibility of gaining valuable insights into the spatial and temporal behavior of entanglement in many-body systems by employing the framework of the information lattice.
The information lattice could indeed provide additional information on dynamical heterogeneity observed in localized systems~\cite{artiaco2022spatiotemporal} and, more generally, on the complex structure of bipartite quantum entanglement in both ergodic and localized systems.
Finally, we expect that the LITE approach will make valuable contributions in the field of quantum computation, such as facilitating classically optimized Hamiltonian simulations on quantum hardware~\cite{keever2023classically,keever2023towards}.

\section*{Acknowledgements}

We thank L.~Herviou for insightful discussions. This work received funding from the European Research Council (ERC) under the European Union’s Horizon 2020 research and innovation program (Grant Agreement No. 101001902), and the Knut and Alice Wallenberg Foundation (KAW) via the project Dynamic Quantum Matter (2019.0068). The computations were enabled by resources provided by the National Academic Infrastructure for Supercomputing in Sweden (NAISS) and the Swedish National Infrastructure for Computing (SNIC) at Tetralith, partially funded by the Swedish Research Council through Grant Agreements No.~2022-06725 and No.~2018-05973.
The research of T.~K.~K. is funded by the Wenner-Gren Foundations.

%################################
\bibliography{biblio}
%################################

%######################################################

\newpage
\onecolumngrid
\appendix

\section{Conservation of local constants of motion}
\label{app:sec:local-conserved-quantities}

A fundamental feature of the LITE approach is the conservation of local constants of motion on scales smaller than the minimization scale, $\ell < \ell_\mathrm{min}$.
Importantly, truncation errors inherent to the algorithm do not compromise this conservation.
Consider a \textit{local} operator $\mathcal{O}$, which can be expressed in the form 
\begin{equation}
    \mathcal{O}=\sum_{n}\mathcal{o}_{n}^{\ell_{\mathrm{min}}-1},
\end{equation}
where $\mathcal{o}_{n}^{\ell_{\mathrm{min}}-1}$ is an operator acting only on the subsystem $\mathcal{C}^{\ell_\mathrm{min}-1}_n$.
By assumption, $\mathcal{O}$ commutes with the Hamiltonian: $[H,\mathcal{O}]=0$.
Then, the expectation value of $\mathcal{O}$ at time $t$ is a constant of the motion:
\begin{equation}
    \braket{\mathcal{O}}_{t}=\sum_{n}\tr[\mathcal{o}_{n}^{\ell_{\mathrm{min}}-1}\rho_{n}^{\ell_{\mathrm{min}}-1}(t)]=\text{const}.
    \label{eq:preservation-of-constant-of-motion}
\end{equation}
In this appendix, we demonstrate that, within the LITE algorithm, $\braket{\mathcal{O}}_{t}$ remains unchanged over time, mirroring the behavior under exact dynamics.

The LITE algorithm comprises two primary steps, as depicted in Fig.~\ref{fig:algorithm} in the main text: local-information minimization and time evolution via Petz recovery maps.
First, let us consider the minimization of local information on level $\ell_{\mathrm{min}}$.
As discussed in Sec.~\ref{sec:approximate_time_evolution} in the main text, minimization under constraints~\eqref{eq:condition1} and \eqref{eq:condition2} does not affect the density matrices on levels $\ell \leq \ell_{\mathrm{min}}-1$; thus, this step does not alter the expectation value~\eqref{eq:preservation-of-constant-of-motion}. 
Second, let us consider the time-evolution step, which corresponds to the integration of the equation of motion~\eqref{eq:subsys_vonNeumann}.
To demonstrate that time integration also preserves local constants of motion, it is sufficient to show that 
\begin{equation}
\label{app:eq:condition-constant-of-motion}
    \sum_{n}\tr[\mathcal{o}_{n}^{\ell_{\mathrm{min}}-1}\partial_{t}\rho_{n}^{\ell_{\mathrm{min}}-1}(t)]=0,
\end{equation}
where $\partial_{t}\rho_{n}^{\ell_{\mathrm{min}}-1}(t)$ is the time derivative,
as defined by the algorithm. 
In the LITE approach, it takes the form [see Eq.~\eqref{eq:subsys_vonNeumann} in the main text]
\begin{multline}
\label{app:eq:equation-of-motion}
    \partial_{t}\rho_{n}^{\ell_{\mathrm{min}}-1}(t)=
    -i\left[H_{n}^{\ell_{\mathrm{min}}-1},\rho_{n}^{\ell_{\mathrm{min}}-1}\right] \\
    -i\mathrm{Tr}_{L}^{r}\left(\left[H_{n-r/2}^{\ell_\mathrm{min}-1+r}-H_{n}^{\ell_{\mathrm{min}}-1},\rho_{n-r/2}^{\ell_\mathrm{min}-1+r}\right]\right) 
    -i\mathrm{Tr}_{R}^{r}\left(\left[H_{n+r/2}^{\ell_\mathrm{min}-1+r}-H_{n}^{\ell_{\mathrm{min}}-1},\rho_{n+r/2}^{\ell_\mathrm{min}-1+r}\right]\right),
\end{multline}
where $\rho_{n}^{\ell_{\mathrm{min}}-1+r}$ are higher-level density matrices.
These are either known \textit{a priori}, for instance, at the initial time $t=0$, or recovered using projected Petz recovery maps (see App.~\ref{app:sec:PPRM}). 
In both cases, the density matrices $\rho_{n}^{\ell_{\mathrm{min}}-1+r}$ preserve lower-level density matrices [see Eqs.~\eqref{app:eq:PPRM-lower-levels_1}-\eqref{app:eq:PPRM-lower-levels_3} below].
By the recursive application of projected Petz recovery maps, we can construct the density matrix $\rho^{L-1}$ acting on the full system, from which the subsystem density matrices at level $\ell_\mathrm{min}-1$ for all $n$ can be obtained by suitable partial trace operations:
\begin{equation}
    \rho_{n}^{\ell_{\mathrm{min}}-1}  =\tr_{\bar{\mathcal{C}}_{n}^{\ell_{\mathrm{min}}-1}}\rho^{L-1}.
\end{equation}
Here $\bar{\mathcal{C}}_{n}^{\ell_{\mathrm{min}}-1}$ is the complement subsystem of $\mathcal{C}_{n}^{\ell_{\mathrm{min}}-1}$, that is, it is the set of all the physical sites that do not belong to the subsystem $\mathcal{C}_{n}^{\ell_{\mathrm{min}}-1}$.
By rewriting Eq.~\eqref{app:eq:equation-of-motion} as
\begin{equation}
    \partial_{t}\rho_{n}^{\ell_\mathrm{min}-1}(t)= -i\tr_{\bar{\mathcal{C}}_{n}^{\ell_{\mathrm{min}}-1}}([H,\rho^{L-1}]),
\end{equation}
we find that
\begin{align}
\label{app:eq:preservation-constant-of-motion}
    \sum_{n}\tr(\mathcal{o}_{n}^{\ell_{\mathrm{min}}-1}\partial_{t}\rho_{n}^{\ell_{\mathrm{min}}-1}(t))=-i\sum_{n}\tr(\mathcal{o}_{n}^{\ell_{\mathrm{min}}-1}[H,\rho^{L-1}])
    =-i\tr(\mathcal{O}[H,\rho^{L-1}])=-i\tr([\mathcal{O},H]\rho^{L-1})=0,
\end{align}
verifying Eq.~\eqref{app:eq:condition-constant-of-motion}.

To complete the proof, we need to demonstrate that the errors due to the finite time-step size used to integrate Eq.~\eqref{app:eq:equation-of-motion} do not affect the preservation of the local conserved quantities.
In the numerical implementation of LITE, we exclusively use Runge-Kutta integration methods (as discussed in App.~\ref{app:sec:runge-kutta}).
Let us write the Runge-Kutta integration scheme applied to Eq.~\eqref{app:eq:equation-of-motion} in the general form
\begin{equation}
    \rho_{n}^{\ell_{\mathrm{min}}-1}(t+\delta t)=\rho_{n}^{\ell_{\mathrm{min}}-1}(t)+\delta t\sum_{i=1}^{K}b_{i}\kappa_{n}^{\ell_{\mathrm{min}}-1,{i}} + O(\delta t^{K})
\end{equation}
where the $b_{i}$ are the Runge-Kutta parameters, the $\kappa_{n}^{\ell_{\mathrm{min}}-1,{i}}$
are derivative functions evaluated for different density matrices~\cite{feagin2012high}, and $K$ is the order of the truncation error.
Importantly, derivatives within the LITE approach are always of the form~\eqref{app:eq:equation-of-motion}. 
Therefore, by the same argument used in Eq.~\eqref{app:eq:preservation-constant-of-motion}, we find that
\begin{equation}
    \sum_{n}\tr(o_{n}^{\ell_{\mathrm{min}}-1}\kappa_{n}^{\ell_{\mathrm{min}}-1,{i}})=0.
\end{equation}
This proves that, by employing Runge-Kutta methods, $\braket{\mathcal{O}}_{t}$ is a constant of the motion up to machine precision (no matter the value of $\delta t$).
Note that, for other integration schemes, such as the Suzuki-Trotter decomposition, local conserved quantities are not guaranteed to be conserved~\cite{klein2022time}.

\section{Details on the Petz recovery maps}
\label{app:sec:PRM}

As discussed in the main text, recovery maps are needed for closing the equations of motion of the subsystems at a given level $\ell^*$ [Eq.~\eqref{eq:subsys_vonNeumann} in the main text], and, consequently, for performing the approximate time evolution within the LITE algorithm. 
The recovery maps employed in the numerical implementation of LITE are required to be optimized for computational efficiency and to satisfy the condition of nonalteration of lower-level density matrices.
Both aspects are discussed in this appendix.

\subsection{Petz recovery maps without error bounds}

Let us consider two (potentially) overlapping regions $A$ and $B$ with corresponding density matrices $\rho_A$ and $\rho_B$.
If the mutual information between the states of $A$ and $B$ vanishes [in equation form, $i(A;B)=0$ with $i$ defined in Eq.~\eqref{eq:mutual_information} in the main text], several analytical expressions can be employed to reconstruct the state of the union region $AB \coloneqq A \cup B$ using the reduced density matrices $\rho_A$ and $\rho_B$.
For instance, three distinct recovery maps yield the same (exact) density matrix $\rho_{AB}$ when $i(A;B)=0$~\cite{zhang2014lower}:
\begin{eqnarray}
\label{app:eq:Petz_map_sqrt_A}
    \rho_{A B} &=&  \rho_A^{1/2}~ \rho_{A \cap B }^{-1/2} ~\rho_B  ~ \rho_{A \cap B }^{-1/2} 
    ~\rho_A^{1/2} \\
    \label{app:eq:Petz_map_sqrt_B}
    &=& \rho_B^{1/2}~ \rho_{A \cap B }^{-1/2} ~\rho_A  ~ \rho_{A \cap B }^{-1/2} 
    ~\rho_B^{1/2} \\
    \label{app:eq:Petz_twisted}
    &=& \exp \left[ \, \ln (\rho_A) + \ln (\rho_B) - \ln (\rho_{A\cap B}) \, \right].
\end{eqnarray}
The last equation corresponds to the twisted Petz recovery map introduced in Eq.~\eqref{eq:Petz_map} of the main text.
Only the twisted Petz recovery map possesses a known error bound when the mutual information $i(A;B) \neq 0$ [see Eq.~\eqref{eq:Petz_error} in the main text].
While having an error bound is generally advantageous, recovering  $\rho_{AB}$ from Eq.~\eqref{app:eq:Petz_twisted} necessitates diagonalizing a larger matrix (with the same dimensions as $\rho_{AB}$) compared to Eqs.~\eqref{app:eq:Petz_map_sqrt_A} and \eqref{app:eq:Petz_map_sqrt_B}.
Indeed, the recovery of $\rho_{AB}$ using Eqs.~\eqref{app:eq:Petz_map_sqrt_A} and \eqref{app:eq:Petz_map_sqrt_B} involves matrix diagonalizations only for the smaller density matrices $\rho_A$, $\rho_B$ and $\rho_{A\cap B}$, which significantly enhances computational efficiency.

\subsection{The projected Petz recovery map}
\label{app:sec:PPRM}

In the approximate time-evolution scheme of LITE, we recover higher-level density matrices from lower-level ones even in the presence of finite (albeit small, see App.~\ref{app:sec:numerical-implementation}) mutual information between subsystems $A$ and $B$.
In the subsystem-lattice framework, this translates to the fact that we recover the density matrices $\rho^{\ell^*+r}_n$ from the density matrices $\rho^{\ell^*}_n$, even though $i^{\ell^*+\ell^\prime} \neq 0$ for $\ell^\prime = 1,\dots,r$.
As a first approximation, we define the recovery map to be implemented in the numerics as
\begin{eqnarray}
\label{app:eq:Petz_numerics}
      \tilde{\rho}_{n+1/2}^{\ell^*+1}= 
    \begin{cases}
       (\rho_n^{\ell^*})^{1/2} ~ (\rho_{n+1/2}^{\ell^*-1})^{-1/2} ~ \rho_{n+1}^{\ell^*} ~ (\rho_{n+1/2}^{\ell^*-1})^{-1/2} ~ (\rho_n^{\ell^*})^{1/2}  & ~~\text{if}~ i_{n}^{\ell^*} < i_{n+1}^{\ell^*} \\
       (\rho_{n+1}^{\ell^*})^{1/2} ~ (\rho_{n+1/2}^{\ell^*-1})^{-1/2} ~ \rho_n^{\ell^*}  ~ (\rho_{n+1/2}^{\ell^*-1})^{-1/2} ~ (\rho_{n+1}^{\ell^*})^{1/2}  & ~~\text{if}~~i_{n}^{\ell^*}> i_{n+1}^{\ell^*},\\
    \end{cases}  
\end{eqnarray}
and if $i^{\ell^*}_n = i^{\ell^*}_{n+1}$ we average over the two choices.
Eq.~\eqref{app:eq:Petz_numerics} can be iterated to obtain $\tilde{\rho}^{\ell^*+r}_n$.
By using Eq.~\eqref{app:eq:Petz_numerics} when $i^{\ell^*+\ell^\prime} \neq 0$ for $\ell^\prime = 1,\dots,r$, we generate erroneous density matrices at level $\ell^* + r$ with uncontrolled error bounds.
The main problem of having such errors is that the density matrices $\tilde{\rho}^{\ell^*+r}_n$ may not preserve the lower-level density matrices; for instance, $\mathrm{Tr}_L^r (\tilde{\rho}^{\ell^*+r}_n) \neq \rho^{\ell^*}_{n+r/2}$.
As a result, errors are introduced on all length scales, causing the algorithm to fail to preserve the local constants of the motion.

To remedy this problem we add a \textit{projection} step, exemplified here for $r=1$.
We compute $\rho^{\ell^*+1}_n$ via the projected Petz recovery map as
\begin{equation}
\label{app:eq:Petz_shift1}
    \rho_{n+1/2}^{\ell^*+1} \coloneqq \tilde{\rho}_{n+1/2}^{\ell^*+1} + \varrho_{n+1/2}^{\ell^*+1},
\end{equation}
where
\begin{equation}
\label{app:eq:Petz_shift2}
     \varrho_{n+1/2}^{\ell^*+1} \coloneqq \left(\rho_n^{\ell^*} - \mathrm{Tr}_R^1( \tilde{\rho}_{n+1/2}^{\ell^*+1} ) \right) \otimes \frac{\mathbb{1}_2}{2}  
     + \frac{\mathbb{1}_2}{2} \otimes \left(\rho_{n+1}^{\ell^*} - \mathrm{Tr}_L^1 ( \tilde{\rho}_{n+1/2}^{\ell^*+1} ) \right) 
    - \frac{\mathbb{1}_2}{2} \otimes \left(\rho_{n+1/2}^{\ell^*-1} - \mathrm{Tr}_L^1\mathrm{Tr}_R^1 ( \tilde{\rho}_{n+1/2}^{\ell^*+1} ) \right) \otimes  \frac{\mathbb{1}_2}{2}.
\end{equation}
Importantly, $ \varrho_{n+1/2}^{\ell^*+1} \to 0$ as $i^{\ell^*}_{n+1/2} \to 0$ and the Petz recovery map becomes exact. 
It is easy to verify that, $\forall~ n$,
\begin{align}
\label{app:eq:PPRM-lower-levels_1}
    \tr^1_R (\rho_{n+1/2}^{\ell^* + 1}) & = \rho_{n}^{\ell^*}, \\
\label{app:eq:PPRM-lower-levels_2}
    \tr^1_L (\rho_{n+1/2}^{\ell^* + 1}) & = \rho_{n+1}^{\ell^*}, \\
\label{app:eq:PPRM-lower-levels_3}
    \tr^r_R (\rho_{n+1/2}^{\ell^* + 1}) & = \tr^r_L (\rho_{n-1/2}^{\ell^*+1}).
\end{align}
The recovered density matrix $\rho_{n+1/2}^{\ell^*+1}$ serves as an approximation to the exact density matrix of the subsystem  $\mathcal{C}_{n+1/2}^{\ell^*+1}$ and it is used to perform the subsystem time evolution on level $\ell^*$.
Note that throughout this work the symbol $\rho$ is used to denote density matrices employed in the time-evolution scheme of LITE.

\section{\label{app:sec:math-details-minimization}Details on the minimization of local information under constraints}

\subsection{\label{app:sec:taylor-expansion} Gradient and Hessian of the von Neumann entropy}

We wish to compute the gradient $\nabla_\rho S$ and the Hessian $\mathscr{H}_\rho $ of the von Neumann entropy at $\rho$, which are defined through the Taylor expansion:
\begin{equation}
\label{eq:entropy_expansion_0}
    S(\rho + \lambda \xi)= S(\rho)+ \mathrm{Tr}\left(\nabla_\rho S \, \lambda \xi \right) + 
\frac{1}{2}\mathrm{Tr}\left(\lambda \xi \,\mathscr{H}_\rho \, \lambda \xi  \right)  + O(\lambda^3),
\end{equation}
where $\lambda \ll 1$ is a small perturbation parameter and $\xi$ is a Hermitian matrix.
For ease of notation, we drop the indices $n$ and $\ell$ in this and the following section.
Let us write 
\begin{equation}
\label{app:eq:density-matrix-eigenvalues}
    \rho = \sum_i \kappa_i \Pi_i
\end{equation}
where $\kappa_i$ are the density-matrix eigenvalues, and $\Pi_i = \vert \psi_i \rangle \langle \psi_i \vert$ are the projectors onto density-matrix eigenstates $\vert \psi_i \rangle$.
Given Eq.~\eqref{app:eq:density-matrix-eigenvalues}, we write the von Neumann entropy as
\begin{equation}
S(\rho + \lambda \xi) = - \sum_i k_i \ln(k_i),
\end{equation}
where $k_i$ are the eigenvalues of the shifted density matrix $\rho+\lambda\xi$, obtained from standard non-degenerate perturbation theory:
\begin{equation}
\label{eq:perturbation_theory_eigvals}
k_i = \kappa_i + \mathrm{Tr}\left(\Pi_i \xi \right) \lambda + \sum_{\substack{i,j\\ i\neq j}} \frac{\mathrm{Tr}\left(\Pi_i \xi \Pi_j \xi \right)}{\kappa_i-\kappa_j} \lambda^2 + O(\lambda^3).
\end{equation}
The Taylor expansion of the von Neumann entropy around $\lambda=0$ reads
\begin{equation}
\label{eq:taylor}
S(\rho +\lambda \xi)= S(\rho)- \sum_i (1+\ln(k_i))\partial_\lambda k_i \bigg \vert_{\lambda=0} \lambda 
 - \frac{1}{2}\sum_i \bigg (\frac{(\partial_\lambda k_i)^2}{k_i} + (1+\ln(k_i) )\partial_\lambda^2 k_i ) \bigg) \bigg \vert_{\lambda=0} \lambda^2 + O(\lambda^3).
\end{equation}
By using Eqs.~\eqref{eq:entropy_expansion_0} and \eqref{eq:perturbation_theory_eigvals} in Eq.~\eqref{eq:taylor}, we find
\begin{equation}
\label{eq:taylor_simplified_0}
S(\rho  + \lambda \xi) =  S(\rho)+ \mathrm{Tr}\left( \nabla_\rho S \, \xi \right) \lambda
- \frac{1}{2}\sum_i \frac{\tilde{\xi}_{ii}^2}{\kappa_i} \lambda^2  - \frac{1}{2} \sum_{\substack{i,j\\ i\neq j}} \frac{\tilde{\xi}_{ij}\tilde{\xi}_{ji}}{\kappa_i-\kappa_j}\ln(\kappa_i) \lambda^2 + O(\lambda^3),
\end{equation}
where 
\begin{equation}
\label{app:eq:entropy-gradient}
    \nabla_{\rho}S = - \ln(\rho) - \mathbb{1}.
\end{equation}
If $\xi$ is traceless, as imposed in the minimization scheme of LITE by the constraints~\eqref{eq:condition1} [$\mathrm{Tr}(\xi) = \mathrm{Tr}(\mathrm{\textbf{P}}\chi) = 0 $], the identity term when inserted in Eq.~\eqref{eq:taylor_simplified_0} vanishes.
Moreover, we defined $\tilde{\xi}_{ij} \coloneqq \langle \psi_i \vert \xi \vert \psi_j\rangle$ and used that
\begin{equation}
\sum_{\substack{i,j\\ i\neq j}} \frac{\tilde{\xi}_{ij}\tilde{\xi}_{ji}}{\kappa_i-\kappa_j}=\sum_i \partial_\lambda^2 k_i= \partial_\lambda^2 \mathrm{Tr}(\rho + \lambda \xi)=0.
\end{equation}
We can further simplify Eq.~\eqref{eq:taylor_simplified_0} by writing 
\begin{equation}
\sum_{\substack{i,j\\ i\neq j}} \frac{\tilde{\xi}_{ij}\tilde{\xi}_{ji}}{\kappa_i-\kappa_j}\ln(\kappa_i)  = \frac{1}{2}\sum_{\substack{i,j\\ i\neq j}} \frac{\tilde{\xi}_{ij}\tilde{\xi}_{ji}}{\delta_{ij}}\ln\left(\frac{1}{2}(\delta_{ij}+\Delta_{ij})\right) - \frac{1}{2}\sum_{\substack{i,j\\ i\neq j}}\frac{\tilde{\xi}_{ij}\tilde{\xi}_{ji}}{\delta_{ij}}\ln\left(\frac{1}{2}(\Delta_{ij}-\delta_{ij})\right),
\end{equation}
where $\delta_{ij}= \kappa_i-\kappa_j$, $\Delta_{ij}= \kappa_i+\kappa_j$, and in the second sum on the right-hand side we have exchanged $i \leftrightarrow j$.
Using $\mathrm{arctanh}(x)= \frac{1}{2}\ln(1+x)-\frac{1}{2}\ln(1-x)$, we arrive at
\begin{equation}
\label{eq:taylor_part2}
\sum_{\substack{i,j\\ i\neq j}} \frac{\tilde{\xi}_{ij}\tilde{\xi}_{ji}}{\kappa_i-\kappa_j}\ln(\kappa_i) = \sum_{\substack{i,j\\ i\neq j}} \frac{\tilde{\xi}_{ij}\tilde{\xi}_{ji}}{\kappa_i-\kappa_j}\mathrm{arctanh}\left(\frac{\kappa_i-\kappa_j}{\kappa_i+\kappa_j}\right).
\end{equation}
We now want to use Eq.~\eqref{eq:taylor_part2} in Eq.~\eqref{eq:taylor_simplified_0}.
Notably, $\lim_{x\rightarrow 0}\frac{\mathrm{arctanh}(x/y)}{x}=\frac{1}{y}$ allows further simplifications.
Then,
\begin{equation}
-\frac{1}{2}\sum_i \frac{\tilde{\xi}_{ii}^2}{\kappa_i}- \frac{1}{2} \sum_{\substack{i,j\\ i\neq j}} \frac{\tilde{\xi}_{ij}\tilde{\xi}_{ji}}{\kappa_i-\kappa_j}\ln(\kappa_i) =- \frac{1}{2} \sum_{ij} \frac{\tilde{\xi}_{ij} \tilde{\xi}_{ji}}{\kappa_i-\kappa_j}\mathrm{arctanh}\left(\frac{\kappa_i-\kappa_j}{\kappa_i+\kappa_j}\right).
\end{equation}
By defining
\begin{equation}
\label{eq:hessian_elements}
    \mathrm{h}_{ij} : =-\frac{\mathrm{arctanh}\left(\frac{\kappa_i-\kappa_j}{\kappa_i+\kappa_j}\right)}{\kappa_i-\kappa_j},
\end{equation}
we arrive at
\begin{equation}
\label{eq:entropy_expansion}
S(\rho +\lambda \xi)= S(\rho)+ \mathrm{Tr}\left(\nabla_\rho S \,  \xi \right) \lambda + 
\frac{1}{2} \sum_{ij} \tilde{\xi}_{ij} \mathrm{h}_{ij} \tilde{\xi}_{ji} \, \lambda^2 + O(\lambda^3).
\end{equation}
The second-order term of the expansion can be rewritten as 
\begin{equation}
\frac{1}{2} \sum_{ij} \tilde{\xi}_{ij} \mathrm{h}_{ij} \tilde{\xi}_{ji} = \frac{1}{2}\mathrm{Tr}\left(\tilde{\xi} \, (\mathrm{h} \star \tilde{\xi})\right),
\end{equation}
where $\star$ represents elementwise multiplication.
Given that $\tilde{\xi}= \sum_{ij} \Pi_i \xi \Pi_j \coloneqq U_\rho^{\dagger} \xi U_\rho$, where $U_\rho $ is the unitary matrix having as columns the eigenvectors of $\rho$ (i.e., $\vert \psi_i \rangle $), and by using the cyclic property of the trace, we obtain 
\begin{equation}
\frac{1}{2} \sum_{ij} \tilde{\xi}_{ij} \mathrm{h}_{ij} \tilde{\xi}_{ji} = \frac{1}{2}\mathrm{Tr}\left(\xi \, U_\rho \left( \mathrm{h} \star U_\rho^{\dagger} \xi U  \right) U_\rho^\dagger \right) = \frac{1}{2}\mathrm{Tr}\left(\xi \mathscr{H}_\rho (\xi)  \right),
\end{equation}
where
\begin{equation}
\label{app:eq:hessian_of_xi}
\mathscr{H}_\rho (\xi) = U_\rho \left( \mathrm{h} \star U_\rho^{\dagger} \xi U \right) U_\rho^{\dagger}
\end{equation}
is the Hessian of the von Neumann entropy at $\rho$, applied on $\xi$.

\subsection{\label{app:sec:newton-method} Newton's optimization and preconditioned conjugate gradient method}

By replacing $\xi = \mathrm{\textbf{P}}\chi$, where $\mathrm{\textbf{P}}$ is the projector defined in Eq.~\eqref{eq:total-projector} of the main text, and using the Hermiticity of the projector $\mathrm{\textbf{P}}$, Eq.~\eqref{eq:entropy_expansion} becomes
\begin{equation}
S(\rho +  \mathrm{\textbf{P}}\chi)= S(\rho)+ \mathrm{Tr}\left( \mathrm{\textbf{P}}\nabla_\rho S \chi \right) + 
\frac{1}{2}\mathrm{Tr}\left(\chi \mathrm{\textbf{P}}\mathscr{H}_\rho  \mathrm{\textbf{P}} \chi \right)  + O(\xi^3).
\end{equation}
Importantly, since $\kappa_i\geq 0$ are the eigenvalues of the density matrix $\rho$ from Eq.~\eqref{eq:hessian_elements} it is clear that all the elements of $\mathrm{h}$ are strictly negative.
Thus, we can apply Newton's method for optimization to find $\chi^*$ such that $S(\rho +  \mathrm{\textbf{P}} \chi^*)$ is maximal.
Formally, the optimal $\chi^*$ is given by
\begin{equation}
 \chi^* = - \left( \mathrm{\textbf{P}} \mathscr{H}_\rho  \mathrm{\textbf{P}} \right)^{+} \mathrm{\textbf{P}} \nabla_\rho S,
\end{equation}
where $(\sbullet)^+$ denotes the pseudoinverse~\cite{ben2003generalized}.
In practice, it is more convenient to solve the related linear equation system
\begin{equation}
\label{eq:conj_grad}
    \mathrm{\textbf{P}} \mathscr{H}_\rho  \mathrm{\textbf{P}} \, \chi^* = - \mathrm{\textbf{P}} \nabla_\rho S,
\end{equation}
by virtue of numerical methods such as the \textit{preconditioned conjugate gradient method}~\cite{barrett1994templates}.
To ensure fast convergence, we multiply Eq.~\eqref{eq:conj_grad} from the left by a preconditioning matrix $Q$ approximately equal to the pseudoinverse $ \left( \mathrm{\textbf{P}} \mathscr{H}_\rho  \mathrm{\textbf{P}} \right)^+ $.
Indeed, while computing the pseudoinverse of $\mathrm{\textbf{P}} \mathscr{H}_\rho  \mathrm{\textbf{P}}$ is nontrivial, the computation of the following superoperator is straightforward:
\begin{equation}
\label{eq:inverse_hessian}
\mathscr{H}_\rho^{-1}\, \sbullet = U_\rho \left(\mathrm{h}^{\star -1} \star U_\rho^{\dagger} \sbullet U_\rho \right) U_\rho^{\dagger},
\end{equation}
where $\mathrm{h}^{\star -1}$ is the elementwise inverse of the matrix $\mathrm{h}$ [see Eq.~\eqref{eq:hessian_elements}].
Hence, we use as preconditioning matrix
\begin{equation}
\label{app:eq:precondition}
    Q=\mathrm{\textbf{P}} \mathscr{H}_\rho^{-1} \mathrm{\textbf{P}}.
\end{equation}

In the numerical simulations, to minimize the information at level $\ell_{\mathrm{min}}$, we solve Eq.~\eqref{eq:conj_grad} with precondition~\eqref{app:eq:precondition} for each density matrix $\rho_n^{\ell_{\mathrm{min}}}$ individually. 
Since the lower-level density matrices on $\ell < \ell_\mathrm{min}$ and information currents on $\ell \leq \ell_\mathrm{min}$ are kept fixed, the order in which we minimize does not influence the result.
We initialize $\chi_n^{\ell_{\mathrm{min}}}$ as the difference between the density matrix obtained from a projected Petz recovery map (see App.~\ref{app:sec:PPRM}) from lower levels and the density matrix $\rho_n^{\ell_{\mathrm{min}}}$ before minimization.
We keep $\rho_n^{\ell_{\mathrm{min}}}$ fixed, and consequently $\mathscr{H}_\rho$ and $\nabla_\rho S$, while we iteratively update $\chi^{\ell_\mathrm{min}}_n$ until Eq.~\eqref{eq:conj_grad} is satisfied.
Since the conjugate gradient method is unstable against small perturbations, it is impossible in practice to converge within a desired tolerance $w$. 
To solve this issue, we modify our iterative scheme by adding a damping factor as
\begin{equation}
    (\chi_n^{\ell_{\mathrm{min}}})_{j+1} = (\chi_n^{\ell_{\mathrm{min}}})_{j} + \varepsilon \biggl( (\chi_n^{\ell_{\mathrm{min}}})_{j+1} - (\chi_n^{\ell_{\mathrm{min}}})_{j} \biggr),
\end{equation}
where $j$ is the iteration index.
If $\varepsilon=1$, the damping is not present; if $\varepsilon=0$, the density matrix is never updated.
The optimal value of $\varepsilon$ to converge in the minimal number of iterations should be empirically found on a case-by-case basis.
Here, we set $w = 10^{-5}$ and $\varepsilon=0.9$.

\section{Details on the numerical implementation}
\label{app:sec:numerical-implementation}

As described in Sec.~\ref{sec:approximate_time_evolution} in the main text, the LITE algorithm comprises two fundamental stages.
First, it entails the time evolution of subsystems by means of the projected Petz recovery maps (see Appendix~\ref{app:sec:PPRM}).
During this time evolution, the subsystem scale $\ell^*$ on which time evolution is performed dynamically increases until it reaches the predefined maximum scale $\ell_{\mathrm{max}}$.
Second, once $\ell^* = \ell_\mathrm{max}$ and information has accumulated at $\ell_\mathrm{max}$, information is minimized at scale $\ell_{\mathrm{min}}$.
The minimization of information ensures that density matrices at levels $\ell < \ell_\mathrm{min}$ and the global information current from lower levels to $\ell_\mathrm{min}$ remain fixed.
In this appendix, we revisit these steps from the perspective of their numerical implementation.

\subsection{Asymptotically time-invariant initial states}
\label{app:sec:asymptotic-initial-states}

\begin{figure}
    \centering
    \includegraphics[scale=0.6]{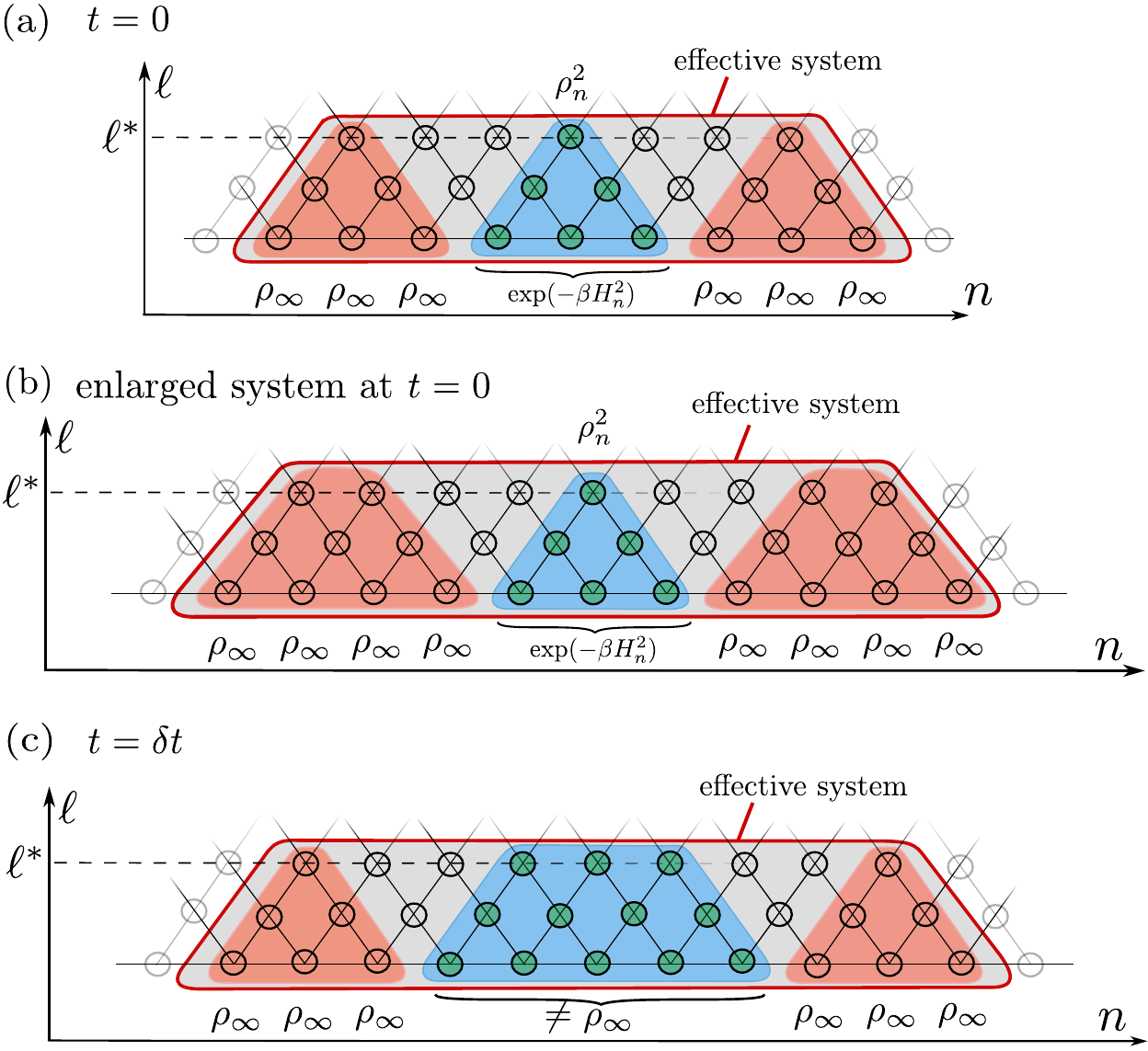}
    \caption{Schematic of the effective system used in the numerical simulations with asymptotically time-invariant initial states and $r=1$. The color intensity of the information-lattice points quantifies the local information located on a given site. (a) The effective system at $t=0$ comprises a central region (blue area) deviating from the infinite-temperature state and containing local information (as indicated by the green dots). On each boundary, there are $\ell_\mathrm{init}+1$ infinite-temperature physical sites (red areas). If $i^{\ell_\mathrm{init}}_n < q_{\ell^*}$ we set $\ell^*=\ell_\mathrm{init}$ (otherwise, we need to extend the effective system to higher levels, see Fig.~\ref{app:fig:update_ell}). (b) Before performing a time-evolution step, we enlarge the effective system by adding $r$ infinite-temperature physical sites at each boundary. (c) We perform a time-evolution step of length $\delta t$ by numerically integrating the subsystem von Neumann equations (Eq.~\eqref{eq:subsys_vonNeumann} in the main text). We then check whether after the time step some physical sites that were in the red areas deviate from the infinite-temperature state. We remove unnecessary physical sites at the boundaries of the effective system in order to always keep $\ell^*+1$ infinite-temperature physical sites at the boundaries.}
    \label{app:fig:enlarge_system}
\end{figure}

The LITE approach for time evolution is applicable across a wide range of initial states. 
For example, it can be used for studying the transport properties of finite-size systems initialized in specific finite-temperature states, such as domain-wall states.
Furthermore, in scenarios involving translation-invariant Hamiltonians, the method can be efficiently used for performing time evolution starting from translation-invariant initial states, as explored in Ref.~\onlinecite{klein2022time} for diffusive systems.
Nonetheless, the focus of this work lies in the development of a comprehensive framework for capturing the time evolution of generic Hamiltonians.
This inclusivity encompasses non-translation-invariant and disordered Hamiltonians.
Consequently, it becomes convenient to initialize the dynamics in states that commute with the Hamiltonian asymptotically in space. 
As, by construction, the state of the asymptotic region does not change in time, we refer to these states as \textit{asymptotically time invariant}.
This feature allows us to simulate infinitely extended systems, as detailed in the following.
Thanks to the absence of boundaries, one can obtain a robust extrapolation of the time-dependent behavior of diffusion coefficients, as the propagation of conserved quantities over time and space remains unaffected by boundary reflections.

Out of the family of asymptotically time-invariant states, we consider those given in Eq.~\eqref{eq:init_state} in the main text, reproduced here for completeness
\begin{equation}
\label{app:eq:initial-state-1}
    \rho_\mathrm{init} = \left (\bigotimes_{m<n-\ell/2} \rho_{m,\infty} \right)~ \otimes \mathcal{\rho }_{n,\mathrm{init}}^\ell \otimes ~ \left( \bigotimes_{m>n+\ell/2} \rho_{m,\infty}\right),
\end{equation}
where $\rho_{m,\infty}=\mathbb{1}_{d}/d$ is the infinite-temperature single-site density matrix located on the physical site $m$ with Hilbert space dimension $d$.
In particular, we can choose (as introduced in the main text, Eq.~\eqref{eq:init_state_thermal_part})
\begin{equation}
\begin{split}
\label{app:eq:initial-state-2}
    \rho_{n,\mathrm{init}}^\ell= \frac{1}{Z}\exp \left( - \beta H_n^\ell\right), \\
    Z = \mathrm{Tr}\left(\exp\left(-\beta H_n^\ell\right)\right).
\end{split}
\end{equation}
Since infinite-temperature density matrices are invariant under time evolution (as can be easily seen from Eq.~\eqref{eq:subsys_vonNeumann} in the main text), by initializing the system in the state~\eqref{app:eq:initial-state-1}, we only need to solve the subsystem time evolution for those subsystems in the central region whose state deviates from the infinite-temperature density matrix.
This sets an \textit{effective system} on which time evolution is performed. 
Since information spreads over time, the effective region needs to be enlarged during time evolution.

To avoid any boundary-induced distortions in the dynamics, we perform the following procedure.
We initialize the effective system at $t=0$ as composed of $3(\ell_\mathrm{init}+1)$ physical sites, where $\ell_\mathrm{init}$ is the correlation scale of $\rho^{\ell_\mathrm{init}}_{n,\mathrm{init}}$ deviating from the infinite-temperature state.
As depicted in Fig.~\ref{app:fig:enlarge_system}(a), while the central region (blue area) is in the state $\rho^{\ell_\mathrm{init}}_{n,\mathrm{init}} \neq \mathbb{1}_{d^{\ell_\mathrm{init}+1}}/d^{\ell_\mathrm{init}+1}$ and contains local information (green filled circles), the boundary states (red area) are at infinite temperature and do not contain local information (open circles).
The states in the gray areas are obtained by recovery maps of lower-level density matrices.
Given the presence of the central region deviating from infinite temperature, they are, in general, not thermal.
Let us assume that local information on sites $(n,\ell_\mathrm{init})$ is smaller than the threshold value $q_{\ell^*}$ (see App.~\ref{app:sec:runge-kutta}), such that we can perform time evolution on level $\ell^* = \ell_\mathrm{init}$.
This assumption makes the following discussion valid for any level $\ell^*$ on which time evolution is performed and for any extent of the effective system.
Fig.~\ref{app:fig:enlarge_system}(b) shows that, before performing the time-evolution step, we enlarge the effective system by $r$ infinite-temperature physical sites at each boundary.
This ensures that we can recover all the necessary density matrices on level $\ell^* + r$ and perform time evolution on all subsystems on level $\ell^*$ that, after the time-evolution step, can deviate from the infinite-temperature state.
Notice that the additional higher-level density matrices up to level $\ell^*$ are infinite-temperature density matrices by construction.
We perform the time-evolution step at $\ell^*$ as prescribed by Eq.~\eqref{eq:subsys_vonNeumann} in the main text.
We then check whether at the later time $\delta t$ some physical sites that were in the red areas at $t=0$ are in a state that deviates from the infinite temperature one, as sketched in Fig.~\ref{app:fig:enlarge_system}(c) by the broadening of the blue area.
If so, we need to enlarge the effective system. 
Note that at most $r$ physical sites may deviate from the infinite-temperature state after a time-evolution step. 
To prevent any boundary-induced distortion, we keep $\ell^*+1$ physical sites at infinite temperature at each end of the effective system.
Therefore, if needed we remove excess infinite-temperature physical sites at the boundaries.
Numerically, we determine whether a physical-site state is at infinite temperature by computing the norm of the difference between the actual state and $\mathbb{1}_{d}/d$ and check if it is smaller than $p$.
We set $p=10^{-12}$.

\subsection{Runge-Kutta integration scheme}
\label{app:sec:runge-kutta}

\begin{figure}
    \centering
    \includegraphics[scale=0.6]{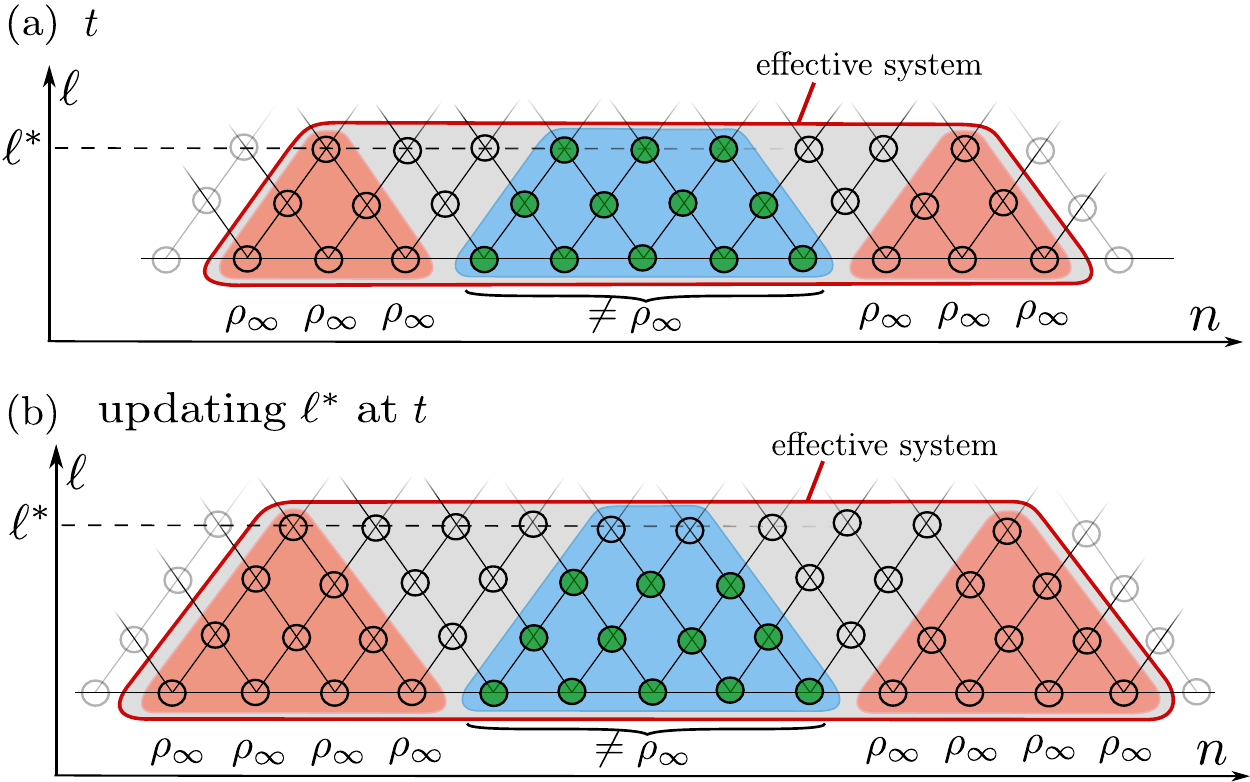}
    \caption{Schematic of the effective system used in the numerical simulations with asymptotically time-invariant initial states and $r=1$ (a) before and (b) after updating $\ell^*$.
    The intensity of the green color of the information-lattice points quantifies the local information located on a given site. 
    To update the effective system in the case $i^{\ell^*}_n > q_{\ell^*}$ for any $n$, as sketched by the green dots in (a), we add $r$ infinite-temperature physical sites at each boundary and increase $\ell^*$ by $r$, as sketched in (b).
    }
    \label{app:fig:update_ell}
\end{figure}

We time evolve each subsystem density matrix $\rho^{\ell^*}_n$ by integrating the subsystem's equation of motion [see Eq.~\eqref{eq:subsys_vonNeumann} in the main text] by means of Runge-Kutta integration methods:
\begin{equation}
    \rho_{n}^{\ell^*}(t+\delta t) = \rho_{n}^{\ell^*}(t)+\delta t\sum_{i=1}^{K}b_{i}\kappa_{n}^{\ell^*,{i}} + O(\delta t^{K}).
\end{equation} 
Here the $b_{i}$ are the Runge-Kutta parameters, the $\kappa_{n}^{\ell^*,{i}}$
are the derivative functions given by Eq.~\eqref{app:eq:equation-of-motion} evaluated for different density matrices~\cite{feagin2012high}, and $K$ is the order of the truncation error.
We use an adaptive fifth-order Runge-Kutta method with an embedded fourth-order method used to estimate the time-step error~\cite{fehlberg1969low}.
Such adaptive methods are characterized by a dynamic time step set by the time-step error $\epsilon$; we set the time-step error to be $\epsilon<10^{-8}$.

Runge-Kutta methods, although particularly well suited for the LITE approach since they preserve the local constants of motion (see App.~\ref{app:sec:local-conserved-quantities}), can fail due to the presence of small eigenvalues of $\rho^{\ell^*}_n(t)$.
In fact, in this case, the matrices needed to compute the derivative functions $\kappa_n^{\ell^*,i}$ can have negative eigenvalues.
Since the subsystem time evolution~\eqref{eq:subsys_vonNeumann} is only defined for positive semidefinite matrices, this can cause the failure of the Runge-Kutta integration step.
The smallest eigenvalues typically increase when there is a flow of information from small to large scales. 
Therefore, this failure is more likely to occur early in the time evolution or when there is suppressed flow of information from small to large scales (as might happen in disordered systems).

In the simulations presented in this article for the asymptotically time-invariant initial states in Eqs.~\eqref{app:eq:initial-state-1} and \eqref{app:eq:initial-state-2}, small eigenvalues can be remedied by shifting the density matrices to be time evolved, $\rho^{\ell^*}_n$, by the infinite-temperature state multiplied by a factor $\alpha$:
\begin{equation}
\label{app:eq:shift}
    \rho^{\ell^*}_{n,\mathrm{shift}}(t) = \frac{1}{1 + \alpha} \left( \rho^{\ell^*}_n(t) + \frac{\alpha}{d^{\ell^*+1}} \mathbb{1}_{d^{\ell^*+1}} \right).
\end{equation}
In fact, given that the boundaries of the system are in the infinite-temperature state by construction, such a shift does not create additional local information and clearly removes close-to-zero eigenvalues, thus improving numerical stability.
After shifting the density matrices, we perform the projected Petz recovery map and the Runge-Kutta time-evolution step.
At the later time $t + \delta t$, we shift back the density matrices:
\begin{equation}
    \rho^{\ell^*}_{n}(t+\delta t) = \left( 1 + \alpha \right) \rho^{\ell^*}_{n,\mathrm{shift}}(t+\delta t) - \frac{\alpha}{d^{\ell^*+1}} \mathbb{1}_{d^{\ell^*+1}}.
\end{equation}

Unfortunately, there is no general guarantee that, by performing the projected Petz recovery map and the Runge-Kutta time-evolution step on the shifted density matrices in Eq.~\eqref{app:eq:shift}, one conserves the same convergence properties as with the unshifted matrices.
In particular, the shift should be chosen on a case-by-case basis depending on the properties of the state of the system.
For instance, for an asymptotically time-invariant initial state in which the boundary state is the Gibbs state (i.e., $\rho_\mathrm{asymp} \propto \exp{(-\beta H^{\ell_\mathrm{init}}_n)}$), a convenient choice would be to substitute in Eq.~\eqref{app:eq:shift} the identity matrix with $\exp{(-\beta H^{\ell_\mathrm{init}}_n)}$.
We note that the positivity issue is not present for other integration schemes, such as the Suzuki-Trotter decomposition~\cite{klein2022time}, which, however, has the important drawback of not preserving the local constants of the motion.

After each time step, we evaluate whether to dynamically update the level $\ell^*$ at which time evolution is performed.
If any subsystem state at level $\ell^*$ has acquired more information than a threshold value $q_{\ell^*}$ we increase $\ell^*$ by $r$.
In this work, we set $q_{\ell^*} = 10^{-10}$.
As depicted in Fig.~\ref{app:fig:update_ell}, when using an asymptotically time-invariant initial state, this implies increasing the level $\ell^*$ by $r$ and enlarging the effective system by $r$ physical sites on both boundaries.

\subsection{Minimization of local information}
\label{app:sec:minimization}

The minimization of local information is performed when two conditions are satisfied: (i) the scale on which time evolution is performed $\ell^*$ equals $\ell_\mathrm{max}$; (ii) the total amount of information on scale $\ell_\mathrm{max}$, $i^{\ell_\mathrm{max}}= \sum_n i^{\ell_\mathrm{max}}_n$, exceeds a threshold percentage value, $q_{\mathrm{max}}$, of the total information in the system, $i_{\mathrm{tot}} = \sum_{\{\mathrm{all~lattice~points}\}} i_n^{\ell}$.
The minimization is performed on each subsystem-lattice site $(n,\ell_\mathrm{min})$ individually.
Because of the conservation of lower-level density matrices and currents, changing the order of the sites on which minimization is performed does not alter the result.
After minimization, the effective-system level is reduced to $\ell_\mathrm{min}$; time evolution is then performed on $\ell_\mathrm{min}$.
A pseudocode for the minimization is given in Algorithm~\ref{app:alg:minimization}.

\begin{Algorithm}
\begin{center}
  \begin{varwidth}{\linewidth}
\begin{codebox}
  \Procname{$\proc{Minimization}(\{ \rho_n^{\ell_\mathrm{max}}\},q_{\mathrm{max}})$}
     \li $i_{\mathrm{tot}} = \sum_{\{\mathrm{all~lattice~points}\}} i_n^{\ell}$
    \Comment compute total information in the system
    \li  $i^{\ell_\mathrm{max}} = \sum_{n} i_n^{\ell_{\mathrm{max}}}$
    \Comment compute information at level $\ell_{\mathrm{max}}$
    \li \If $ i^{\ell_\mathrm{max}} \geq q_{\mathrm{max}} \, i_{\mathrm{tot}}$ :
    \li \Do $ \{\rho_n^{\ell_{\mathrm{min}}}\} = \proc{Reduce-To-$\ell_\mathrm{min}$}( \{\rho_n^{\ell_{\mathrm{max}}}\} )$ 
    \Comment obtain the density matrices at level $\ell_{\mathrm{min}}$ by partial trace
    \li $ \{\rho_n^{\ell_{\mathrm{min}}}\}  \leftarrow \proc{Minimize}( \{\rho_n^{\ell_{\mathrm{min}}}\} )$ 
    \Comment minimize the information of each density matrix at level $\ell_{\mathrm{min}}$
    \li \Return $ \{\rho_n^{\ell_{\mathrm{min}}}\} $
    \li \Else :
    \li \Return $ \{\rho_n^{\ell_{\mathrm{max}}}\} $
\end{codebox}
  \end{varwidth}
\end{center}
\caption{Pseudo-code for the minimization of local information.
We use the notation $\{ \rho_n^{\ell} \}$ to indicate the set of the subsystem density matrices at fixed $\ell$ and $\forall~ n$.
We pass to the function $\proc{Minimization}$ all the density matrices at level $\ell_\mathrm{max}$ and the numerical parameter $q_\mathrm{max}$ (notice that $q_\mathrm{max}$ is set at the beginning of the simulation and never changes).
First, the total information in the system and on level $\ell_\mathrm{max}$ are computed.
If the information on level $\ell_\mathrm{max}$, $i^{\ell_\mathrm{max}}$, exceeds $q_\mathrm{max}$ times the total information in the system, $i_\mathrm{tot}$, \proc{Reduce-To-$\ell_\mathrm{min}$} is called.
This function finds all density matrices at level $\ell_\mathrm{min}$ by suitable partial-trace operations on the ones at level $\ell_\mathrm{max}$.
In addition, it removes excess infinte-temperture physical sites in order to keep only $\ell_\mathrm{min}+1$ infinite-temperature physical sites at each boundary of the effective system.
Then, \proc{Minimize} minimizes the information for each density matrix at level $\ell_\mathrm{min}$ while keeping all lower-level density matrices (on $\ell < \ell_\mathrm{min}$) and information currents (on $\ell \leq \ell_\mathrm{min}$) fixed.}
\label{app:alg:minimization}
\end{Algorithm}

\subsection{Flow of the LITE algorithm}

The flow of the two-level scheme of the LITE approach for time evolution is detailed in Algorithm~\ref{app:alg:LITE}.
In addition, in Table~\ref{tab:parameters} we summarize the numerical parameters and the numerical values employed in this work.

\begin{Algorithm}
\begin{center}
  \begin{varwidth}{\linewidth}
\begin{codebox}
  \Procname{$\proc{Time-evolve}(\{ \rho_n^{\ell^*}\},q_{\ell^*}, p, q_{\mathrm{max}})$}
  \li $\proc{Enlarge-System}( \{ \rho_n^{\ell^*}\} )$
   \Comment enlarge the system by $r$ sites on each end
  \li $ \{ \rho_n^{\ell^*} \}\leftarrow  \{ (\rho^{\ell^*}_n + \alpha \, \mathbb{1}_{d^{\ell^*+1}}/d^{\ell^*+1}) / (1 + \alpha) \}$
  \Comment shift all density matrices at level $\ell^*$ to remove close-to-zero eigenvalues
    \li $\{ \rho_{n+r/2}^{\ell^*+r} \} = \proc{PPRM}(\{\rho_n^{\ell^*}\})$
    \Comment compute all density matrices at level $\ell^* + r $ via projected Petz recovery maps
    \li \For $(n,\ell^*) \in \{ (n,\ell^*) \} $ : 
      \li \Do
         $\rho_n^{\ell^*} \leftarrow  $ $\proc{Runge-Kutta}( \rho_n^{\ell^*}, \rho_{n-r/2}^{\ell^*+r}, \rho_{n+r/2}^{\ell^*+r} )$ 
         \Comment perform a time-evolution step for all density matrices at level $\ell^*$
    \End
    \li $\proc{Compute-Observables}( \{ \rho_n^{\ell^*} \}  )$
    \li  \{ $\rho_n^{\ell^*} \}\leftarrow \{ ( 1 + \alpha ) \rho^{\ell^*}_{n} - \alpha \, \mathbb{1}_{d^{\ell^*+1}} / d^{\ell^*+1} \}$     \Comment unshift all density matrices at level $\ell^*$
    \li $\proc{Remove-Additional-Sites}( \{ \rho_n^{\ell^*}\},p )$
    \Comment remove unnecessary infinite-temperature physical sites at the boundaries
     \li \If (any $i_n^{\ell^*} > q_{\ell^*} $ \textbf{and}  $\ell^* \neq \ell_{\mathrm{max}}$): 
    \li \Do  $\ell^* \leftarrow \ell^* +r $ 
    \Comment update $\ell^*$ to $\ell^* +r$ 
    \li $ \{ \rho_n^{\ell^*+r} \} \leftarrow \proc{PPRM}( \{\rho_n^{\ell^*} \} )$ \Comment compute all density matrices at level $\ell^* +r$ via projected Petz recovery maps
    \li \Return  $ \{ \rho_n^{\ell^*+r} \} $
    \li \ElseIf $\ell^* = \ell_{\mathrm{max}}$:
        \li \Do \Return $\proc{Minimization}(\{\rho_n^{\ell_\mathrm{max}}\},q_{\mathrm{max}})$ \Comment check whether to minimize information at level $\ell_\mathrm{min}$
    \li \Else :
    \li \Return  $ \{ \rho_n^{\ell^*} \} $
\end{codebox}
  \end{varwidth}
\end{center}
\caption{Pseudo-code for the two-level scheme of LITE.
We use the notation $\{ \rho_n^{\ell} \}$ and $\{ (n,\ell) \}$ to indicate the set of the subsystem density matrices and subsystem labels at fixed $\ell$ and $\forall~ n$.
$\proc{Enlarge-System}$ and $\proc{Remove-Additional-Sites}$ take care of enlarging the effective system and removing unnecessary infinite-temperature physical sites at the boundaries, respectively (see Fig.~\ref{app:fig:enlarge_system}).
$p$ is a threshold parameter used when comparing the density matrices of the physical sites at the boundaries of the effective system to infinite-temperature density matrices (see text).
$\proc{PPRM}$ applies projected Petz recovery maps to recover higher-level density matrices.
$\proc{Runge-Kutta}$ performs one time-evolution step by numerically integrating the subsystems' equations of motion.
$\proc{Compute-Observables}$ computes relevant physical observables on the time-evolved density matrices.
Finally, if any information-lattice point possesses local information higher than $q_{\ell^*}$ and the time-evolution level $\ell^*$ is smaller than $\ell_\mathrm{max}$, the effective system is enlarged and updated to level $\ell^*+r$ (see Fig.~\ref{app:fig:update_ell});
if $\ell^*=\ell_\mathrm{max}$ the function $\proc{Minimize}$ is called and checks whether the minimization at level $\ell_\mathrm{min}$ should be performed; if neither of these situations is satisfied, the density matrices at level $\ell^*$ are returned.}
\label{app:alg:LITE}
\end{Algorithm}

\begin{table}[]
    \centering
    \begin{tabular}{c c c c}
    \hline \hline
    Parameter & Description & Prescription & Numerical value \\
    \hline
         $\ell_\mathrm{min}$ & Minimization level & As large as possible & 3,4,\dots,8 \\
         $\ell_\mathrm{max}$ &  Maximum level at which time evolution is performed & As large as possible ($\ell_\mathrm{max} > \ell_\mathrm{min}$) & 4,5,\dots,9 \\
         $q_{\mathrm{max}}$ & Information threshold to activate the minimization & On a case-by-case basis & $0.5$\% -$2$\% \\
         $w$ & Convergence tolerance of the minimization & As small as possible & $10^{-5}$\\
         $\varepsilon$ & Damping factor of the minimization & On a case-by-case basis & $0.9$ \\
         $p$ & Threshold to enlarge the effective system & As small as possible  & $10^{-12}$ \\
         $q_{\ell^*}$ & Threshold to update $\ell^*$ & As small as possible & $10^{-10}$ \\
         $\epsilon$ & Runge-Kutta error &  As small as possible  & $< 10^{-8}$ \\
         \hline
    \end{tabular}
    \caption{Numerical parameters of the LITE algorithm.
    These parameters are set at the beginning of the simulation and never change.
    }
    \label{tab:parameters}
\end{table}

\subsection{Parallelization}

The numerically costly operations of the LITE algorithm are three: the projected Petz recovery map (see App.~\ref{app:sec:PPRM}), the time-evolution step via the adaptive Runge-Kutta integration method (see App.~\ref{app:sec:runge-kutta}), and the local-information minimization on level $\ell_\mathrm{min}$ (see App.~\ref{app:sec:minimization}).
All these operations are based on matrix-diagonalization or matrix-multiplication routines; thus, they significantly profit from OpenMP parallelization~\cite{dagum1998openmp}.
Since these operations have to be carried out for each density matrix individually, it is beneficial to additionally use an MPI multiprocessing approach~\cite{gabriel2004openmpi}.
However, notice that different parts of the algorithm require different inputs: for instance, assuming $r=1$, to recover via the projected Petz map the density matrix at subsystem lattice site $(n,\ell^*+1)$, we require the lower-level density matrices at $(n-1/2, \ell^*)$ and $(n+1/2,\ell^*)$; whereas to numerically solve the subsystem time-evolution equation at $(n-1/2,\ell^*)$, we require the higher-level density matrices at both $(n-1,\ell^*+1)$ and $(n,\ell^*+1)$. 
Thus, different (independent) MPI processes have to be interfaced and exchange density matrices throughout the algorithm execution.

%###########################################################

\end{document}